\documentclass[a4paper,11pt]{article}
\usepackage[utf8]{inputenc}
\usepackage{amsmath, amssymb}
\usepackage{graphicx}
\usepackage{geometry}
\usepackage{pdflscape} 
\usepackage{array}
\usepackage[version=4]{mhchem} 
\usepackage[colorlinks=true, urlcolor=blue, linkcolor=blue, citecolor=blue]{hyperref}

\usepackage{booktabs} 
\usepackage[super,sort&compress,comma]{natbib} 
\usepackage{multirow}
\usepackage{longtable} 
\usepackage{amssymb}   
\usepackage{xcolor}    

\usepackage{svg}       
\setsvg{
    inkscapeexe={"C:/Program Files/Inkscape/bin/inkscape.exe"},
    inkscapeversion=1,  
    inkscapeopt={-D} 
}
\usepackage{float}     

\newcommand{\stb}{\textcolor{green!60!black}{\textbf{Stable}}}
\newcommand{\qst}{\textcolor{orange}{\textbf{Quasi}}}
\newcommand{\ust}{\textcolor{red}{\textbf{Unstable}}}

\usepackage{microtype} 
\DeclareUnicodeCharacter{8201}{} 

\title{VibroML: an automated toolkit for high-throughput vibrational analysis and dynamic instability remediation of crystalline materials using machine-learned potentials}

\author{
    Rogério Almeida Gouvêa$^{1,*}$ \and Gian-Marco Rignanese$^{1,2,*}$
}

\date{%
    \small
    $^{1}$Institute of Condensed Matter and Nanosciences, Université Catholique de Louvain, Louvain-la-Neuve, Belgium.\\
    $^{2}$WEL Research Institute, Avenue Pasteur 6, Wavre, Belgium.\\[2ex]
    *Corresponding authors: \texttt{rogeriog.em@gmail.com}; \texttt{gian-marco.rignanese@uclouvain.be}\\
}

\begin{document}
\maketitle
\begin{abstract}
    While machine-learned interatomic potentials (MLIPs) accelerate phonon dispersion calculations, merely identifying dynamical instabilities in computationally predicted materials is insufficient; automated pathways to resolve them are required. We introduce VibroML, an open-source Python toolkit driven by foundational MLIPs that shifts the paradigm from stability verification to automated structural remediation. VibroML employs an energy-guided genetic algorithm that vastly outperforms traditional soft-mode following, efficiently navigating the potential energy surface to uncover diverse, dynamically stable polymorphs. As 0~K harmonic stability does not guarantee macroscopic viability, an automated molecular dynamics workflow evaluates finite-temperature structural retention. VibroML also couples with ProtoCSP, our combinatorial structure prediction engine, to stabilize frustrated crystal topologies via targeted alloying, successfully rescuing functional perovskite networks like \ce{Cs2KInI6} and \ce{KTaSe3}. Demonstrating broader applicability, we mined the Alexandria database—where $\sim$50\% of quaternary and 99.5\% of quinary elemental combinations lack any structural entries—to identify thousands of abandoned, high-symmetry stoichiometries. Deploying ProtoCSP's `cold start' retrieval and VibroML's evolutionary search on a sample, we successfully identified dynamically stable low-symmetry candidates. Through integrated structural remediation, thermal validation, and systematic compositional exploration, VibroML enables a comprehensive deep-screening approach, yielding physically sound structural propositions that far surpass standard high-throughput workflows.
\end{abstract} 

\textbf{Keywords:} materials discovery, polymorph mapping, dynamical stability, machine-learned interatomic potentials (MLIPs), high-throughput screening, soft mode displacement, phase transition pathways, molecular dynamics 

\section{Introduction}

The discovery of new materials with specific properties is a key driver of innovation for many technologies, from renewable energy to electronics. Historically, this process was slow and relied on experimental trial and error. Today, computational materials science and high-throughput screening (HTS) have changed this, allowing researchers to quickly analyze a vast number of potential materials. This approach has significantly shortened the materials development cycle that typically takes decades~\cite{shahzadAcceleratingMaterialsDiscovery2024,joressApplicationsHighThroughput2022}. This revolution has been fueled by increased computing power and accurate methods like Density Functional Theory (DFT), which enables the automated calculation of material properties for thousands of candidates in a single study~\cite{chenAcceleratingComputationalMaterials2024}. However, for a computationally predicted material to be considered a viable candidate for experimental synthesis, it must satisfy fundamental principles of stability. While thermodynamic stability (often approximated by a low formation energy relative to competing phases) is a necessary condition, it is not sufficient. A material must also be dynamically stable, meaning its crystal lattice must be resilient to small atomic perturbations~\cite{bartelReviewComputationalApproaches2022,rudinGeneralizationSoftPhonon2018}. This property is governed by the collective vibrational modes of the lattice, known as phonons.

The presence of phonon modes with imaginary frequencies (unstable or ``soft'' modes) at any point in the Brillouin zone signifies a dynamical instability; the structure will spontaneously distort along the eigenvector of this mode to find a new, lower-energy configuration. Therefore, a complete phonon dispersion calculation is an essential check for any crystal structure~\cite{bartelReviewComputationalApproaches2022,xiaoHighthroughputComputationalScreening2025,griesemerAcceleratingPredictionStable2023}. Neglecting this check can lead to a high rate of ``false positives'' in high-throughput screening (HTS) campaigns, where thermodynamically favorable but dynamically unstable structures are erroneously identified as promising candidates.

A major obstacle in incorporating dynamical stability evaluation in HTS is the significant computational expense of ab-initio phonon calculations. Whether using the finite-displacement supercell approach or Density Functional Perturbation Theory (DFPT), these gold-standard DFT methods scale poorly for large screening campaigns. Machine-learned interatomic potentials (MLIPs) effectively mitigate this bottleneck by acting as fast and accurate surrogates~\cite{xiaoHighthroughputComputationalScreening2025,shiotaUniversalNeuralNetwork2024}. Trained on vast databases of high-fidelity DFT data, including energies, forces, and stresses, these models learn to replicate the quantum mechanical potential energy surface (PES). Once trained, an MLIP can perform predictions with near-DFT accuracy at a computational cost that is orders of magnitude lower, approaching the speed of classical force fields~\cite{wangLeadFreeAntimonyBasedPerovskite2024}. The key recent advancement in MLIP architecture is the integration of E(3)-equivariance, wherein the network is built using operations like tensor products of spherical harmonics. These operations are mathematically guaranteed to respect the symmetries of 3D Euclidean space, thus enforcing this physical constraint directly within the model~\cite{batznerE3equivariantGraphNeural2022}. Equivariant models were shown to be capable of reaching state-of-the-art accuracy while being trained on datasets that were up to three orders of magnitude smaller than those required by previous invariant models~\cite{batznerE3equivariantGraphNeural2022,batatiaMACEHigherOrder2022}.

The architectural breakthroughs in E(3)-equivariant MLIPs, particularly their remarkable data efficiency and expressive power, set the stage for the next major paradigm shift in the field: the development of ``foundation models'' for atomistic simulation~\cite{choiPerspectiveFoundationModels2025}. This progression is exemplified by influential models such as the pioneering NequIP~\cite{batznerE3equivariantGraphNeural2022}, the scalable, strictly local design of Allegro~\cite{musaelianLearningLocalEquivariant2023}, and the MACE architecture~\cite{batatiaMACEHigherOrder2022}, which improved speed and accuracy by simplifying higher body-order message passing. Furthermore, recent state-of-the-art eSEN and UMA models~\cite{fuLearningSmoothExpressive2025, woodUMAFamilyUniversal2025} from FAIRChem, address the critical need for energy conservation in molecular dynamics simulations. The application of foundational MLIPs has been extensively validated for calculating phonon properties and thermal transport, where they have consistently demonstrated the ability to reproduce DFT-level results with outstanding efficiency~\cite{fuLearningSmoothExpressive2025,arabhaRecentAdvancesLattice2021,loewUniversalMachineLearning2025,guoLatticeDynamicsModeling2025,batatiaFoundationModelAtomistic2025, hanBenchmarkingUniversalMachine2025}, although fine-tuning might still be required for certain materials and applications~\cite{choiPerspectiveFoundationModels2025,radovaFinetuningFoundationModels2025,liuFineTuningUniversalMachineLearned2025}.

The need for high-throughput dynamical stability assessment has never been more urgent, driven by the recent explosion of candidate structures produced by massive computational screenings and advanced generative artificial intelligence. For instance, high-throughput compositional surveys leveraging MLIPs can now relax tens of millions of crystal structures, creating vast pools of thermodynamically favorable but dynamically unverified candidates \cite{parackalScreening39Billion2026}. Simultaneously, state-of-the-art generative frameworks, such as Open Materials Generation (OMatG) which employs stochastic interpolants, are rapidly accelerating the design of inorganic crystals \cite{hoellmerOpenMaterialsGeneration2025}. However, generating a thermodynamically plausible structure does not guarantee physical viability. A recent large-scale evaluation, PhononBench, demonstrated that across several leading generative models, the average dynamical stability rate of proposed crystals is a mere 25.8\%~\cite{hanPhononBenchLargeScalePhononBased2025}. This widespread limitation highlights that generative algorithms and massive screening pipelines frequently propose structures trapped in shallow, dynamically unstable saddle points, underscoring the critical need for automated remediation workflows.

We present in this paper the VibroML package, which leverages these leading-edge foundational MLIPs to fill the need for the automated verification and remediation of dynamical instability of promising structures. VibroML systematically explores their PES to identify dynamically stable polymorphs taking inspiration from the automatic phase-transition pathway (PTP) search algorithm~\cite{togoEvolutionCrystalStructures2013}. The package integrates a suite of sophisticated algorithms for identifying and resolving soft phonon modes, providing a comprehensive list of optimized, lower-energy structures. Additionally, it enables the exploration of PTPs with the nudged elastic band (NEB) method and validates thermal stability through molecular dynamics (MD), tracking key metrics like root-mean-square deviation (RMSD), radial distribution function (RDF), volume fluctuations, and symmetry retention.

Furthermore, true materials discovery often necessitates moving beyond known, fixed stoichiometries into the complex chemical spaces of novel solid solutions and entirely unmapped elemental combinations. To address this, we introduce ProtoCSP, a Crystal Structure Prediction (CSP) tool designed for seamless integration with VibroML. When a predicted baseline material suffers from geometric frustration and dynamical instability, the two engines operate iteratively: once VibroML maps the competing lower-symmetry polymorphs of the unstable base compound, ProtoCSP systematically generates combinatorial ordered superstructures across these identified phases to model targeted alloying and evaluate their thermodynamic stability. VibroML may then be used to evaluate the dynamical stability of these alloyed configurations. Crucially, this integrated framework extends beyond localized chemical engineering; it features a database-driven prototype retrieval module to initialize structural guesses for completely novel chemistries. By intelligently retrieving and transmuting prototypes for unmapped stoichiometries, and subsequently applying VibroML's automated remediation, the toolkit provides a systematic pathway to populate the vast, unexplored ``white spaces'' of multi-component chemistry.

Here, we first benchmark the remediation tools on known compounds and evaluate MLIP fidelity, including their reproduction of DFT artifacts. We then demonstrate the automated pipeline's predictive power in resolving geometric frustration within novel complex perovskites via targeted alloying. Finally, expanding to a database-wide scale, we showcase the toolkit's broad applicability by mining the Alexandria database to systematically populate compositional white spaces and remediating high-symmetry candidates in unexplored chemical systems that previously failed to reach the thermodynamic convex hull.

\section{Results}

To evaluate the methods implemented in VibroML, we used the following initial structures with known polymorphs: \ce{LiF} ($Pm\overline{3}m$), \ce{CsPbI3} ($Pm\overline{3}m$), \ce{HfO2} ($Fm\overline{3}m$) and \ce{Bi2Sn2O7} ($Fd\overline{3}m$). The four workflows for instability remediation were used with the MACE-OMAT-0 (medium) model~\cite{batatiaFoundationModelAtomistic2025,mace_foundations_repo}. These structures were chosen based on their varying complexity and the availability of known, well-explored polymorphs in the literature, with \ce{HfO2} and \ce{Bi2Sn2O7} being the subject of focused studies \cite{qiCompetingPhasesHfO2025, rahimPolymorphExplorationBismuth2020}. Using these structures in the soft phonon remediation workflow to determine their polymorphs facilitates the verification of our results which are presented in Table~\ref{tab:stability_remediation}.

\begin{table}[htbp]
    \centering
    \caption{\textbf{Benchmark of VibroML instability remediation algorithms across four test systems using MACE-OMAT-0 (medium) model as MLIP.} The table compares the relative energies ($\Delta E$) of the most relevant polymorphs identified starting from high-symmetry unstable phases. Energies are reported in meV/atom relative to the initial unstable structure (set to 0.0). $\Delta E_{f,\text{MACE}}$ denotes the energy predicted by the MLIP, while $\Delta E_{f,\text{init}}$ and $\Delta E_{f,\text{relax}}$ correspond to DFT verification values for the initial and relaxed structures, respectively. Dynamical stability is assessed by the lowest phonon frequency ($\omega_{\text{min}}$) and the fraction of soft modes in the phonon band structure, with \textbf{Stability} indicating the outcome: \stb{} indicates $\omega_{\text{min}} > -1.5$ THz and soft modes fraction $< 2 \%$; \qst{} indicates one criterion failed; \ust{} indicates both failed. The columns under "Remediation Algorithm" indicate which method successfully recovered the specific polymorph: Genetic Algorithm (GA), Single Mode Following (TRAD), Coupled Mode Following (ALL), and Stochastic Search (RAND).}
    \label{tab:stability_remediation}

\setlength{\tabcolsep}{0pt} 
\renewcommand{\arraystretch}{1.3} 
\footnotesize
\begin{tabular*}{\textwidth}{@{\extracolsep{\fill}} l c c c c c c c c c }
    \toprule
    \textbf{Compound} & 
    \multirow{2}{*}{\textbf{Space Group}} & 
    \textbf{$\Delta E_{f,\text{MACE}}$} & 
    \textbf{$\Delta E_{f,\text{init}}$} & 
    \textbf{$\Delta E_{f,\text{relax}}$} & 
    \multirow{2}{*}{\textbf{Stability}} & 
    \multicolumn{4}{c}{\textbf{Remediation Algorithm}} \\
    \cmidrule(lr){7-10}
    
    \textbf{(\# atoms)} & & (meV/at) & (meV/at) & (meV/at) & & \textbf{GA} & \textbf{TRAD} & \textbf{ALL} & \textbf{RAND} \\
    \midrule
    
    \textbf{LiF} & $Pm\bar{3}m$ & 0.0 & 0.0 & 0.0 & \ust & ref & ref & ref & ref \\
    \textbf{(2)} & $Pnma$ & -276.8 & -275.4 & -275.4 & \stb & \checkmark & & & \\
     & $F\bar{4}3m$ & -292.5 & -289.9 & -289.9 & \stb & \checkmark & & & \\
     & $P4_2/mnm$ & -296.4 & -294.5 & -294.5 & \stb & \checkmark & & & \\
     & $Fm\bar{3}m$ & -297.2 & -296.1 & -296.1 & \stb & \checkmark & \checkmark & \checkmark & \checkmark \\
     & $P6_3mc$ & -303.1 & -300.7 & -300.6 & \stb & \checkmark & & \checkmark & \checkmark \\
     & $P1$ & -4088.3 & -3121.4 & -- & \ust & & & & \checkmark \\
     & $Pm$ & -7347.9 & -7691.9 & -- & \ust & & & & \checkmark \\
     & $Cm$ & -14042.8 & -15170.5 & -- & \ust & & & & \checkmark \\
    \midrule
    
    \textbf{CsPbI$_3$} & $Pm\bar{3}m$ & 0.0 & 0.0 & -5.4 & \ust & ref & ref & ref & ref \\
    \textbf{(5)} & $C2/m$ & -18.1 & -21.5 & -21.5 & \qst & & \checkmark & & \\
     & $Cm$ & -23.0 & -23.0 & -23.5 & \stb & & \checkmark & \checkmark & \\
     & $Pnma$ & -29.3 & -29.1 & -27.9 & \stb & \checkmark & & \checkmark & \\
     & $P\bar{1}$ & -29.9 & -28.1 & -28.5 & \stb & & & & \checkmark \\
     & $P2_1/m$ & -33.3 & -26.2 & -46.7 & \stb & \checkmark & & & \\
     & $Cmcm$ & -34.9 & -26.7 & -26.8 & \stb & \checkmark & & & \\
     & $Pnma$ & -44.2 & -37.9 & -38.0 & \stb & \checkmark & & & \\
    \midrule
    
    \textbf{HfO$_2$} & $Fm\bar{3}m$ & 0.0 & 0.0 & -5.4 & \qst & ref & ref & ref & ref \\
    \textbf{(12)} & $P4_2/nmc$ & -12.2 & -38.0 & -39.7 & \stb & & \checkmark & \checkmark & \checkmark \\
     & $P2_1/m$ & -18.5 & -68.0 & -68.8 & \stb & \checkmark & & & \checkmark \\
     & $Pnma$ & -21.7 & -66.0 & -66.4 & \qst & \checkmark & & & \checkmark \\
     & $Pbcn$ & -38.5 & -79.4 & -79.7 & \stb & \checkmark & \checkmark & & \\
     & $P2_1/c$ & -48.5 & -94.4 & -95.1 & \stb & \checkmark & & & \checkmark \\
    \midrule
    
    \textbf{Bi$_2$Sn$_2$O$_7$} & $Fd\bar{3}m$ & 0.0 & 0.0 & -10.2 & \ust & ref & ref & ref & ref\\
    \textbf{(88)} & $P2_12_12_1$ & +20.7 & +6.1 & +5.4 & \stb & \checkmark & & & \\
     & $P2_1$ & +12.7 & +1.0 & +0.5 & \qst & \checkmark & & & \checkmark \\
     & $Cc$ & -12.6 & -23.4 & -24.1 & \stb & & \checkmark & \checkmark & \\
     & $Pna2_1$ & -12.7 & -23.7 & -24.0 & \stb & \checkmark & & \checkmark & \\
     & $P2_1$ & -13.1 & -24.1 & -24.4 & \stb & \checkmark & & \checkmark & \\
     & $P2_12_12_1$ & -13.7 & -24.5 & -24.8 & \stb & \checkmark & & \checkmark & \checkmark \\
     & $Cc$ & -14.1 & -25.7 & -25.8 & \stb & \checkmark & \checkmark & \checkmark & \\
    \bottomrule
\end{tabular*}
\end{table}

\subsection{Performance of Remediation Algorithms}
\label{sec:performance_remediation}

We first assessed the efficiency and exploratory power of the different remediation workflows. As summarized in Table~\ref{tab:stability_remediation}, the \texttt{traditional}, \texttt{traditional\_all}, and \texttt{opt\_random} algorithms could not obtain dynamically stable polymorphs with lower energies than those found by the genetic algorithm (GA). Additionally, the structural diversity generated by the genetic algorithm surpasses that of the other algorithms, as evidenced by the high density of checkmarks in the GA column. For instance, in \ce{LiF}, the GA successfully identified several dynamically stable polymorphs, whereas traditional and coupled mode following found only one and two phases, respectively. The structural diversity achieved by the GA and stochastic methods is remarkable. A complete quantitative breakdown of the unique polymorphs generated by each algorithm across the four compounds is provided in the Supplementary Information (Table~\ref{tab:unique_structures}). 
Furthermore, the complete output from all optimizations is openly available in our supporting data repository at \url{https://github.com/rogeriog/VibroML_SupportData}.

Moreover, even when alternative algorithms successfully identified the lowest-energy polymorph using the MLIP potential, they generally required a larger supercell than the GA method. For example, the lowest energy $Cc$ space group of \ce{Bi2Sn2O7}, was found using a $1 \times 2 \times 2$ supercell (352 atoms) with the GA method, whereas the \texttt{ALL} method required a $2 \times 2 \times 2$ supercell (704 atoms) to find it, and the \texttt{TRAD} method required a massive $2 \times 2 \times 4$ supercell (1408 atoms). This reliance on massive supercells for traditional mode-following causes severe computational bottlenecks and relaxation stagnation on extrapolated MLIP potential energy surfaces. A detailed comparison of execution times and the specific automated heuristics implemented in VibroML to mitigate these challenges are provided in the Supplementary Information (Section~\ref{sec:comp_efficiency}).  Notably, while our benchmarks reflect sequential execution on a single GPU, VibroML's population-based algorithms are inherently parallelizable, offering near-linear scaling on multi-GPU architectures.

Interestingly, the stochastic nature of the \texttt{opt\_random} method allowed it to uncover edge cases that mode-following algorithms missed. In the case of \ce{LiF}, the \texttt{opt\_random} method identified structures with energies significantly lower than the experimental ground state, although these structures presented severe steric clashes (see Figure~\ref{fig:lif_anomaly_structures}). Remarkably, subsequent DFT verification confirmed these low energies, indicating that the MACE potential was not ``hallucinating'' but rather faithfully reproducing a numerical artifact inherent in the DFT level of theory (likely due to pseudopotential failure at short interatomic distances).

We conducted a detailed quantitative analysis of the DFT results for these \ce{LiF} polymorphs, which is presented in the Supplementary Information (Section \ref{sec:lif_anomaly}). While the older MACE-MP model—trained primarily on near-equilibrium data—failed to reproduce this artifact and predicted a physically expected repulsive barrier, the foundation models trained on extensive non-equilibrium datasets (MACE-OMAT, eSEN, and UMA) faithfully reproduced the DFT anomaly. This confirms that while these advanced models offer superior fidelity to the ground-truth PES, they inevitably inherit the specific limitations and artifacts of the underlying level of theory used in their training data. 

It is important to highlight that these non-physical configurations are separated from the global minimum by a sizeable energy barrier, making them unlikely to be accessed during standard gradient-based optimization or molecular dynamics simulations at moderate temperatures. However, the \texttt{opt\_random} algorithm in VibroML circumvents these barriers by stochastically sampling the PES with minimal constraints, thoroughly uncovering these specific, high-energy anomalies in simple systems like the two-atom unit cell of \ce{LiF}. We have not observed this behavior when applying \texttt{opt\_random} to systems with larger unit cells nor when using the other remediation algorithms, which operate mainly with the soft eigenmodes.

\subsection{Structural Fidelity and Relaxation}
\label{sec:structural_fidelity}
Beyond algorithmic efficiency, a critical metric for the success of any structural remediation algorithm driven by an MLIP is how closely the generated structures resemble the true \textit{ab initio} local minima. Table~\ref{tab:stability_remediation} presents the relevant polymorphs identified during our search, comparing the initial DFT verification energies (column $\Delta E_{f,\text{init}}$)---calculated directly on the MLIP-optimized geometries---with the fully relaxed DFT energies (column $\Delta E_{f,\text{relax}}$). For the vast majority of the identified polymorphs, the energy difference between the unrelaxed MLIP geometry and the fully relaxed DFT geometry is exceptionally small, often falling within 0.1 to 1.0~meV/atom revealing outstanding structural fidelity.

These small differences indicate that the MACE-OMAT-0 model is optimizing the crystal structures to geometries that are virtually indistinguishable from the true DFT local minima, with the energy drop remaining within standard DFT precision. None of the structures changed phase during relaxation, showing they represent true local minima or saddle points of the PES.

There are a few notable exceptions that warrant mention. For example, the $P2_1/m$ phase of \ce{CsPbI3} experiences a significant relaxation drop from -26.2~meV/atom to -46.7~meV/atom upon DFT optimization. Similarly, the high-symmetry starting phases of \ce{CsPbI3}, \ce{HfO2}, and \ce{Bi2Sn2O7} show minor relaxation drops (between 5.4 and 10.2~meV/atom), though the structures remain in the initial space group. This highlights that while the MLIP accurately maps the overall PES, it can occasionally underestimate the steepness of specific relaxation pathways for certain configurations.

\subsection{Thermodynamic Hierarchy and Polymorph Ordering}
\label{sec:thermodynamic_hierarchy}

Beyond locating the structural minima, a reliable surrogate model must accurately reproduce the thermodynamic hierarchy of the discovered polymorphs. A comparison of the relative energies predicted by MACE ($\Delta E_{f,\text{MACE}}$) against the fully relaxed DFT values ($\Delta E_{f,\text{relax}}$) in Table~\ref{tab:stability_remediation} shows that the foundational model captures the macroscopic energetic trends excellently, with only minor rank swapping among closely competing phases. 

In the case of \ce{LiF}, for the physically valid structures, the MLIP preserves the DFT energetic sequence perfectly ($Pm\overline{3}m > Pnma > F\overline{4}3m > P4_2/mnm > Fm\overline{3}m > P6_3mc$), correctly identifying the $P6_3mc$ structure as the lowest-energy physically viable polymorph. Note that the anomalous steric clash phases ($P1$, $Pm$, and $Cm$) are excluded from this physical hierarchy but follow the same massive energetic drop in both the MLIP predictions and the DFT verification.

The other compounds exhibit similar predictive success but with slight variations in the exact energetic ordering. For \ce{CsPbI3}, MACE correctly separates the high-energy and low-energy structures, the $Cmcm$ and the first $Pnma$ phases swap ranks due to energy differences of barely 1.1~meV/atom. On the other hand, the distinct DFT relaxation of the $P2_1/m$ phase shifts it from a mid-tier stability ranking in the MLIP to the energy minimum in DFT. This phase, unlike the displacive octahedral tilting that generates the other 3D perovskite polymorphs (e.g., $Pnma$), represents a reconstructive transformation into a 1D edge-sharing columnar motif, which, to the best of our knowledge, is previously unreported~\cite{fadlaFirstprinciplesInvestigationStability2020, wangEnergeticsStructuresPhase2019}. This demonstrates VibroML's ability to not only map shallow displacive polymorphs but also to identify pathways toward complete structural reconstructions.

For \ce{HfO2}, the MLIP identifies a monoclinic $P2_1/c$ phase as the ground state, with the only discrepancy being a minor swap between the $P2_1/m$ and $Pnma$ phases, which are separated by only 2.4~meV/atom in the DFT validation. Importantly, our MLIP-driven search successfully recovered the well-documented fundamental polymorphs of hafnia, including the high-temperature cubic ($Fm\overline{3}m$), tetragonal ($P4_2/nmc$), and monoclinic ground state ($P2_1/c$) phases, as well as the technologically critical ferroelectric $Pca2_1$ (oIII) and $Pmn2_1$ (oIV) phases \cite{lvPhysicalOriginHafniumbased2024}. Furthermore, VibroML identified the $Pbcn$ structure and a $Pnma$ phase. These results align closely with recent first-principles mapping by Qi and Rabe \cite{qiCompetingPhasesHfO2025}, who highlighted $Pbcn$ as a key intermediate reference structure and identified the $Pmna$ symmetry (an alternate setting of $Pnma$) as a highly stable period-2 superlattice formed by two oIV cells with opposite polarization. Notably, because VibroML's stochastic and genetic algorithms are not constrained to specific mode-coupling pathways from a single high-symmetry reference, our workflow also uncovered several stable lower-symmetry basins (e.g., $Fdd2$ and $I4_1/amd$) outside the primary flat-band distortions recently reported, demonstrating the broad exploratory power of our approach.

Finally, the hierarchical agreement for \ce{Bi2Sn2O7} is exceptional given the complexity of its 88-atom unit cell. The MLIP accurately predicts the energy penalty of the high-energy $P2_12_12_1$ and $P2_1$ phases relative to the starting $Fd\overline{3}m$ structure, with a single deviation being a negligible flip between $Pna2_1$ and the first $Cc$ phase (a difference of a mere 0.1~meV/atom in DFT). Crucially, our automated MLIP-driven search successfully reproduces the intricate polymorph landscape mapped by Rahim et al. using a rigorous but labor-intensive first-principles phonon mode-mapping approach \cite{rahimPolymorphExplorationBismuth2020}. Starting from the high-temperature cubic pyrochlore aristotype ($Fd\overline{3}m$), VibroML identified the dynamically stable room-temperature ground state possessing $Cc$ symmetry, aligning perfectly with their newly proposed $\alpha_{new}$ structural model. Furthermore, our exploration recovered the specific, novel metastable phases newly identified in their work, namely the $\delta$-\ce{Bi2Sn2O7} ($P2_12_12_1$) and $\epsilon$-\ce{Bi2Sn2O7} ($Pna2_1$) structures. While Rahim et al. relied on an iterative, step-by-step ab initio mapping along imaginary harmonic modes to traverse the complex PES, VibroML's unconstrained genetic and stochastic algorithms arrived at this same diverse ensemble of low-symmetry basins---including additional intermediate $P2_1$ phases and others---leveraging the speed of MLIPs. This demonstrates the remarkable capability of foundational models to accelerate the discovery and phase mapping of complex systems.

In Tables~\ref{tab:fullcomp_MLIPs_phonon_evals}, \ref{tab:relative_energies_mlip_dft}, and \ref{tab:phase_rankings} (Supplementary Information), we present a comprehensive comparison of the relative energies and phase rankings for the identified polymorphs across the MACE-OMAT, MACE-MP, eSEN, and UMA foundational models, benchmarked against fully relaxed DFT values. Across all four compounds, the foundational models demonstrate a remarkable consensus in identifying and ranking the most energetically favorable structural motifs (Table~\ref{tab:phase_rankings}). For instance, in the complex \ce{Bi2Sn2O7} system, all models uniformly identify the $Cc$, $P2_1$, and $P2_12_12_1$ phases as significant minima, with their relative energy rankings remaining tightly clustered and generally mirroring the DFT hierarchy. Similarly, for the structurally simpler \ce{LiF} system, MACE-OMAT, eSEN, and UMA flawlessly reproduce the exact energetic sequence observed in DFT ($Pm\overline{3}m > Pnma > F\overline{4}3m > P4_2/mnm > Fm\overline{3}m > P6_3mc$). When deviations from the DFT sequence do occur, they mostly involve closely competing phases; for example, in \ce{CsPbI3}, all MLIPs rank the $P2_1/m$ phase as a mid-tier stability structure, whereas distinct relaxation pathways in DFT propel it to the global minimum among candidates.

While the qualitative ranking order is largely preserved across the models, the absolute magnitude of the stabilization energy ($\Delta E$) relative to the high-symmetry reference varies significantly depending on the engine (Table~\ref{tab:relative_energies_mlip_dft}). Foundation models like eSEN and UMA frequently predict deeper energetic wells compared to the MACE models. Crucially, a comparison with the fully relaxed DFT energies reveals that these deeper predictions are often more accurate. In \ce{HfO2}, for example, the ground state $P2_1/c$ phase is stabilized by $-48.5$~meV/atom according to MACE-OMAT, whereas eSEN and UMA predict much steeper stabilizations of $-87.2$ and $-86.1$~meV/atom, respectively, which align much closer to the true DFT relaxation value of $-95.1$~meV/atom. However, this trend is not universal; for the 88-atom \ce{Bi2Sn2O7} unit cell, all tested MLIPs uniformly underpredict the stabilization energies of the lowest phases (predicting drops of roughly $-14$ to $-16$~meV/atom compared to the DFT drop of roughly $-24$ to $-25$~meV/atom).
Despite these variations in absolute energy magnitudes, MACE-OMAT proves to be equally capable of successfully navigating the PES to locate critical low-energy polymorphs. 

From a practical standpoint, MACE-OMAT is highly advantageous for high-throughput pipelines due to its computational efficiency. A direct benchmark on \ce{HfO2} reveals that while MACE and MACE-MP compute phonon dispersions in roughly 1 minute, eSEN and UMA require approximately 3 and 4.6 minutes (2.8$\times$ and 4.4$\times$ slower), respectively. For molecular dynamics simulations, the performance gap widens significantly: eSEN and UMA are 7.7$\times$ and 11.3$\times$ slower than MACE, respectively (detailed in Supplementary Section~\ref{sec:mlip_cost_comparison}). This significant reduction in computational cost, combined with its reliable phase ranking and structural fidelity, demonstrates that MACE-OMAT is an exceptionally efficient and robust engine for driving high-throughput instability remediation workflows without sacrificing exploratory power.

\subsection{Dynamical stability evaluation}
\label{sec:dynamical_stability}

A core capability of VibroML is not just the discovery of structurally diverse polymorphs, but the automated evaluation of their dynamical stability as a high-throughput screening tool. To systematically categorize the vast number of generated structures, we employed a three-tier classification system based on our observations---\textbf{Stable}, \textbf{Quasi-Stable}, and \textbf{Unstable}---based on the harmonic phonon spectra computed by the MLIPs. This classification relies on two primary metrics extracted from the phonon band structure: the lowest identified phonon frequency ($\omega_{\text{min}}$) and the overall fraction of soft modes across the sampled Brillouin zone. A structure is classified as \textbf{Stable} if $\omega_{\text{min}} > -1.5$~THz and the softness fraction is less than 2\%. If only one of these criteria is met, it is designated as \textbf{Quasi-Stable}, while failing both results in an \textbf{Unstable} classification. 

The implementation of finite tolerance metrics is a deliberate design choice motivated by the limitations of employing MLIPs for the evaluation of phonon band structures. While a strict 0.0~THz cutoff is theoretically ideal for a perfectly converged equilibrium structure, enforcing it within an automated MLIP-driven screening pipeline may yield an unrealistic rate of false positives. This over-penalization stems from two compounding factors. First, technical artifacts inherent to the finite-displacement supercell method, such as minor violations of the acoustic sum rule or interpolation errors at non-commensurate wavevectors, frequently produce spurious imaginary modes near the $\Gamma$-point or Brillouin zone boundaries \cite{pallikaraPhysicalSignificanceImaginary2022}. 

Second, and more critically, MLIPs are statistical interpolators. When a high-throughput workflow generates novel, structurally diverse polymorphs, the model inevitably encounters out-of-distribution atomic environments. While MLIPs predict energies and forces with high fidelity, phonon frequencies depend strictly on the second derivative of the PES. Because second derivatives are mathematically hypersensitive to minor topological fluctuations or high-frequency noise in the neural network's regression landscape, slight interpolative inaccuracies often manifest as spurious, shallow imaginary pockets along flat acoustic or optical bands~\cite{fuLearningSmoothExpressive2025,kamathImpactSpuriousImaginary2026}. Therefore, the -1.5~THz threshold acts in our experiments as a robust filter to bypass this numerical noise and ASR-induced bowing without masking true physical instabilities, which typically present as deeply negative modes exceeding this threshold. Concurrently, the 2\% softness fraction tolerance accommodates the spurious flat-band softening characteristic of MLIP interpolation artifacts, ensuring that only materials with pervasive, physically meaningful lattice instabilities are classified as unstable.

Applying these criteria to our benchmark dataset (Table~\ref{tab:stability_remediation} and Table~\ref{tab:fullcomp_MLIPs_phonon_evals}) reveals that the MLIPs successfully distinguish between deep energetic minima and transitional states. As expected, the high-symmetry reference phases (e.g., $Pm\overline{3}m$ \ce{LiF}, $Pm\overline{3}m$ \ce{CsPbI3}, $Fd\overline{3}m$ \ce{Bi2Sn2O7}) are overwhelmingly classified as \textbf{Unstable}, characterized by deeply negative frequencies (often exceeding -4.0 to -6.0~THz) and high soft mode fractions. Conversely, the lowest-energy polymorphs recovered by VibroML's remediation algorithms consistently fall into the \textbf{Stable} category. For instance, the lowest-energy $Cc$ phase of \ce{Bi2Sn2O7}, the $P2_1/c$ phase of \ce{HfO2}, and the physically viable $Pnma$ and $Fm\overline{3}m$ phases of \ce{LiF} exhibit frequencies well above the -1.5~THz cutoff and near-zero softness fractions. This complete remediation of the lattice instabilities can be visually confirmed in Supplementary Figure~\ref{fig:phonon_dispersions_main}, which directly contrasts the imaginary phonon bands of the initial high-symmetry structures with the fully stabilized dispersions of their predicted ground-state polymorphs.

The \textbf{Quasi-Stable} category frequently captures intermediate saddle points or phases that might be stabilized at finite temperatures through anharmonic effects. For example, the $C2/m$ phase of \ce{CsPbI3} and the $P2_1$ and $Pbcn$ phases of \ce{HfO2} occasionally trigger a ``Quasi'' classification across different engines. As shown in the supplementary comprehensive benchmark, while the four foundational models largely agree on the broad thermodynamic hierarchy, their exact mapping of these shallow saddle points varies slightly, causing a phase to be deemed \textbf{Stable} by MACE-MP but \textbf{Quasi-Stable} by eSEN. 

It is important to acknowledge that while harmonic phonon calculations via MLIPs are highly indicative of structural viability, rigorous, ground-truth confirmation of dynamical stability ultimately requires explicit \textit{ab-initio} phonon calculations. However, performing such explicit first-principles calculations on the hundreds of complex, low-symmetry unit cells generated during a genetic or stochastic search is computationally prohibitive and beyond the scope of this automated screening workflow. Fortunately, recent large-scale benchmarking studies have demonstrated that modern foundational MLIPs serve as excellent proxies for \textit{ab-initio} dynamical stability assessments. As established by Loew et al.\cite{loewUniversalMachineLearning2025}, foundational models are increasingly ``ready for phonons'', capturing the necessary force constant matrices with high fidelity. Furthermore, in our own recent work, Bakhsh et al.\cite{bakhshSymmetryStabilityStructural2026} explicitly utilized the VibroML toolkit to successfully map phonon-driven phase transitions and evaluate the dynamical stability of \ce{Cs2KInI6} compared to DFPT calculations, demonstrating that the MLIP-driven workflow achieves accuracy rivaling explicit \textit{ab-initio} methods. Therefore, while VibroML is fundamentally focused on rapid polymorph mapping and structural remediation rather than definitive first-principles phonon validation, the stability classifications provided by these integrated MLIPs offer a highly reliable, physically grounded heuristic. This ensures that the candidate pool forwarded for downstream high-fidelity computational or experimental screening consists of strong candidates for thermodynamical and dynamical stability.

\subsection{Thermal stability evaluation}
\label{sec:thermal_stability}
While harmonic phonon calculations enable assessment of dynamical stability at 0~K, realistic material performance is governed by finite-temperature effects and anharmonicity. To validate the stability of the polymorphs identified during the PES exploration, we subjected them to our automated MD screening workflow as shown in Table~\ref{tab:md_thermal_summary}. The comprehensive results across the four foundational MLIPs---evaluated against thresholds for RMSD, volume fluctuation, RDF retention, and symmetry preservation---are summarized in Table~\ref{tab:fullcomp_MLIPs_MD_evals}. The MD workflow provides insights into kinetic trapping and dynamic phase transitions, highlighted by the ``Phase Retained'' metric.

\begin{table}[htbp]
    \centering
    \caption{\textbf{Thermal stability benchmarking of identified polymorphs via MACE-OMAT MD.} This table mirrors the phases in Table 1, reporting time-averaged trajectory metrics under finite-temperature NPT MD simulations (300 K). The overall \textbf{Verdict} classifies a structure as \stb\ (at least 3 criteria passed) or \ust\ (2 or fewer passed). The individual metrics are: 
    (1) \textbf{RMSD (\AA)}: Root-mean-square deviation from initial lattice sites (Pass: $< 1.0$~\AA); 
    (2) \textbf{Vol (\%)}: Maximum cell volume fluctuation relative to the early production average (Pass: $< 6\%$); 
    (3) \textbf{RDF}: Pearson correlation between the initial Radial Distribution Function and the ensemble-averaged final RDF (Pass: $> 0.5$); and 
    (4) \textbf{Phase Retained}: Evaluates if a time-averaged structure from the final 50 frames preserved the initial space group within a tolerance sweep. ``Yes'' indicates the symmetry was retained, while ``No'' indicates a spontaneous phase transition (with the new space group noted) or loss of crystallinity.}
    \label{tab:md_thermal_summary}
    \small
    \setlength{\tabcolsep}{4pt}
    \renewcommand{\arraystretch}{1.2}
    \begin{tabular}{ll c c c c l}
        \toprule
        \textbf{Compound} & \textbf{Phase} & \textbf{Verdict} & \textbf{RMSD (\AA)} & \textbf{Vol (\%)} & \textbf{RDF} & \textbf{Phase Retained} \\
        \midrule
        \textbf{LiF} & $Pm\bar{3}m$ & \ust & 1.579 & 28.3 & -0.011 & No ($F\bar{4}3m$) \\
                     & $Pnma$       & \ust & 2.258 & 48.0 & 0.001  & No ($P6_3/mmc$) \\
                     & $F\bar{4}3m$ & \ust & 3.631 & 9.0  & -0.012 & No ($Fm\bar{3}m$) \\
                     & $P4_2/mnm$   & \ust & 0.464 & 31.4 & -0.027 & No ($Fm\bar{3}m$) \\
                     & $Fm\bar{3}m$ & \stb & 0.174 & 2.9  & 0.281  & Yes \\
                     & $P6_3mc$     & \ust & 0.333 & 17.7 & 0.034  & No ($P6_3/mmc$) \\
                     & $P6_3/mmc$   & \ust & 1.917 & 5.6  & 0.370  & No ($Fm\bar{3}m$) \\
                     & $P1$         & \ust & \multicolumn{3}{c}{\textit{Explosion (FAILED\_UNSTABLE)}} & N/A \\
                     & $Pm$         & \ust & \multicolumn{3}{c}{\textit{Explosion (FAILED\_UNSTABLE)}} & N/A \\
                     & $Cm$         & \ust & \multicolumn{3}{c}{\textit{Explosion (FAILED\_UNSTABLE)}} & N/A \\
        \midrule
        \textbf{CsPbI$_3$} & $Pm\bar{3}m$ & \ust & 0.793 & 4.9  & 0.332 & No ($P1$) \\
                           & $Pnma$       & \ust & 1.176 & 12.4 & 0.335 & Yes \\
                           & $P\bar{1}$   & \ust & 2.989 & 8.1  & 0.670 & Yes \\
                           & $Pm$         & \stb & 0.834 & 3.8  & 0.692 & Yes \\
                           & $P2_1/m$     & \ust & 5.136 & 12.9 & 0.539 & Yes \\
                           & $Cmcm$       & \ust & 1.123 & 22.2 & 0.356 & No ($P2_1/m$) \\
        \midrule
        \textbf{HfO$_2$} & $Fm\bar{3}m$ & \ust & 0.453 & 3.1  & 0.189 & No ($Pca2_1$) \\
                         & $P4_2/nmc$   & \ust & 0.530 & 8.4  & 0.564 & No ($P2_1/c$) \\
                         & $P2_1/m$     & \ust & 1.756 & 9.0  & 0.518 & Yes \\
                         & $Pnma$       & \stb & 0.101 & 9.9  & 0.772 & Yes \\
                         & $Pbcn$       & \ust & 1.051 & 17.6 & 0.606 & No ($P2_1/c$) \\
                         & $P2_1/c$     & \stb & 1.216 & 4.0  & 0.511 & Yes \\
        \midrule
        \textbf{Bi$_2$Sn$_2$O$_7$} & $Fd\bar{3}m$   & \stb & 0.423 & 4.5 & 0.463 & Yes \\
                                   & $P2_12_12_1$   & \stb & 0.359 & 5.1 & 0.675 & No ($Imma$) \\
                                   & $P2_1$         & \ust & 6.865 & 9.8 & 0.705 & No ($P1$) \\
                                   & $Cc$           & \stb & 0.242 & 0.7 & 0.874 & No ($Fd\bar{3}m$) \\
                                   & $Pna2_1$       & \stb & 0.350 & 6.2 & 0.713 & No ($Fd\bar{3}m$) \\
        \bottomrule
    \end{tabular}
\end{table}

For the \ce{LiF} system, the MD evaluations show that the workflow serves as a critical filter for numerical artifacts inherited from DFT training data, while simultaneously confirming experimentally known ground states. %
The simulations identify the cubic $Fm\overline{3}m$ phase as the most stable configuration, passing all MD criteria with minimal RMSD and volume fluctuations, aligning with experimental observations of its thermodynamic stability at ambient conditions. Conversely, the other presented structures fail the stability checks, undergoing spontaneous transitions and ultimately relaxing toward the $Fm\overline{3}m$ global minimum under the effects of temperature. Furthermore, as discussed in the Performance of Remediation Algorithms section, the \ce{LiF} PES contains non-physical steric clash anomalies (e.g., $P1$, $Pm$, $Cm$) at highly negative energies. When subjected to MD, the vast majority of these artifact structures undergo catastrophic structural failure (labeled as \textit{FAILED\_UNSTABLE}), triggering massive volume expansions and a complete loss of crystallinity. This highlights the role of finite-temperature screening: even if an MLIP predicts a deep, spurious local minimum at 0~K, the MD workflow may be able to filter out these false positives by leveraging their lack of physical cohesion under realistic dynamic conditions.

The application of our automated MD workflow to \ce{CsPbI3} reveals clear limitations when dealing with highly anharmonic, soft materials. These limitations manifest not only in the evaluation of stable ground states but also in the predicted kinetic pathways of unstable phases. For instance, in Table~\ref{tab:fullcomp_MLIPs_MD_evals}, while all four MLIPs correctly flag the high-temperature cubic $Pm\overline{3}m$ phase as dynamically unstable at 300~K, they disagree on its structural degradation pathway: MACE-MP and eSEN successfully capture the spontaneous transition to the expected $Pnma$ ground state, whereas MACE-OMAT and UMA predict distortions into alternative low-symmetry phases ($P1$ and $P2_1/m$, respectively). Furthermore, there is noticeable disagreement across the models regarding the finite-temperature stability of the known $Pnma$ ground state itself. Seemingly low-symmetry phases, such as $Pm$ and $Pc$, occasionally pass the stability checks or emerge as transition endpoints, creating a highly convoluted evaluation landscape.

We attribute the difficulty of the stability workflow for this material to challenging characteristics of halide perovskites, which align closely with recent perspectives on MLIP limitations for this class of materials~\cite{bianMachineLearningPotentials2026}. First, the highly anharmonic, flat potential energy surface of \ce{CsPbI3} causes the lattice to naturally form transient, low-symmetry local correlation patterns even within high-symmetry phases~\cite{baldwinDynamicLocalStructure2024}. Consequently, our automated stability criteria (such as rigid RMSD thresholds) may overly penalize these legitimate, highly dynamic octahedral tilts, misidentifying natural structural flexibility as phase degradation. Second, models trained predominantly on PBE-level data often fail to adequately capture the delicate dispersion forces required for halide perovskites, artificially softening the octahedral tilting modes~\cite{braeckeveltAccuratelyDeterminingPhase2022, bianMachineLearningPotentials2026}. Finally, while foundational models generally excel at mapping displacive transitions (e.g., cubic to $Pnma$), the transition to the 1D edge-sharing $P2_1/m$ phase is a reconstructive transition involving significant bond breaking. Accurately modeling these reconstructive degradation pathways remains a recognized challenge for current MLIPs due to the difficulty of sampling such rare-event, high-barrier configuration spaces~\cite{bianMachineLearningPotentials2026}.

For deep, well-defined energetic minima, the MD predictions strongly align with experimental thermodynamic hierarchies. For instance, the monoclinic ground state of hafnium oxide (\ce{HfO2}, $P2_1/c$) maintains excellent structural integrity across the simulations, exhibiting sub-angstrom RMSD values and space group retention. Conversely, high-symmetry aristotypes like the cubic ($Fm\overline{3}m$) and tetragonal ($P4_2/nmc$) phases---which experimentally require extreme temperatures ($>$2000~K), doping, or strain to stabilize \cite{lvPhysicalOriginHafniumbased2024}---rapidly fail the ambient MD stability criteria. Our simulated trajectories capture these unstable phases spontaneously distorting into the global minimum $P2_1/c$ phase or the metastable, technologically critical ferroelectric $Pca2_1$ phase. This dynamic relaxation mimics the martensitic quenching pathways used to experimentally trap the ferroelectric properties of \ce{HfO2}, demonstrating that the MLIPs can reproduce kinetic pathways driven by finite-temperature anharmonicity.

In the complex \ce{Bi2Sn2O7} system, experimental characterization shows that the high-temperature cubic $\gamma$-phase ($Fd\overline{3}m$, $>$900~K) distorts upon cooling into the room-temperature $\alpha$-phase ($Cc$)~\cite{shannonPolymorphismBi2Sn2O71980, lewisExhaustiveSymmetryApproach2016}. While our 0~K harmonic phonon mode-mapping correctly identifies the $Cc$ ground state alongside other metastable low-symmetry phases such as $\delta$-\ce{Bi2Sn2O7} ($P2_12_12_1$) and $\epsilon$-\ce{Bi2Sn2O7} ($Pna2_1$) reported in recent computational explorations~\cite{rahimPolymorphExplorationBismuth2020}, finite-temperature MD reveals the kinetic shallowness of these energetic basins. 
Structures initialized in the $Cc$, $Pna2_1$, and specific $P2_1$ and $P2_12_12_1$ configurations frequently pass basic thermodynamic and RDF checks but fail to retain their initial phase. Instead, anharmonic thermal fluctuations provide sufficient energy to overcome local barriers, triggering solid-solid transitions into the highly symmetric $Fd\overline{3}m$ aristotype. This demonstrates the toolkit's ability to identify dynamic stabilization effects, where anharmonicity at finite temperatures stabilizes higher-symmetry phases over shallow local minima.

In Table~\ref{tab:fullcomp_MLIPs_MD_evals} the results of the MD stability analysis are compared across the four foundational MLIPs examined in this study. Overall, while the tested MLIPs generally agree on the macroscopic thermal stability of deep minima, minor discrepancies in the kinetic evaluation of transition states are observable. The discrepancies are more evident in the \ce{CsPbI3} system, where the MLIPs disagree on the stability of the $Pnma$ and $P2_1/m$ phases, reflecting variations in how each model extrapolates the PES curvature away from equilibrium. Ultimately, the automated extraction of RMSD, RDF, and symmetry metrics provides a meaningful, high-throughput curation filter. While experimental agreement should be verified case-by-case, our results suggest that MLIP-driven MD stability analysis with current foundational models can offer critical guidance by flagging dynamically unstable structures and providing an overview on possible kinetic pathways to be evaluated via more computationally intensive methods or experimental synthesis. 

\subsection{Evaluating Phase Transition Pathways}

Another critical application of these MLIPs is the evaluation of phase transition pathways. We utilized the NEB module incorporated in VibroML to determine the minimum energy paths (MEPs) from the initial high-symmetry phases to the most stable polymorphs found with DFT. These paths were driven entirely by the MLIPs and subsequently verified with single-point DFT calculations to evaluate the true energy landscape. 

The results, detailed in the Supplementary Information (Section \ref{sec:neb_results}), show that the state-of-the-art MLIPs successfully capture complex transition mechanisms. However, a performance trade-off was observed among the architectures: MACE generally provided the highest energy fidelity along the paths, eSEN excelled in capturing local force gradients, and UMA proved to be a robust all-rounder, occasionally finding kinetically more favorable transition channels (lower barriers) that other models missed.

Notably, the NEB benchmarking also highlighted a nuanced limitation in how foundation models generalize to novel low-symmetry phases. While our Table~\ref{tab:stability_remediation} analysis demonstrates that the MLIPs find configurations virtually identical in energy to the DFT ground states, DFT verification of the NEB endpoints frequently revealed residual forces (0.07 to 0.21~eV/\AA). This suggests a phenomenon of ``softening'' in the out-of-distribution regions of the MLIP PES \cite{fuLearningSmoothExpressive2025,anamComprehensiveAssessmentBenchmark2025}. The models accurately locate the energetic depth of the local minimum but underestimate the stiffness of the surrounding potential well, effectively flattening it. Consequently, the exact geometric coordinate where the MLIP gradient reaches zero sits slightly off-center relative to the steeper, narrower DFT well, manifesting as residual forces without a significant energy penalty.

\subsection{Automated Discovery of Higher-Dimensional Perovskite Networks via the VibroML-ProtoCSP Pipeline}

The utility of VibroML extends beyond diagnosing instabilities and mapping novel polymorphs; it provides actionable pathways to proactively engineer materials. As demonstrated, VibroML excels at navigating the PES to discover dynamically stable ground states for a fixed stoichiometry. However, in high-throughput computational design, a highly desirable crystal topology—such as a functional 3D network—is frequently found to be geometrically frustrated. In these cases, VibroML correctly predicts its collapse into a non-functional, lower-dimensional phase. Rather than abandoning the sought-after framework, we developed the coupled VibroML-ProtoCSP pipeline to rescue the topology by systematically nudging the material's chemistry via targeted alloying. This integrated workflow operates in three distinct stages:

\begin{enumerate}
\item \textbf{Base Polymorph Mapping:} VibroML maps the potential energy surface of the pure, unalloyed base structure to identify all competing lower-symmetry phases.
\item \textbf{Combinatorial Structure Generation:} For a target grid of alloyed compositions, ProtoCSP automatically generates multiple ordered superstructures across all the previously identified base polymorphs. Because the optimal crystallographic sites for substitutional atoms are unknown \textit{a priori}, ProtoCSP evaluates various configurations with an MLIP to pinpoint the lowest-energy atomic ordering for every composition-polymorph pair and proceeds to compute their formation energy.
\item \textbf{Stability Tracking:} The dynamical and thermodynamic stability of these optimized structures are evaluated to track how the energetic hierarchy of the polymorphs evolves across the compositional space.
\end{enumerate}

We applied this pipeline to two specific perovskite systems identified in our previous works as highly promising yet geometrically frustrated: the lead-free halide double perovskite \ce{Cs2KInI6}~\cite{bakhshSymmetryStabilityStructural2026} and the chalcogenide perovskite \ce{KTaSe3}~\cite{babuMEIDNetMultimodalGenerative2026}. In both cases, the baseline stoichiometric compounds suffer from severe structural instabilities, driving them to collapse from their ideal 3D forms into lower-dimensional or non-perovskite phases. This loss of corner-sharing connectivity induces charge-carrier localization and increased bandgaps degrading the material's optoelectronic performance.

\subsubsection{Alloying Strategies for Higher-Dimensional Networks in \ce{Cs2KInI6}}
Lead-free double perovskites (elpasolites) with the general formula $A_2MM'X_6$ have emerged as highly promising candidates for photovoltaic applications due to their expanded chemical space and improved thermal and moisture stability compared to traditional hybrid perovskites. In our previous work by Bakhsh et al.~\cite{bakhshSymmetryStabilityStructural2026}, we utilized VibroML to reveal that the high-symmetry cubic phase ($Fm\overline{3}m$) of \ce{Cs2KInI6} is dynamically unstable. While this cubic phase possesses an optimal direct bandgap of 1.24~eV, employing the VibroML GA method yielded 42 lower-symmetry stable phases. Crucially, the most energetically stable of these phases (e.g., $I\overline{4}2m$ and $Cmc2_1$) decompose into non-perovskite configurations characterized by tetrahedrally coordinated indium. 

Geometric formability factors quantify this instability. While the global Goldschmidt tolerance factor ($t \approx 0.877$) initially suggests a stable overall fit, the localized octahedral factor ($\mu$) of the \ce{InI6} sub-lattice fails. The \ce{In^{3+}} cation (0.80~\AA)~\cite{shannonRevisedEffectiveIonic1976} is too small to adequately support the large \ce{I^-} anion cage (2.20~\AA), resulting in $\mu \approx 0.364$, which falls well below the theoretical stability threshold of 0.414. 

To resolve this geometric mismatch while preserving the chemical properties of the original perovskite, we utilized ProtoCSP to generate a finely tuned, partial solid solution by alloying with a sodium-bromine based perovskite: \ce{Cs2(K_{1-x}Na_x)In(I_{1-y}Br_y)6}. By substituting \ce{I^-} with the smaller \ce{Br^-} anion and replacing \ce{K^+} with the smaller \ce{Na^+} cation on the B-site, the global tolerance factor pushes toward an ideal $t \approx 0.946$, while the local octahedral factor improves to $\mu \approx 0.408$ at the fully substituted limit. To systematically map this space, ProtoCSP generated ordered superstructures varying the B-site substitution $x$ across 0 to 0.4, and the halide anion ratio $y$ up to $2/3$. This dual-site partial alloying scheme is designed to simultaneously shrink the octahedral volume to improve the local fit for indium and slightly reduce the average B-site radius, elevating the global tolerance factor to preserve the 3D double-perovskite network.

\begin{figure}[h]
    \centering
    \includegraphics[width=0.8\textwidth]{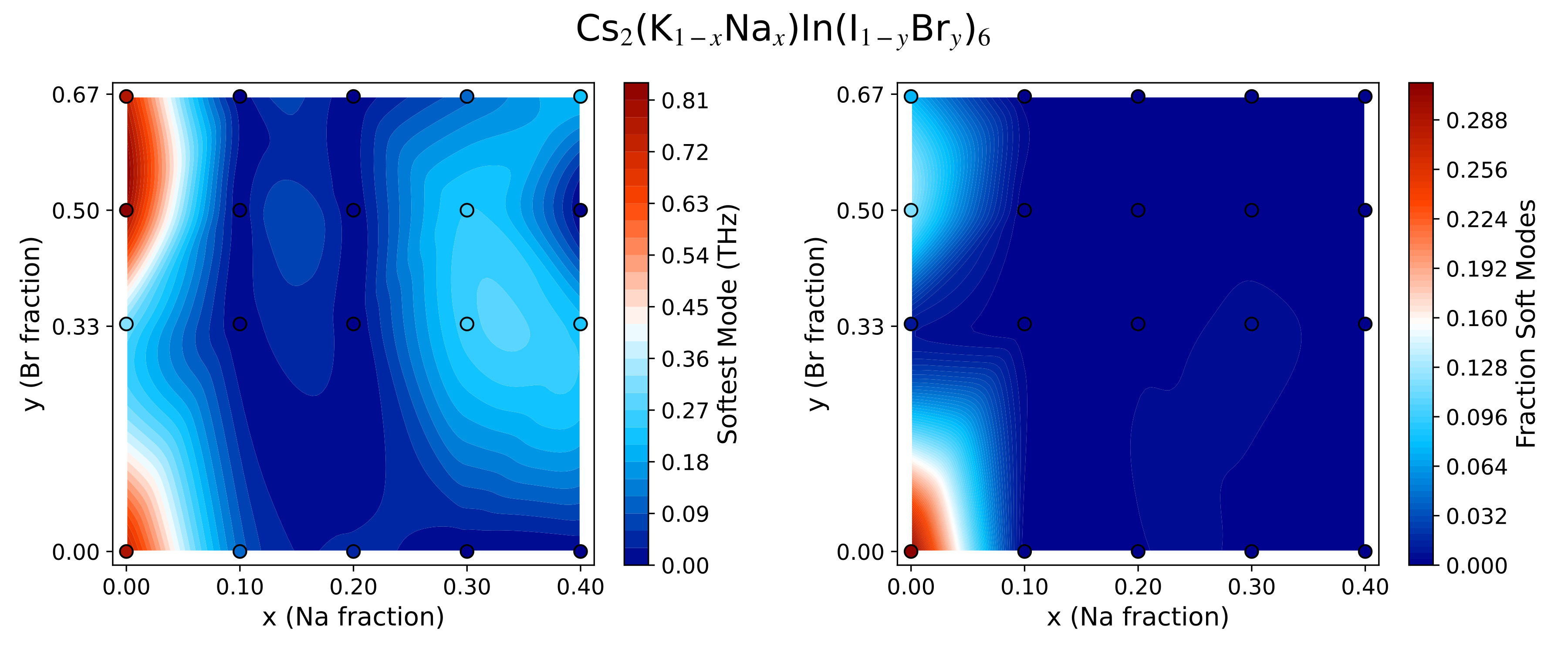}
    \caption{Evolution of dynamical stability across the \ce{Cs2(K_{1-x}Na_x)In(I_{1-y}Br_y)6} solid solution space. The heatmaps display the lowest identified optical phonon frequency (left) and the percentage of soft phonon modes across the sampled Brillouin zone (right) as functions of the B-site Na fraction ($x$) and the halide Br fraction ($y$). The maps highlight the critical role of B-site substitution ($x \ge 0.1$) in suppressing imaginary modes and dynamically stabilizing the 3D double-perovskite framework.}
    \label{fig:cs2kini6_phonon_heatmaps}
\end{figure}

Phonon spectra computations for the generated compositions demonstrate a dramatic stabilization of the high-symmetry lattice, driven primarily by the B-site substitution. As shown in the heatmaps in Figure~\ref{fig:cs2kini6_phonon_heatmaps}, the pure base compound ($x=0.0$, $y=0.0$) is highly unstable, characterized by a deeply negative softest mode of -0.75~THz and a massive fraction of soft modes across the sampled Brillouin zone (31.5\%).

A systematic analysis of the compositional space reveals several key insights into the stabilization mechanism. First, attempting to relieve the geometric mismatch solely by replacing iodine with the smaller bromine anion ($x=0.0$, $y > 0$) is insufficient to stabilize the $Fm\overline{3}m$ phase. While the overall fraction of soft modes drops to between 1\% and 12\%, the most unstable optical modes remain deeply imaginary (ranging from -0.31~THz to -0.80~THz). This indicates that the 3D framework remains highly susceptible to octahedral tilting and subsequent collapse. In contrast, the introduction of the smaller Na cation at the K site proves to be the definitive driver for dynamical stability. Even a modest 10\% substitution ($x=0.1$) across any halogen ratio abruptly shifts the softest modes to near-zero values (all greater than -0.11~THz) and effectively eliminates the soft phonon fraction. This stabilizing effect reaches an optimal plateau around a B-site substitution of $x=0.2$. Within this regime, the 3D double-perovskite framework is dynamically stable across the entire sampled Br concentration range.

\begin{figure}[h]
    \centering
    \includegraphics[width=0.8\textwidth]{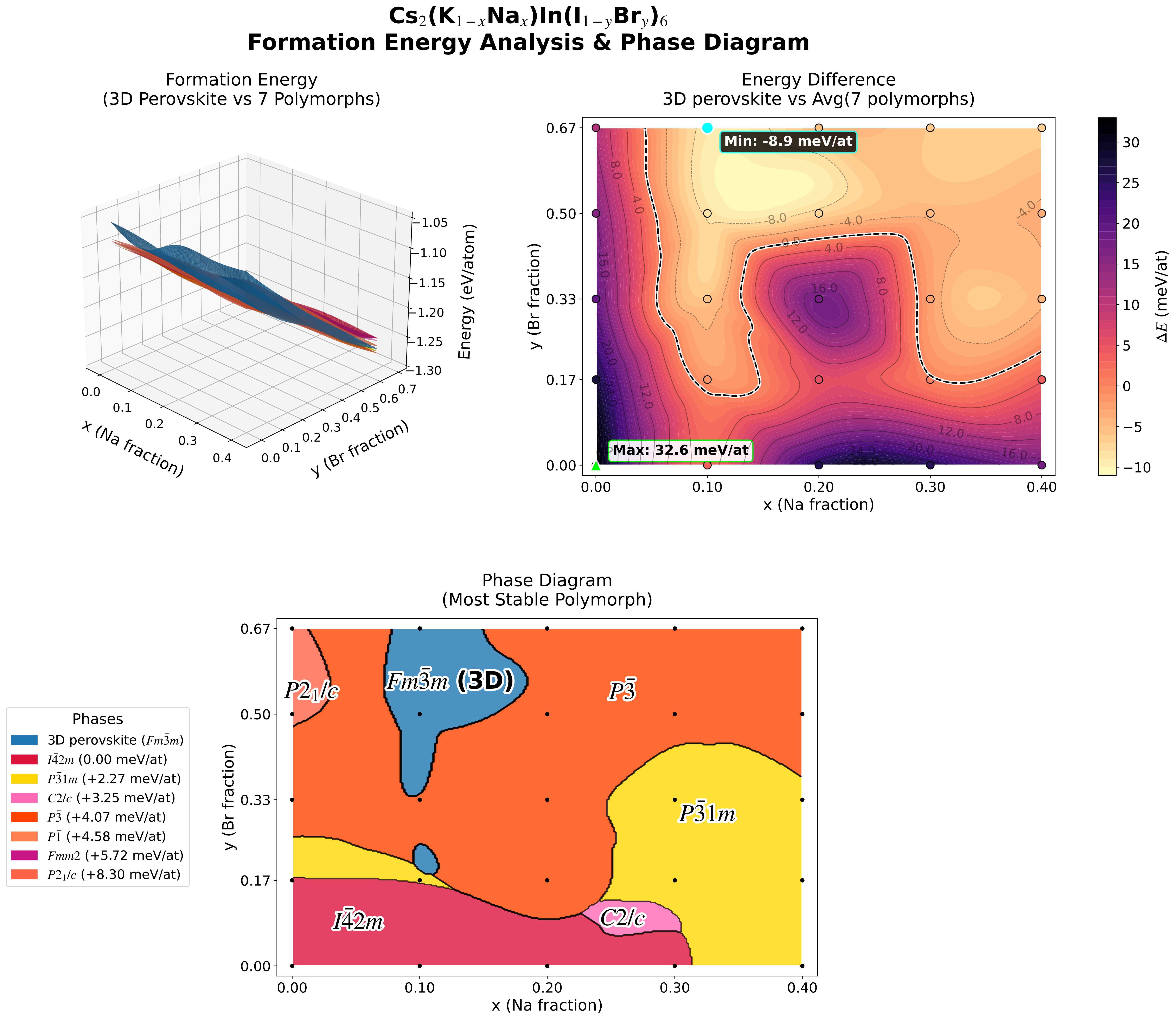}
    \caption{Thermodynamic stability analysis and phase evolution of the \ce{Cs2(K_{1-x}Na_x)In(I_{1-y}Br_y)6} solid solution. (\textbf{Top left}) 3D surface plot of the formation energies, illustrating the energetic descent of the target 3D $Fm\overline{3}m$ perovskite (blue surface) relative to competing lower-symmetry polymorphs as a function of B-site Na ($x$) and halide Br ($y$) substitutions. (\textbf{Top right}) Contour map showing the energy difference ($\Delta E$) between the $Fm\overline{3}m$ phase and the average formation energy of the competing polymorphs. The dashed contour line denotes the thermodynamic crossover point ($\Delta E = 0$). (\textbf{Bottom}) Predicted compositional phase diagram displaying the absolute lowest-energy structural phase at each coordinate, revealing a distinct stability window where the targeted 3D double perovskite network emerges as the thermodynamic ground state.}
    \label{fig:cs2kini6_formation_energy_panel}
\end{figure}

The resulting energetic hierarchy is visualized in the formation energy 3D surface plot (Figure~\ref{fig:cs2kini6_formation_energy_panel}, top left). The blue surface, representing the 3D $Fm\overline{3}m$ perovskite, clearly intersects and descends below the clustered energy surfaces of the competing polymorphs as the fractions of sodium ($x$) and bromine ($y$) increase. To quantify this thermodynamic crossover, we mapped the energy difference ($\Delta E$) between the 3D perovskite and the average formation energy of the seven competing polymorphs (Figure~\ref{fig:cs2kini6_formation_energy_panel}, top right). In the pristine \ce{Cs2KInI6} limit ($x=0.0$, $y=0.0$), the 3D network is penalized, sitting at a unfavorable maximum of +32.6~meV/atom relative to the polymorph average. However, the synergistic dual-site substitution induces a steep stabilization gradient. The dashed contour line explicitly demarcates the thermodynamic tipping point ($\Delta E = 0$), revealing a substantial parameter space where the 3D framework becomes energetically preferred, reaching a peak stabilization of -8.9~meV/atom near $x=0.1$ and $y=0.67$. 

While outperforming the average of the competing phases is a strong indicator of structural viability, true thermodynamic stability requires the 3D perovskite to be the absolute global minimum on the potential energy surface. By projecting the absolute lowest formation energy among all considered phases at each coordinate, we constructed a phase diagram for the \ce{Cs2(K_{1-x}Na_x)In(I_{1-y}Br_y)6} solid solution (Figure~\ref{fig:cs2kini6_formation_energy_panel}, bottom). The phase diagram illustrates the structural evolution driven by the relief of geometric frustration. The iodine-rich base composition is entirely dominated by the non-perovskite $I\overline{4}2m$ and $P\overline{3}1m$ phases. As halide substitution increases without B-site modification, the system transitions into lower-symmetry monoclinic ($P2_1/c$) and trigonal ($P\overline{3}$) regimes. 

Ultimately, the phase diagram confirms the existence of a distinct, targeted stability window for the ideal 3D double perovskite. The $Fm\overline{3}m$ phase emerges as thermodynamic ground state within a specific compositional ``canyon'', localized primarily around a B-site sodium fraction of $x \approx 0.10$ to $0.15$ and intermediate-to-high bromine fractions ($y \geq 0.33$). This exact alignment between the dynamically stable regions identified in our phonon analysis and the thermodynamically favored regions mapped in the phase diagram  validates the VibroML-ProtoCSP pipeline to uncover promising materials. 

\subsubsection{Alloying Strategies for Higher-Dimensional Networks in \ce{KTaSe3}}

The chalcogenide \ce{KTaSe3}, predicted from our multimodal inverse design pipeline with MEIDNet~\cite{babuMEIDNetMultimodalGenerative2026}, presents a distinct stabilization challenge. In its ideal cubic form, the \ce{TaSe6} octahedra are highly unstable and break coordination to form \ce{TaSe5} polyhedra, which predominantly assemble into 1D chains that dominate the lowest-energy polymorphs. This structural collapse is driven by a severe geometric mismatch: the \ce{Ta^{5+}} cation is drastically undersized for the \ce{Se^{2-}} cavity, yielding a local octahedral factor of $\mu \approx 0.323$ and a Bartel's tolerance factor of $\tau \approx 4.817$.

To counteract this collapse and target a 2D layered motif, we explored the solid solution space \ce{(K_{1-x}Ba_x)(Ta_{1-x}Zr_x)(Se_{1-y}O_y)3}. Substituting a fraction of selenium with oxygen introduces a distinct thermodynamic tension arising from the difference between the short \ce{Ta-O} bonds and the longer \ce{Ta-Se} bonds, acting as a chemical wedge to encourage the lattice to segregate into rigid 2D sheets of Ta-X polyhedra. Concurrently, the aliovalent substitution of monovalent potassium with divalent barium introduces a higher-charge cation to prevent delamination, while the charge-balancing substitution of tantalum with the larger zirconium (0.72~\AA) partially relieves internal geometric stress. 

\begin{figure}[h]
    \centering
    \includegraphics[width=0.8\textwidth]{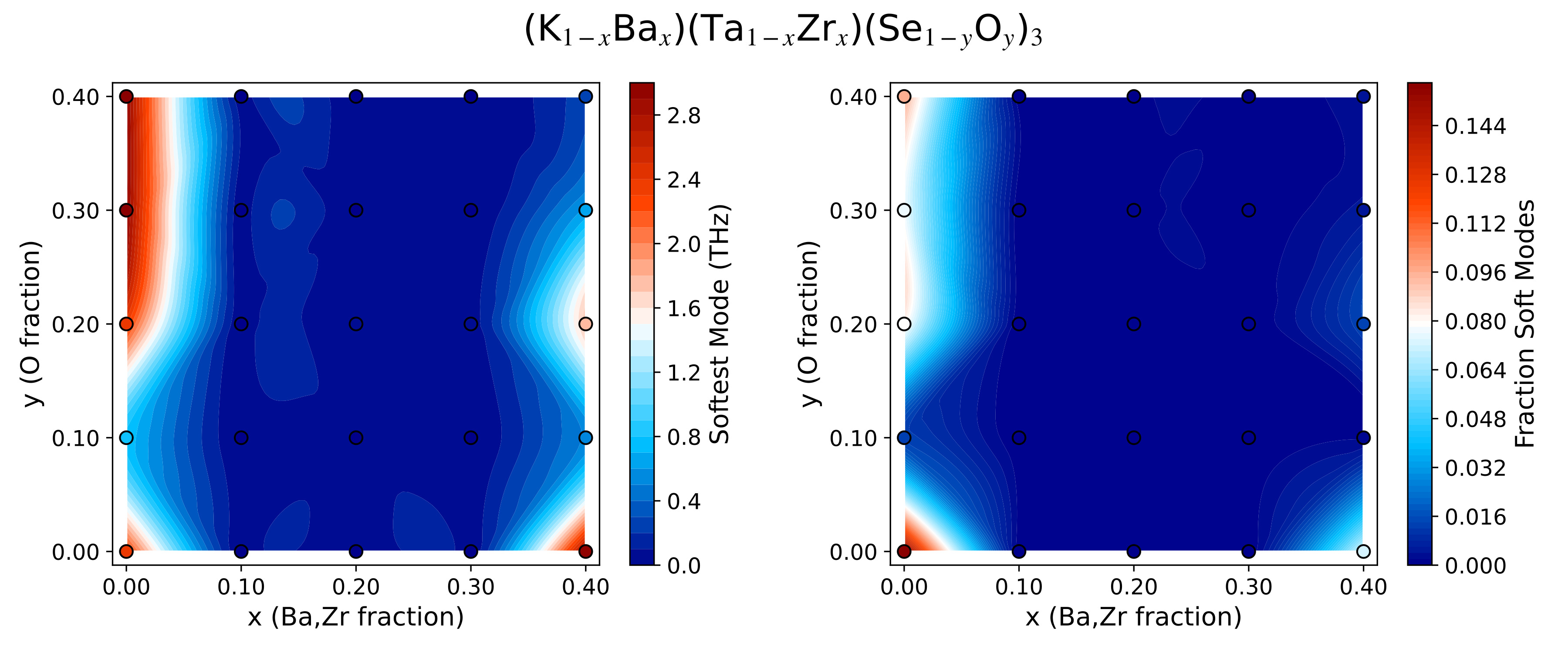}
    \caption{Dynamical stability map for the targeted cubic \ce{(K_{1-x}Ba_x)(Ta_{1-x}Zr_x)(Se_{1-y}O_y)3} solid solution. The heatmaps illustrate the lowest phonon frequency (left) and the fraction of soft modes (right) across varying levels of aliovalent Ba/Zr cation co-substitution ($x$) and O anion substitution ($y$). A distinct dynamical stability window emerges at intermediate cation doping levels ($0.1 \le x \le 0.3$), outside of which the structure undergoes severe geometric collapse.}
    \label{fig:ktase3_phonon_heatmaps}
\end{figure}

We deployed our pipeline across localized doping ranges, testing cation ($x$) and anion ($y$) substitutions at discrete values from 0 to 0.4. The resulting phonon spectra (Figure~\ref{fig:ktase3_phonon_heatmaps}) reveal a highly nuanced stability landscape. The pure baseline compound ($x=0.0$, $y=0.0$) exhibits severe dynamical instability, marked by a deeply negative softest mode of -2.32~THz and a substantial soft mode fraction of 16.0\%, corroborating the predicted geometric collapse.

Attempting to induce stabilization purely through anion substitution---replacing selenium with the significantly smaller, more electronegative oxygen ($x=0.0$, $y > 0$) proves counterproductive. Rather than rigidifying the lattice, the introduction of the isolated \ce{Ta-O} bond-length mismatch exacerbates the instability, driving the softest optical mode even deeper to -2.93~THz at $y=0.3$ and $0.4$. This indicates that without the supporting structural pillar of a higher-charge A-site cation, the bond-length discrepancy simply accelerates the collapse of the \ce{TaX6} framework. 

Conversely, the aliovalent cation co-substitution ($x > 0$) acts as a profound structural stabilizer, but only within a strict, intermediate compositional window. The introduction of the \ce{Ba^{2+}} and \ce{Zr^{4+}} cations instantly suppresses the soft modes. Moving into the $x=0.1$ to $x=0.3$ doping regime resolves the deep lattice instabilities across all tested oxygen fractions. Within this optimal plateau, the softest modes are negligible. However, this dynamical stability is bounded. Pushing the cation substitution to $x=0.4$ triggers a catastrophic re-emergence of instability, sending the softest mode plummeting back to -2.89~THz for the pure selenide, and maintaining deep imaginary modes across the oxygen-doped variants. This boundary suggests an upper limit to how much local electronic and steric disruption the pseudo-cubic topology can absorb before the excessive accumulation of \ce{Zr} and \ce{Ba} breaks the corner-sharing connectivity entirely.

\begin{figure}[htbp]
    \centering
    \includegraphics[width=1.0\textwidth]{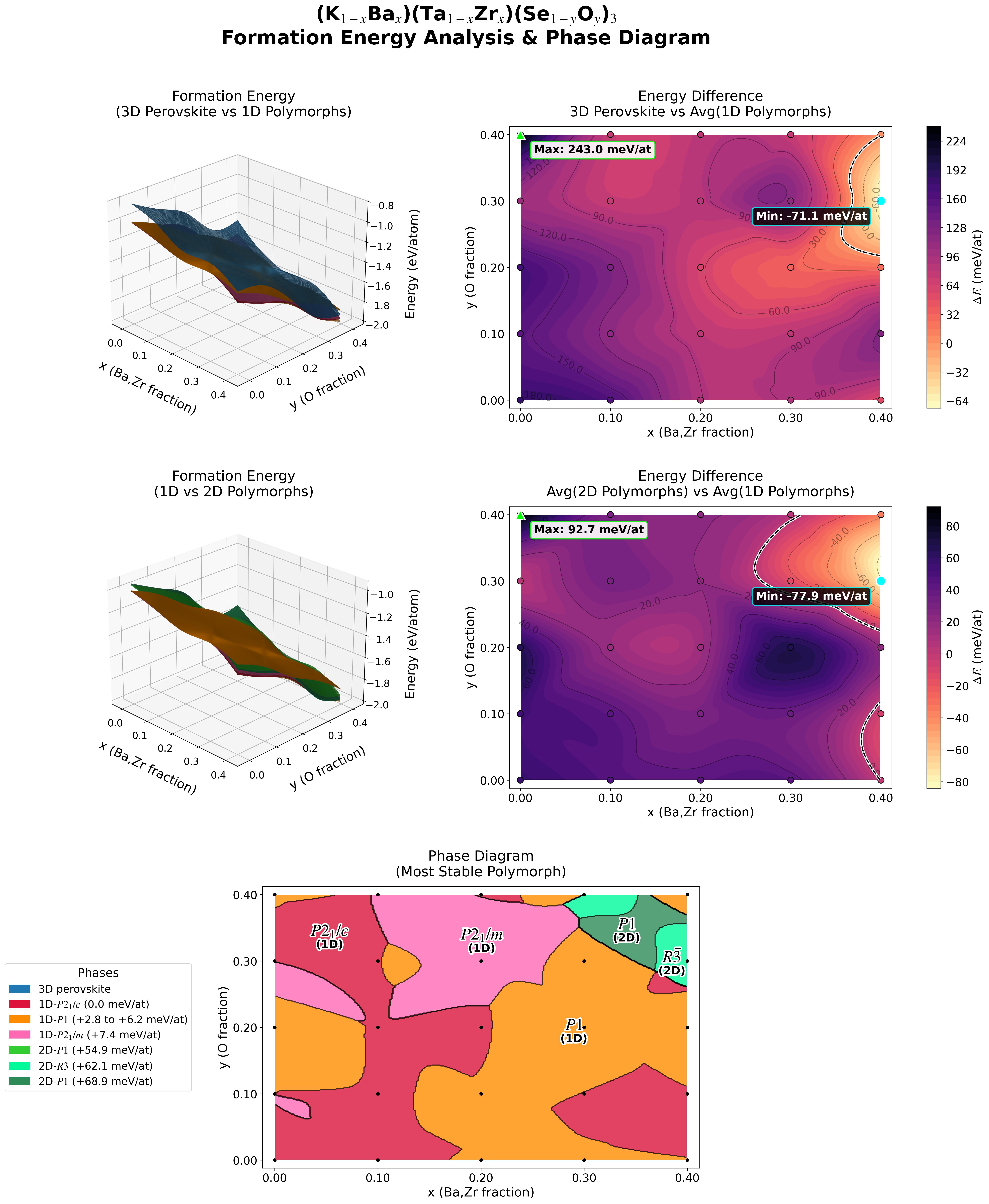}
    \caption{Thermodynamic stability analysis and structural evolution of the \ce{(K_{1-x}Ba_x)(Ta_{1-x}Zr_x)(Se_{1-y}O_y)3} solid solution. (\textbf{Top row}) Energy difference mapping between the targeted 3D perovskite framework and the average formation energy of competing 1D chain polymorphs. (\textbf{Middle row}) Thermodynamic comparison between the targeted 2D layered polymorphs and the 1D chain reference, revealing a deep stabilization gradient driven by O and Ba/Zr co-doping. Dashed contour lines indicate the $\Delta E = 0$ thermodynamic crossover points. (\textbf{Bottom}) The absolute compositional phase diagram, demonstrating the successful arrest of total dimensional collapse: while 1D non-perovskite configurations dominate the unalloyed space, targeted high-level co-doping ($x \ge 0.3$, $y \ge 0.9$) successfully shifts the thermodynamic ground state to the desired 2D layered motif.}
    \label{fig:ktase3_formation_energy_panel}
\end{figure}

Having mapped the dynamical stability boundaries, we evaluated the thermodynamic hierarchy to determine if the targeted alloying successfully prevents total structural collapse. We compared the formation energies of the optimized target frameworks against the 1D chain and 2D layered ordered superstructures mapped from the base \ce{KTaSe3} polymorphs, summarizing the comprehensive landscape in Figure~\ref{fig:ktase3_formation_energy_panel}.

We first compared the target 3D perovskite directly against the dominant 1D chain polymorphs (Figure~\ref{fig:ktase3_formation_energy_panel}, top row). In the pristine \ce{KTaSe3} limit ($x=0$, $y=0$), the 3D network suffers a massive thermodynamic penalty, sitting approximately +243.0~meV/atom higher in energy than the 1D average, confirming its overwhelming propensity to decompose. However, as the concentrations of both the aliovalent cations and oxygen increase, the energy surfaces converge. The thermodynamic difference map reveals a distinct crossover region at high substitution levels where the 3D phase dips below the average energy of the 1D polymorphs, reaching a local stabilization minimum of -71.1~meV/atom. While this is a remarkable energetic recovery for the 3D framework, it must be evaluated in the context of the competing 2D phases. 

We then evaluated the energy difference between the 2D layered polymorphs and the 1D chains (Figure~\ref{fig:ktase3_formation_energy_panel}, middle row). Initially, configurations attempting to form 2D networks reside significantly higher in energy than the 1D chains, presenting a maximum penalty of +92.7~meV/atom in the un-alloyed limit. Yet, the introduction of the \ce{Ta-O} chemical wedge alongside the inter-layer \ce{Ba^{2+}} charge bridge induces a much steeper and deeper stabilization gradient than observed for the 3D phase. The dashed contour line outlines a massive thermodynamic valley where the 2D polymorphs become significantly more stable than the 1D chains, pushing the energy difference to a deep minimum of -77.9~meV/atom at the high-doping extreme. 

The final structural outcome is captured in the phase diagram mapping the lowest formation energy among all considered phases (Figure~\ref{fig:ktase3_formation_energy_panel}, bottom). As expected from the initial geometric mismatch, the lower-doping regions of the solid solution are entirely dominated by various 1D non-perovskite configurations (predominantly $P\overline{1}$ and $P2_1/c$ symmetries). However, the phase diagram confirms the success of our alloying hypothesis: at the highest tested co-doping regimes (Ba/Zr fraction $x \geq 0.3$ and O doping $y \geq 0.9$), the 1D chains are thermodynamically defeated. In this targeted stability window, the 2D layered frameworks (specifically the $P\overline{1}$ and $R\overline{3}$ phases) emerge as the thermodynamic ground states. This multi-tiered thermodynamic analysis proves that while the geometric frustration in \ce{KTaSe3} is too severe to recover a perfectly cubic 3D perovskite, intelligent compositional engineering can successfully arrest the complete dimensional collapse, guiding the discovery of a thermodynamically and dynamically stable 2D structural phase with preserved active connectivity.

The phases determined to be thermodynamically optimal for both \ce{Cs2KInI6} and \ce{KTaSe3}---namely, \ce{Cs2K_{0.9}Na_{0.1}InI3Br3} ($x=0.1, y=0.5$) and \ce{K_{0.6}Ba_{0.4}Ta_{0.6}Zr_{0.4}Se_{2.1}O_{0.9}} ($x=0.4, y=0.3$)---were subjected to a final validation step. Using VibroML, we evaluated both their 0~K dynamical stability and 300~K thermal stability. Both alloyed frameworks were confirmed to be dynamically stable, presenting clean phonon band structures with no imaginary soft modes. Furthermore, 10~ps molecular dynamics simulations at ambient temperature confirmed their structural resilience. Both candidate materials maintained tight volume fluctuations ($<3\%$), minimal atomic drift (RMSD $\le 0.6$~\AA), and strict preservation of their initial structural motifs, showing no signs of spontaneous collapse or phase segregation. The detailed phonon dispersion plots and MD stability metrics are provided in Supplementary Section~\ref{sec:final_validation}.

\newpage
\subsection{Automated Discovery in Sparsely Populated Compositional Regions}

Current computational databases house millions of relaxed crystal structures, yet vast regions of compositional space, specifically complex quaternary systems and beyond, remain unexplored. In these ``white spaces,'' high-throughput screening pipelines frequently attempt a single, high-symmetry structural guess. 
This often leads to a false negative: the chemical space is deemed unviable, when in reality, the material was simply trapped in a frustrated, high-symmetry state during the simulation. These candidate materials are not necessarily unstable; they are just waiting to be relaxed into their true, lower-energy configurations. To demonstrate how the VibroML pipeline can be deployed at scale to rescue these ``failed'' stoichiometries and populate empty compositional regions, we engineered a targeted data-mining and automated remediation workflow. 

First, we scanned the massive Alexandria dataset to evaluate the convex hull coverage of quaternary and quinary systems. We specifically filtered for chemical spaces that had been virtually abandoned: systems containing exactly one generated material candidate that failed to reach the thermodynamic hull ($\Delta E_{\text{hull}} > 0$). To ensure these targets were prime candidates for symmetry-breaking remediation rather than inherently unphysical chemistries, we applied specific heuristic filters. Candidate structures had to reside within a plausible energy window above the hull (between 0.05 and 0.40~eV/atom), possess small unit cells ($\le 20$ atoms) for rapid phonon tractability, and exhibit high crystallographic symmetry (Space Group $\ge 160$). We additionally excluded systems containing heavy, rare-earth or radioactive elements to maintain broad technological relevance. This scan revealed the scale of the limitation, identifying 6,345 quaternary and 229 quinary systems abandoned primarily because their high-symmetry representations, such as cubic space groups 225 and 216, were dynamically unstable.

From this large reservoir of frustrated materials, we randomly sampled a representative subset of 40 abandoned systems (28 quaternary and 12 quinary) to submit to the automated VibroML remediation pipeline, utilizing the MACE-OMAT-0 MLIP and the energy-guided genetic algorithm (\texttt{GA}). The results, highlighted in Table~\ref{tab:whitespace_summary}, confirmed our central hypothesis: the vast majority of these abandoned candidates were severely dynamically unstable, characterized by pervasive soft phonon modes. VibroML's evolutionary search successfully broke these rigid spatial symmetries and navigated the previously uncharted potential energy landscapes. For individual stoichiometries, the algorithm identified dozens of unique, low-symmetry local minima. For example, in the quaternary \ce{Cs2MoHI6} and quinary \ce{BaYScGeO6} systems, the GA uncovered 106 and 81 unique polymorphs, respectively. By resolving the dynamical instabilities, VibroML facilitated massive thermodynamic relaxations, lowering the formation energies by -221.2~meV/atom and -138.9~meV/atom for these respective compounds. This proves that VibroML can systematically pull ``failed'' database entries drastically closer to, and potentially onto, the multi-component convex hull.

To extend this discovery capability into entirely unexplored regions, we coupled VibroML with ProtoCSP's ``cold start'' prototype retrieval module to investigate absolute white spaces. Our baseline analysis reveals that standard high-throughput screening has left massive voids in multi-component chemistry, even when restricting the search to 52 technologically relevant main-group and transition elements. Within this constrained chemical sandbox, 49.1\% of all mathematically possible four-element combinations (over 133,000 systems) and 99.5\% of five-element combinations (nearly 2.6 million systems) had no entries in the Alexandria database.
From this vast expanse of unexplored chemistry, we selected 10 entirely unmapped double-perovskite (elpasolite) stoichiometries (e.g., \ce{K2BiSbF6}, \ce{Li2CoZnI6}, \ce{Na2CrFeCl6}) and evaluated them using two distinct generation strategies. In the first approach, we manually forced the stoichiometries into the ideal cubic elpasolite template ($Fm\overline{3}m$). Unsurprisingly, unless the ionic radii were well matched, this forced high-symmetry initialization resulted in major dynamical instability, with soft modes affecting up to 40\% of the Brillouin zone. VibroML performed profound structural reconstructions to stabilize the lattices, yielding massive energy drops, such as a remarkable -380.1~meV/atom stabilization for \ce{Li2CoZnI6}.

In the second approach, we utilized ProtoCSP's native database retrieval, allowing it to autonomously search the LeMat-Bulk dataset for the most chemically similar prototype and perform elemental transmutation. This intelligent retrieval frequently bypassed the idealized cubic geometry entirely, selecting lower-symmetry prototypes most of the time. Consequently, the initial dynamical instabilities of the ProtoCSP-generated structures were drastically reduced. Remarkably, for several highly frustrated chemistries (such as \ce{Na2BiNbI6} and \ce{Na2CoCrBr6}), ProtoCSP's baseline transmutation yielded configurations that were dynamically stable out-of-the-box (Table~\ref{tab:whitespace_summary}), requiring no subsequent remediation. 

This dual capability demonstrates that the integrated ProtoCSP and VibroML pipeline is a highly versatile engine for accelerated materials discovery. Whether intelligently retrieving stable low-symmetry approximations from scratch or breaking through the ``trapped'' high-symmetry states of failed entries, the toolkit provides a robust, automated pathway to populate the vast chemical white spaces of computational materials science. The complete energetic and vibrational data for all 50 sampled systems is available in Supplementary Table~\ref{tab:full_whitespace_data}.

\begin{table}[htbp]
    \centering
    \caption{\textbf{Representative remediation of high-symmetry candidates in sparsely populated compositional spaces.} This table highlights the performance of the VibroML + ProtoCSP pipeline across previously abandoned database entries and unmapped ``white space'' chemistries. To demonstrate the value of intelligent prototype retrieval, the same elpasolite targets are evaluated by forcing a rigid cubic baseline versus allowing ProtoCSP to transmute an optimal candidate structure. $\Delta E$ denotes the formation energy stabilization relative to the initial structure. $\omega_{min}^{\text{init}}$ and $\omega_{min}^{\text{final}}$ represent the softest identified phonon mode (in THz) before and after GA optimization, while $\zeta_{\text{init}}$ indicates the initial percentage of the Brillouin zone affected by imaginary modes.}
    \label{tab:whitespace_summary}
    \small
    \renewcommand{\arraystretch}{1.2}
    \begin{tabular}{@{}llcccc@{}}
        \toprule
        \textbf{System Category} & \textbf{Composition} & \textbf{Polymorphs} & \begin{tabular}{@{}c@{}}\textbf{$\Delta E$} \\ \textbf{(meV/at)}\end{tabular} & \begin{tabular}{@{}c@{}}\textbf{$\omega_{min}^{\text{init}}$} \\ \textbf{[$\zeta_{\text{init}}$]}\end{tabular} & \textbf{$\omega_{min}^{\text{final}}$} \\
        \midrule
        \multirow{2}{*}{\textbf{Abandoned (Quaternary)}} 
        & \ce{Cs2MoHI6} & 106 & -221.2 & -10.33 \ [10.2\%] & 0.00 \\
        & \ce{Rb2SiNiI6} & 82 & -120.7 & -2.01 \ [11.9\%] & -0.02 \\
        \midrule
        \multirow{2}{*}{\textbf{Abandoned (Quinary)}} 
        & \ce{BaYScGeO6} & 81 & -138.9 & -7.36 \ [17.7\%] & 0.00 \\
        & \ce{BaYTaNbO6} & 133 & -260.5 & -6.05 \ [19.1\%] & 0.00 \\
        \midrule
        \multirow{3}{*}{\textbf{White Space (Forced Cubic)}} 
        & \ce{Li2CoZnI6} & 76 & -380.1 & -5.11 \ [38.8\%] & 0.00 \\
        & \ce{Na2BiNbI6} & 100 & -326.3 & -2.38 \ [39.9\%] & -0.27 \\
        & \ce{Na2CoCrBr6} & 92 & -138.4 & -2.20 \ [31.0\%] & 0.00 \\
        \midrule
        \multirow{3}{*}{\textbf{White Space (Transmuted)}} 
        & \ce{Li2CoZnI6} & - & - & -0.04 \ [0.53\%] & - \\
        & \ce{Na2BiNbI6} & - & - & 0.00 \ [0.13\%] & - \\
        & \ce{Na2CoCrBr6} & - & - & 0.00 \ [0.05\%] & - \\
        \bottomrule
    \end{tabular}
\end{table}

\newpage

\section{Discussion}
In this work, we introduced VibroML, an open-source Python toolkit that fundamentally shifts the computational materials design paradigm from merely diagnosing dynamical instabilities to actively remediating them. By leveraging state-of-the-art E(3)-equivariant foundational MLIPs—such as MACE, eSEN, and UMA—VibroML achieves near-DFT accuracy at a fraction of the computational cost, providing a holistic framework for the rapid discovery and validation of novel crystals.

Our comprehensive benchmarking demonstrates the effectiveness of this integrated approach across four key pillars:

\begin{itemize}
    \item[\textbf{I.}] \textbf{Automated Structural Remediation:} VibroML's diverse suite of optimization algorithms, most notably its energy-guided evolutionary genetic algorithm, successfully navigates complex potential energy landscapes. It consistently identifies a rich diversity of dynamically stable, lower-energy polymorphs that traditional mode-following techniques frequently miss.
    
    \vspace{0.2cm}
    
    \item[\textbf{II.}] \textbf{Finite-Temperature Validation:} Beyond static 0~K optimization, VibroML's automated molecular dynamics (MD) workflows successfully filter out metastable, kinetically trapped phases and numerical DFT artifacts, demonstrating strong agreement with experimental observations. Concurrently, the integrated nudged elastic band (NEB) module accurately maps phase transition pathways, providing critical kinetic insights while highlighting the nuanced predictive capabilities of current MLIPs.
    
    \vspace{0.2cm}
    
    \item[\textbf{III.}] \textbf{Targeted Compositional Engineering:} To address geometrically frustrated materials where a highly desirable topology collapses under internal strain, the coupled VibroML-ProtoCSP pipeline systematically nudges the local chemistry via targeted alloying. We demonstrated this predictive power by successfully rescuing the functional 3D and 2D networks of the complex \ce{Cs2KInI6} and \ce{KTaSe3} perovskite systems.
    
    \vspace{0.2cm}
    
    \item[\textbf{IV.}] \textbf{Systematic Population of Uncharted Chemistries:} Standard high-throughput screening has left staggering voids in multi-component chemistry. By applying ProtoCSP's intelligent ``cold start'' prototype retrieval and VibroML's structural remediation to a representative sample of abandoned chemistries, we demonstrated that this automated engine can smartly populate these white spaces. This comprehensive deep-screening approach successfully rescues ``failed'' database entries trapped in high-symmetry geometric frustration, yielding physically sound structural propositions that surpass the capacity of standard workflows.
    
\end{itemize}

By transforming dynamical stability assessment from a slow, manual verification step into a rapid, automated remediation engine, VibroML significantly accelerates the materials discovery cycle. We envision coupling this toolkit with state-of-the-art generative AI models to systematically conquer the search space for robust, experimentally viable functional materials.

\section{Methods}
\subsection{Model selection}
The core of the VibroML workflow relies on the accuracy of the underlying PES. While the application of foundational MLIPs has been broadly validated for phonon properties \cite{loewUniversalMachineLearning2025}, early generations of universal models were noted to suffer from systematic phonon frequency underestimation, or ``softening'', due to the biased sampling of near-equilibrium states in their training data \cite{anamComprehensiveAssessmentBenchmark2025,dengSystematicSofteningUniversal2025}.

To mitigate this issue and ensure reliable dynamical stability assessments, VibroML integrates the latest iterations of equivariant architectures, specifically MACE-OMAT-0 (medium)~\cite{batatiaFoundationModelAtomistic2025, GitHubACEsuitMacefoundations}, eSEN-30M-OMA~\cite{fuLearningSmoothExpressive2025}, and UMA-m-1p1~\cite{shiotaUniversalNeuralNetwork2024}. The training data for these models includes comprehensive samplings of non-equilibrium configurations from the OMat24 database~\cite{barroso-luqueOpenMaterials20242024}, allowing them to better capture the interatomic force constants and high-frequency modes essential for accurate vibrational analysis. It is important to note that the accuracy of such foundation models is inherently tied to the level of theory of their training data, which for these models is primarily the PBE functional~\cite{radovaFinetuningFoundationModels2025}. While this may not capture certain complex electronic effects with the highest possible accuracy, it provides an excellent and well-established baseline for structural relaxation and vibrational properties, representing an excellent trade-off between speed and accuracy for the discovery phase.

\subsection{Phonon calculation and dynamical instability remediation algorithms}
The primary function of VibroML is to assess the dynamical stability of a crystal structure by calculating its phonon properties. This is achieved through a direct interface with established lattice dynamics codes included in the Atomic Simulation Environment (ASE) \cite{hjorthlarsenAtomicSimulationEnvironment2017}, employing the finite-displacement supercell method and using the foundational MLIP potential as the force calculator. Details on the calculation of phonon dispersions with the aforementioned method are provided elsewhere \cite{alfePHONProgramCalculate2009, togoFirstprinciplesPhononCalculations2023}. VibroML automatically processes the phonon dispersion data along a series of q-points along high-symmetry paths in the Brillouin zone of the crystal structure to identify any phonon modes with imaginary frequencies, which are conventionally represented as negative values on dispersion plots. The toolkit extracts the key information for the most unstable mode(s) in the high-symmetry reciprocal points, including its frequency, its wavevector (q-point), and its eigenvector, which describes the collective atomic displacement pattern of the instability. This information serves as the starting point for the automated instability remediation workflows. Upon detection of a soft mode, one of several optimization strategies can be deployed depending on the method chosen by user, controlled via the \texttt{--method} command-line flag.

\subsubsection{Single soft-mode following (\texttt{--method traditional})}
The search is guided by the eigenmode of the identified softest mode or a mode provided by the user via a precomputed YAML phonon band structure, specifying the mode band index and q-point (in that case, \texttt{--yaml-input}, \texttt{--band\_idx} and \texttt{--q} must be specified). The smallest commensurate supercell to the q wavevector is used by default. Multiple distorted structures are then generated from predefined values of displacement amplitude along the eigenmode and unit cell deformations. This is intended to limit structural relaxation back into its initial unstable local minimum. Finally, a new phonon calculation is performed on the relaxed structure to check if the instability has been resolved. This process is repeated for the number specified in \texttt{soft\_mode\_max\_iterations} or until no soft modes are present in any of the special K-points.

\subsubsection{Coupled soft-mode following (\texttt{--method traditional\_all})}
Sometimes certain polymorphs can only be found by considering the interaction of multiple unstable modes. To this end, this method identifies all soft modes present in the high-symmetry points. It then systematically pairs the softest mode, the one with lowest (``negative'') frequency, with every other identified soft mode. For each pair, it explores different ratios of displacements and also swaps the two modes as the primary displacement. This thorough, exhaustive search ensures that all potential interactions between different instability pathways are explored, increasing the chance of finding the true ground state. If \texttt{soft\_mode\_max\_iterations} is greater than 1 and the lowest-energy structure from this initial search still presents soft modes, it is fed back to the algorithm for another iteration covering all soft modes still present.
\subsubsection{Evolutionary search via genetic algorithm (\texttt{--method ga})}
This method uses a genetic algorithm (GA) to find stable structures by treating the problem as a global optimization challenge. Inspired by Darwinian evolution, a GA excels at navigating the complex energy landscapes of materials and has been widely employed for crystal structure prediction (CSP)~\cite{oganovCrystalStructurePrediction2009,abrahamImprovedRealspaceGenetic2008,curtisGAtorFirstPrinciplesGenetic2018,omeeCrystalStructurePrediction2024}. 
It works by ``evolving'' a population of candidate structures toward lower energy and dynamic stability. The process involves creating an initial population of structures and then iteratively improving them. The algorithm uses operators like crossover and mutation to create new generations of structures from the most ``fit'' (lowest energy) individuals. This continuous cycle of improvement allows the method to efficiently find the most stable structure while also exploring a wide range of possible polymorphs. Additional details on the GA implementation are provided in Supplementary Information (Section~\ref{sec:ga_si}).

\subsubsection{Stochastic search via random displacements (\texttt{--method opt\_random})}
This approach explores the PES by applying random Cartesian displacements to the atoms and random perturbations to the unit cell. The magnitude of these random perturbations is controlled by user-defined bounds. This method avoids the constraints of a specific distortion pathway given by soft eigenmodes, but still works in generations, taking the lowest energy structure as the reference for the next iteration. This method allows to uncover novel stable configurations that might be missed by more directed search techniques, making it a valuable tool for broad, exploratory materials discovery.

\subsection{Testing the dynamical instability remediation algorithms}

We validated our remediation workflows by applying them to four benchmark compounds. Our objective was to establish a fair and rigorous baseline that would expose the true exploratory power and computational scaling of each method. As detailed in Supplementary Table~\ref{tab:sets_remedalgorithms}, we applied standardized settings across the board; consequently, the inherent algorithmic differences and system sizes yielded significantly different execution times. For the large \ce{Bi2Sn2O7} system, we constrained the initial supercell size and volume expansion threshold to prevent memory exhaustion during phonon computations. However, to maintain a fair assessment of native scaling, the maximum number of atoms per supercell was left unconstrained.

\subsection{Analysis of Transformation Pathways and Thermal Stability}
Beyond finding stable ground states, VibroML provides workflows for deeper physical analysis of material transformations and behavior under realistic conditions.

\subsubsection{Minimum energy path finding (\texttt{--method neb})}
Understanding the mechanism and kinetic barrier of a phase transition from one crystal structure to another is a central problem in materials science. NEB is the standard computational technique for this purpose, allowing for the determination of the minimum energy path (MEP) and the associated activation energy barrier between the known initial and final states. A major challenge in solid-state NEB is ensuring that the unit cell dimensions and atomic indices of the endpoints are perfectly compatible. VibroML fully automates this preprocessing step, allowing users to input arbitrary unit cells without manual matching. The algorithm then performs a variable-cell NEB calculation, optimizing a combined space of atomic and lattice degrees of freedom with the aim to describe reconstructive phase transitions. Additionally, a climbing-image NEB variation is implemented (\texttt{--method ci-neb}) which may better capture the true transition state and activation barrier. Detailed implementation is provided in Supplementary Information (Section~\ref{sec:neb_SI}).

\subsubsection{Thermal stability validation via molecular dynamics (\texttt{--method md\_stability})}
Phonon analysis based on the harmonic approximation provides a rigorous assessment of dynamical stability at a temperature of 0 K. However, to confirm that a structure remains stable under realistic operating conditions, it is necessary to consider finite-temperature effects, including anharmonic lattice vibrations. VibroML facilitates this through a MD simulation workflow.

The MD stability workflow is as follows:
\begin{itemize}
    \item \textbf{Simulation Setup:} An MD simulation is performed on a supercell of the candidate structure which can be adjusted by the user. The user specifies the target temperature (\texttt{--temp}), pressure (\texttt{--pressure}), and total simulation time (\texttt{--time}) including the time for equilibration. The equations of motion are integrated using the forces derived from the MLIP potential. The simulation is divided into two phases: equilibration and production.
    \item \textbf{Equilibration:} An initial ``equilibration'' phase allows the system to reach thermal equilibrium at the target conditions. The equilibration starts with the canonical statistical ensemble (NVT) in which a thermostat is gradually applied to reach the target temperature. Then a second phase starts which uses the isothermal-isobaric (NPT) ensemble to reach an equilibration volume along with the gradual use of a barostat to reach the target pressure.
    \item \textbf{Production:} This phase also employs the NPT ensemble and now runs for the rest of the duration specified in the simulation. Only the data obtained in this phase is used to evaluate stability.
    \item \textbf{Analysis of Stability Metrics:} The stability of the crystal structure throughout the production run is assessed by monitoring several key metrics derived from the atomic trajectory:
    \begin{itemize}
        \item \textbf{Root Mean Square Deviation (RMSD):} The RMSD of atomic positions with respect to their initial lattice sites is calculated over time. For a stable crystal, the RMSD will fluctuate around a small, constant value, indicating that atoms are vibrating around their equilibrium positions. A continuous, unbounded increase in RMSD is a clear indicator of structural degradation, such as melting or amorphization~\cite{piechotaInitioMolecularDynamics2024}. Quantitative thresholds can be established to distinguish between stable solid, solid-solid transition, and molten states. The default threshold implemented in VibroML is a generous 1~\AA.
        
        \item \textbf{Radial Distribution Function (RDF):} The RDF, $g(r)$, is calculated for the initial structure and compared against the ensemble-averaged RDF computed from the final segment of the trajectory. A stable crystal will retain sharp, well-defined peaks. Meanwhile, a loss of crystallinity is marked by a broadening and smoothing of these peaks into a liquid-like profile~\cite{kazancInvestigationSolidSolid2013}. We quantify this using the Pearson correlation coefficient between the initial and final averaged RDFs; a correlation below 0.5 signals a loss of structural integrity.
        
        \item \textbf{Thermodynamic Properties:} The time evolution of the total potential energy, temperature, cell volume, and the full stress tensor (hydrostatic pressure and shear components) are monitored. Stable, equilibrated systems will exhibit fluctuations around well-defined average values. In our analysis, volume stability is assessed by comparing the fluctuations during the production phase against a reference average from the initial 2 ps of equilibration. A deviation exceeding a threshold (set to 6\% by default) triggers an instability warning.
        
        \item \textbf{Symmetry Retention:} Thermal vibrations at finite temperatures instantaneously break the ideal crystal symmetry, rendering standard space group analysis on snapshot frames ineffective. To address this, VibroML computes a time-averaged structure from the final 50 frames of the production trajectory, effectively filtering out thermal noise to recover the equilibrium lattice positions. The space group of this averaged structure is then compared to the initial structure using the Pymatgen symmetry analyzer~\cite{ongPythonMaterialsGenomics2013} across a sweep of symmetry tolerances (0.01 to 1.0 \AA). If the original space group is not recovered within these tolerances, it indicates a phase transition or amorphization.
    \end{itemize}
\end{itemize}

The evaluation of thermal stability using only MLIPs must be approached cautiously, therefore, all these criteria along with data plots are reported to the user and a confidence level to the stability or instability of the structure is assigned for their evaluation.

\subsection{Automated Structure Generation via ProtoCSP} 

To enable the broad exploration of chemical spaces, including targeted alloying and the prediction of entirely novel compounds, we developed ProtoCSP, an open-source Python package (available at \url{https://github.com/rogeriog/ProtoCSP}) designed to integrate seamlessly with the VibroML workflow. ProtoCSP operates in two primary modes: combinatorial superstructure generation for exploring solid solutions, and database-driven prototype retrieval for ``cold start'' materials discovery.

\begin{itemize}
    \item[\textbf{I.}] \textbf{Combinatorial Superstructure Generation for Alloying:}\\
    When investigating solid solutions to relieve geometric frustration in unstable baseline materials, ProtoCSP systematically explores compositional variations across known structural frameworks. Given a base crystal structure (such as a specific polymorph mapped by VibroML) and a target fractional composition (e.g., \ce{Cs2(K_{0.8}Na_{0.2})In(I_{0.67}Br_{0.33})6}), ProtoCSP automatically generates exact integer-ratio supercells to accommodate the desired stoichiometry. It then enumerates all symmetrically distinct atomic orderings within these supercells~\cite{hartAlgorithmGeneratingDerivative2008,ongPythonMaterialsGenomics2013, morganGeneratingDerivativeSuperstructures2017}. Because the optimal crystallographic sites for substitutional atoms are generally unknown \textit{a priori}, ProtoCSP interfaces directly with an MLIP engine to perform rapid structural relaxations and energy evaluations across the enumerated configurations. This process pinpoints the lowest-energy atomic ordering for every composition-polymorph pair. Finally, these optimized candidates can be forwarded back to VibroML for rigorous dynamical and thermal stability tracking, or routed directly to an \textit{ab initio} workflow for high-fidelity validation.

    \vspace{0.5cm} 

    \item[\textbf{II.}] \textbf{Prototype Retrieval and Substitution (Cold Start):}\\
    To explore entirely new chemical spaces where no initial experimental or theoretical structure exists, ProtoCSP can rapidly generate chemically plausible prototypes from the LeMat-Bulk dataset~\cite{ramlaouiLeMatTrajScalableUnified2025, LeMaterial_LeMat-BulkUnique_2023}. This algorithm proceeds in three stages:

    \begin{enumerate} 
        \item \textbf{Prototype Retrieval and Stoichiometry Matching:} For a target composition (e.g., \ce{SrTiO3}), the algorithm queries an indexed library of anonymized stoichiometries (e.g., \ce{ABC3}) derived from LeMat. It retrieves parent structures that match the target stoichiometry exactly.
        \item \textbf{Chemical Substitution:} Elements from the target composition are mapped to the crystallographic sites of the prototype based on electronegativity sorting. To ensure chemical compatibility, the target elements ($T$) and prototype species ($P$) are ranked by their Pauling electronegativity ($\chi$). The mapping $f: T \to P$ is constructed such that the most electropositive target element occupies the site of the most electropositive prototype species, minimizing the likelihood of placing cations in anion positions.
        \item \textbf{Physical Validity and Scaling:} After substitution, the unit cell lattice vectors are isotropically scaled to prevent atomic overlap. The scaling factor is determined by comparing the sum of the covalent radii of the new species against the original bond distances. Finally, a density check is performed to filter out unphysical configurations before passing the candidate to the stability workflow.
    \end{enumerate}

    For complex alloys ($>$ 4 elements), where exact stoichiometric matches in the database are rare, ProtoCSP employs a fuzzy matching algorithm to identify prototypes with overlapping element sets, followed by random substitution and decorrelation of site occupancy to model solid solutions.
\end{itemize}

\subsection{Ab-initio calculations}

To validate the MLIP predictions and establish a ground-truth PES, DFT calculations were performed using the Vienna \textit{Ab-initio} Simulation Package (VASP) version 5.4.4 \cite{kresseEfficientIterativeSchemes1996}. Exchange-correlation effects were treated within the Generalized Gradient Approximation (GGA) using the Perdew-Burke-Ernzerhof (PBE) functional \cite{perdewGeneralizedGradientApproximation1996}, and core electrons were described using the Projector Augmented Wave (PAW) method \cite{ kresseUltrasoftPseudopotentialsProjector1999}. High-throughput calculation workflows were managed using the Atomate2 library \cite{ganoseMaterialsprojectAtomate2V00132024}, employing the \texttt{DoubleRelaxMaker} workflow for geometry optimizations. A plane-wave energy cutoff of 520~eV was used throughout. Electronic self-consistency was converged to $10^{-6}$~eV, utilizing Gaussian smearing with a width of 0.05~eV to resolve electronic partial occupancies, and spin polarization was enabled for all structures. The Brillouin zone integration was performed using automatically generated $\Gamma$-centered $k$-point meshes. For the high-precision determination of relative polymorph stability and geometry optimization, a density of 6,000 $k$-points per reciprocal atom (kppa) was utilized. Conversely, for the exploration of the PES along interpolation paths and NEB calculations, a density of 3,000 kppa was employed.

\section*{Data availability}
The complete data supporting the findings of this study—including the raw output from the structural optimizations, phonon dispersion evaluations, molecular dynamics simulations, and NEB pathways—are openly available in the Support Data repository at \url{https://github.com/rogeriog/VibroML_SupportData}.

\section*{Code availability}
The source code for the software developed and utilized in this work is open-source and freely available. The VibroML toolkit can be accessed at \url{https://github.com/rogeriog/VibroML}, and the ProtoCSP package is available at \url{https://github.com/rogeriog/ProtoCSP}.

\bibliographystyle{naturemag}
\bibliography{references}

\section*{Acknowledgements}
Computational resources have been provided by the Consortium des Équipements de Calcul Intensif (CÉCI), funded by the Fonds de la Recherche Scientifique de Belgique (F.R.S.-FNRS) under Grant No. 2.5020.11 and by the Walloon Region. Additionally, the present research benefited from computational resources made available on Lucia, the Tier-1 supercomputer of the Walloon Region, infrastructure funded by the Walloon Region under the grant agreement n°1910247.

\section*{Funding}
This study was financed by Université catholique de Louvain (Ref. No. ARH/MKK/01155839).

\section*{Author contributions}
R.G and G.M.R. conceptualized the study. R.G. performed the experiments and data analysis, and wrote the main manuscript. G.M.R. supervised the investigation, reviewed and approved the final manuscript.

\section*{Competing interests}
The authors declare no competing interests.

\newpage
\appendix
\section*{Supplementary Information}
\setcounter{secnumdepth}{2} 

\setcounter{subsection}{0}                 
\renewcommand{\thesubsection}{S.\arabic{subsection}} 

\setcounter{table}{0} 
\renewcommand{\thetable}{S\arabic{table}}
\renewcommand{\theHtable}{SuppTable.\arabic{table}} 

\setcounter{figure}{0}
\renewcommand{\thefigure}{S\arabic{figure}}
\renewcommand{\theHfigure}{SuppFig.\arabic{figure}} 

\subsection{Genetic algorithm method in VibroML}
\label{sec:ga_si}
The GA implementation within VibroML operates as follows:
\begin{itemize}
    \item \textbf{Population and Individuals:} The algorithm maintains a ``population'' of candidate crystal structures, referred to as ``individuals''. The initial population is generated by applying random distortions, which are combinations of atomic displacements along soft mode eigenvectors and cell deformations, to the unstable input structure. The size of this population is a key parameter, controlled by \texttt{--ga\_population\_size}.
    \item \textbf{Fitness Function:} The ``fitness'' of each individual structure is evaluated to guide the evolutionary selection process. In the context of VibroML, fitness is linked to low energy, as we are typically interested in the lowest-energy polymorphs.
    \item \textbf{Selection and Variation Operators:} The algorithm iteratively creates new generations of structures by selecting the fittest individuals from the current population and applying variation operators to them.
    \item \textbf{Heredity (Crossover):} Pairs of high-fitness ``parent'' structures are selected to produce ``offspring''. This is achieved by combining their structural features, such as slicing and merging their atomic coordinates and lattice vectors, to inherit favorable traits.
    \item \textbf{Mutation:} To maintain genetic diversity and prevent premature convergence to a suboptimal local minimum, random changes are introduced into the offspring structures with a certain probability, governed by \texttt{--ga\_mutation\_rate}. These mutations can include random atomic displacements, cell perturbations, or coupling with another soft eigenmode.
\end{itemize}

This cycle of evaluation, selection, crossover, and mutation is repeated for a set number of generations (\texttt{--ga\_generations}). The lowest-energy structure then has its phonon band structure calculated, providing new soft phonon modes to continue the process. Since this method is designed to explore as many polymorphs as possible, even after a dynamically stable structure is found, it repeats the process for the number defined in \texttt{soft\_mode\_max\_iterations}. However, since soft modes are no longer present, it will use the highest-frequency optical modes to break the structure and find new polymorphs. In this way, the population progressively ``evolves'' toward regions of the configuration space corresponding to dynamically stable and low-energy structures. This approach provides a powerful balance between ``exploitation'' (refining promising solutions) and ``exploration'' (discovering new regions of the search space), often converging on the global minimum stable structure efficiently while providing good coverage of the local minima in the PES.

\subsection{Structural Diversity and Polymorph Generation}
\label{sec:structural_diversity}

A key advantage of automated PES exploration is the ability to map a wide array of competing metastable phases. To quantify the exploratory power of the algorithms implemented in VibroML, we tracked the number of unique structures generated by each method across the benchmarked compounds. Structures were classified as unique based on a low energy difference threshold of 0.2~meV/atom ($0.0002$~eV) following relaxation.

The total counts of unique structures identified by each workflow are presented in Table~\ref{tab:unique_structures}. The traditional single-mode (\texttt{TRAD}) and coupled-mode (\texttt{ALL}) following techniques exhibit limited structural diversity, as they are deterministically constrained to specific distortion pathways branching from the initial high-symmetry reference. Conversely, the Evolutionary Genetic Algorithm (\texttt{GA}) and Stochastic Search (\texttt{RAND}) generate vastly more diverse structural ensembles.

It is worth noting that while the \texttt{RAND} method often produces the highest raw count of unique structures (e.g., 929 for \ce{HfO2}), many of these correspond to high-energy, low-symmetry configurations. The \texttt{GA} method provides the most effective balance: it achieves excellent structural diversity while its energy-guided evolutionary pressure ensures the population remains focused within the low-energy, physically viable regions of the PES. This is particularly evident in the highly complex \ce{Bi2Sn2O7} system, where the \texttt{GA} successfully mapped 186 unique structural configurations while efficiently avoiding the pathological stagnation that hampered the other methods.

\begin{table}[H]
    \centering
    \caption{Number of unique structures generated by each VibroML remediation algorithm. Uniqueness is defined by an energy difference threshold of $>0.2$~meV/atom among successfully relaxed candidates.}
    \label{tab:unique_structures}
    \renewcommand{\arraystretch}{1.2}
    \begin{tabular}{@{}lcccc@{}}
        \toprule
        \textbf{Compound} & \textbf{GA} & \textbf{TRAD} & \textbf{ALL} & \textbf{RAND} \\
        \midrule
        \textbf{\ce{LiF}} & 38 & 2 & 10 & 465 \\
        \textbf{\ce{CsPbI3}} & 134 & 3 & 10 & 505 \\
        \textbf{\ce{HfO2}} & 39 & 2 & 2 & 929 \\
        \textbf{\ce{Bi2Sn2O7}} & 186 & 3 & 5 & 5 \\
        \bottomrule
    \end{tabular}
\end{table}

\subsection{Steric clash anomaly: Evaluation of MLIP fidelity to DFT artifacts}
\label{sec:lif_anomaly}

\begin{figure}
    \centering
    \includegraphics[width=0.8\textwidth]{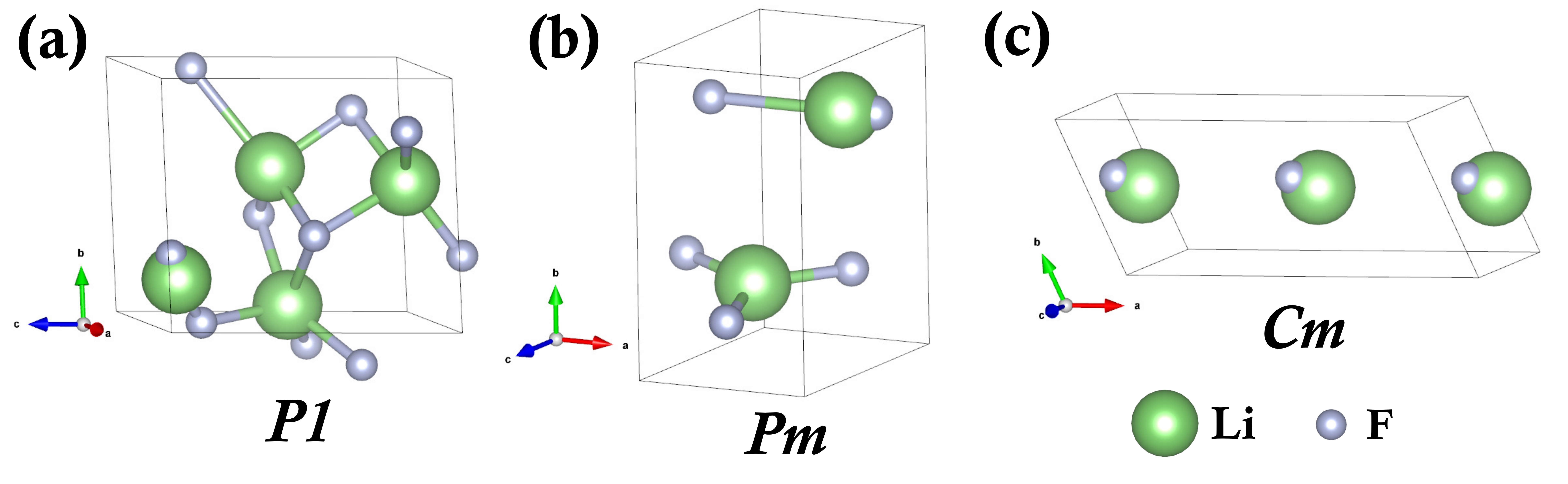}
    \caption{Unphysical LiF polymorphs identified during stochastic PES exploration with VibroML. Structural snapshots of the steric clash anomaly in (a) $P1$, (b) $Pm$ , and (c) $Cm$  space group structures.}
    \label{fig:lif_anomaly_structures}
\end{figure}

During the evaluation of the LiF system, the stochastic search (\texttt{opt\_random}) identified several structures with energies significantly lower than the experimental ground state, yet exhibiting unphysically short interatomic distances. These structures, assigned to $Cm$, $P1$, and $Pm$ symmetries, correspond to a numerical artifact in the DFT PBE PES where the expected Pauli repulsion collapses, effectively allowing a ``partial fusion'' of Li and F atoms, as shown in Figure~\ref{fig:lif_anomaly_structures}. This steric clash anomaly showcases the fidelity of MLIPs to their training data versus physical reality.

To rigorously characterize this anomaly, we employed a linear interpolation approach. We selected the hexagonal $P6_3mc$ polymorph as our reference anchor, as it was the lowest-energy physically valid structure identified by VibroML and verified by DFT, consistent with previous literature reports \cite{hortonAcceleratedDatadrivenMaterials2025} (Materials Project ID: mp-1185301). We then generated a transformation pathway consisting of 7 linearly interpolated images connecting this stable $P6_3mc$ phase to each of the three anomalous polymorphs ($Cm$, $P1$, and $Pm$). We subsequently evaluated the energy and forces along these paths using both the MLIPs and the reference DFT method.

\begin{figure}[H]
    \centering
    \includegraphics[width=\textwidth,  height=10cm,keepaspectratio=false]{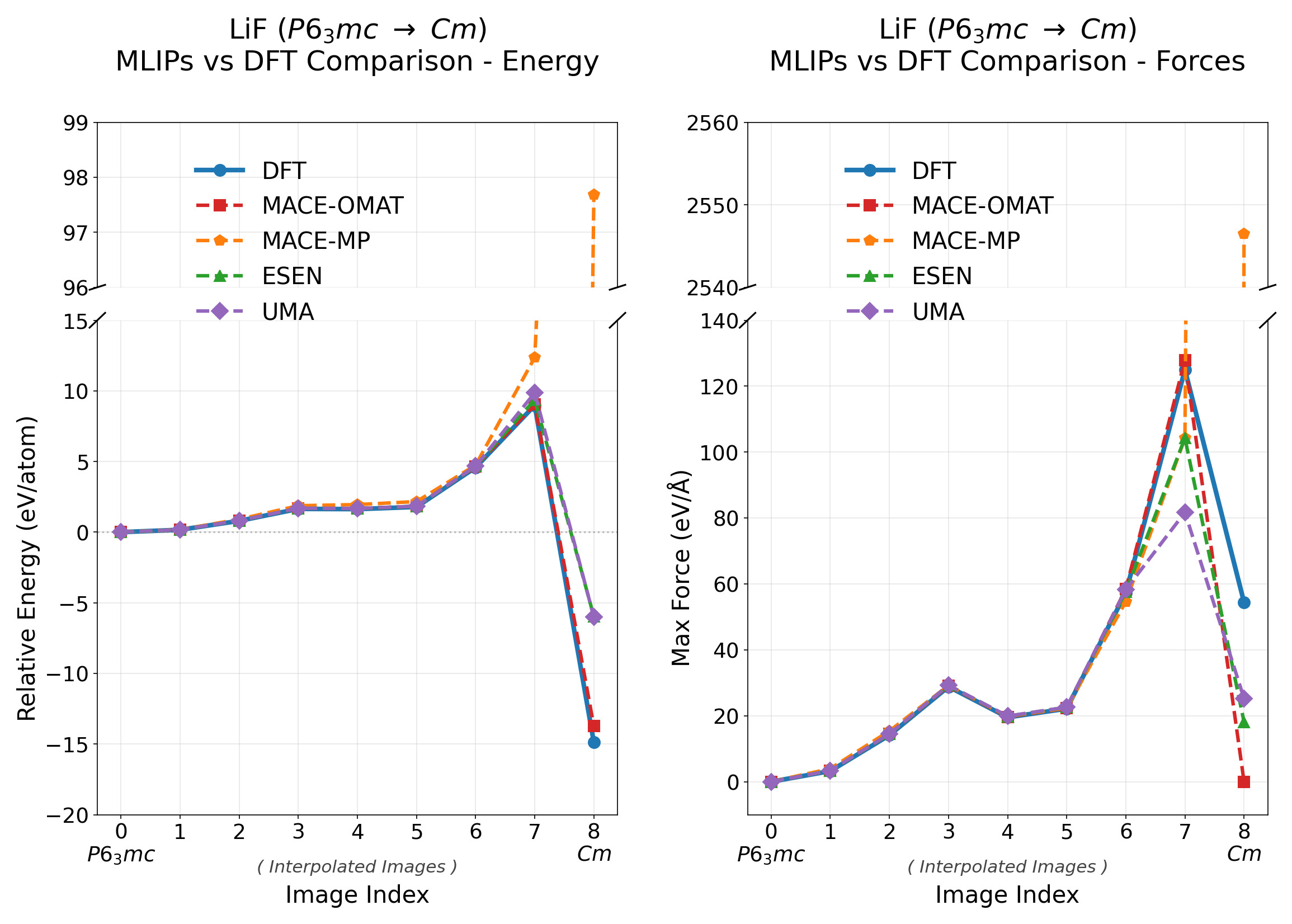}
    \vspace{-0.8cm}
    \caption{Detailed energy and force profiles for the steric clash anomaly in LiF along the $Cm$ transition path. Comparison of MACE-MP, MACE-OMAT, eSEN, and UMA (colored dashed lines) against the DFT reference (solid blue line).}
    \label{fig:lif_anomaly_cm}
\end{figure}

\begin{figure}[H]
    \centering
    \includegraphics[width=\textwidth, height=10cm,keepaspectratio=false]{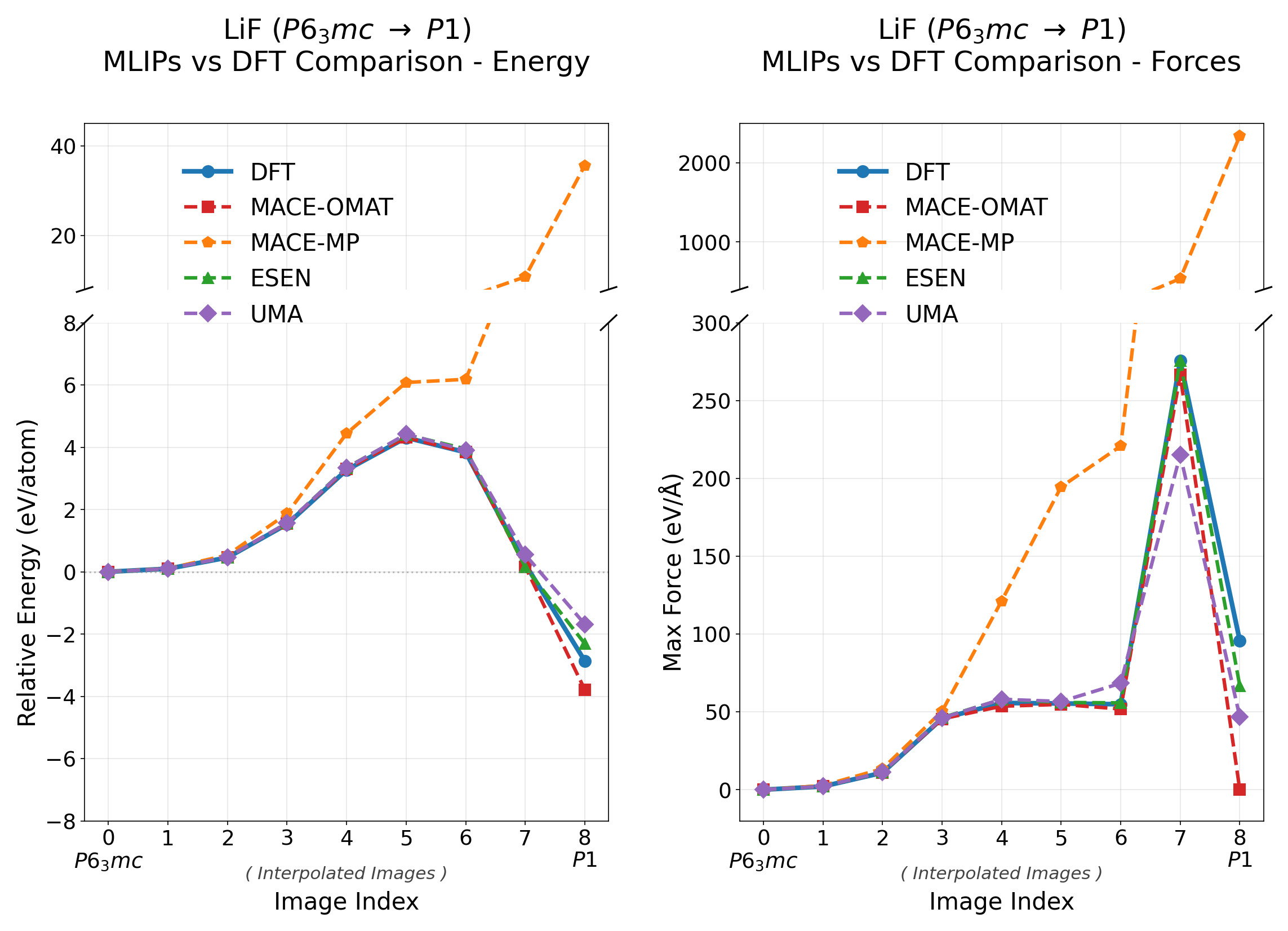}
    \vspace{-0.8cm}
    \caption{Detailed energy and force profiles for the steric clash anomaly in LiF along the $P1$ transition path. Comparison of MACE-MP, MACE-OMAT, eSEN, and UMA (colored dashed lines) against the DFT reference (solid blue line).}
    \label{fig:lif_anomaly_p1}
\end{figure}

\begin{figure}[H]
    \centering
    \includegraphics[width=\textwidth, height=10cm, keepaspectratio=false]{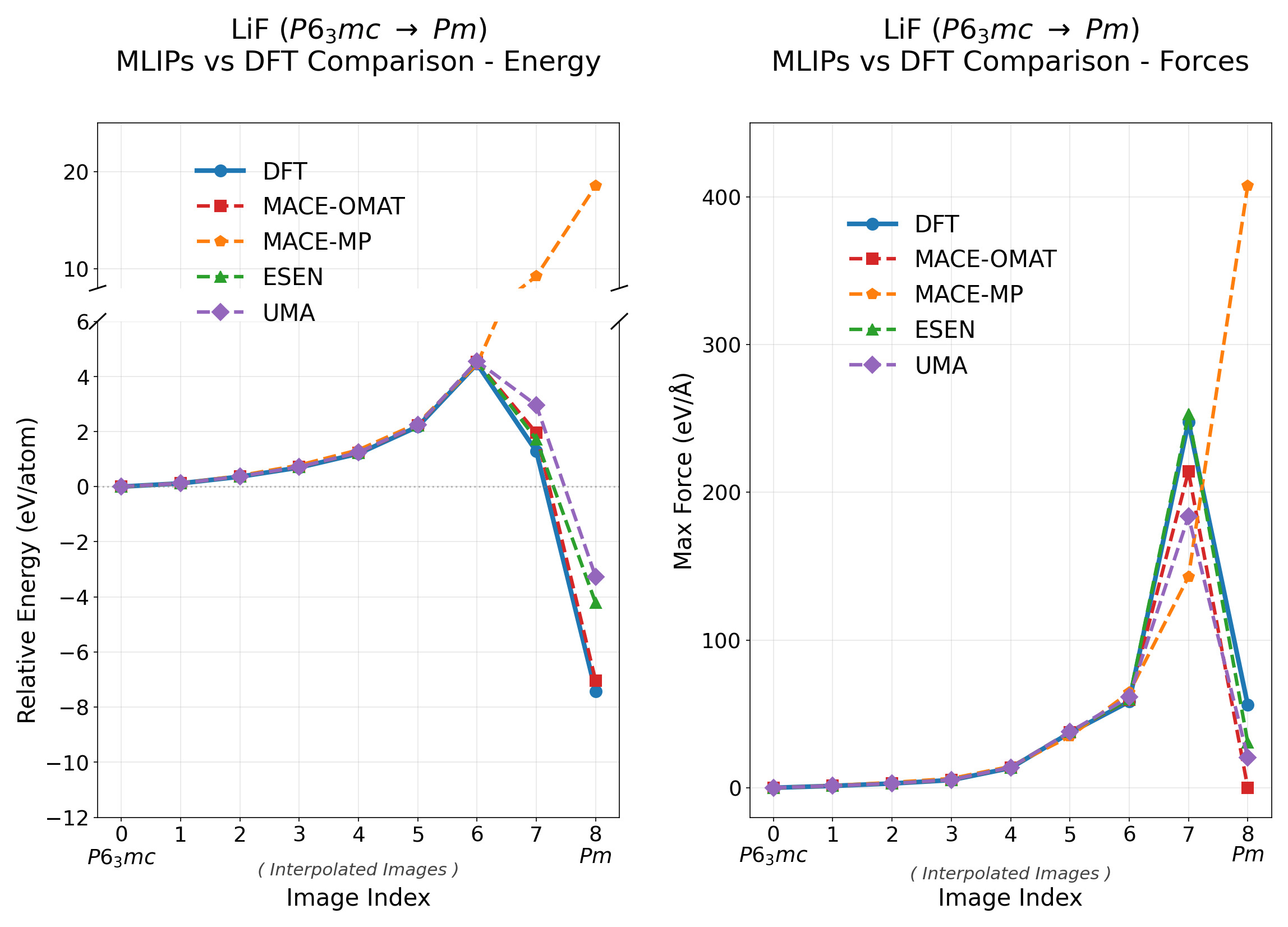}
    \vspace{-0.8cm}
    \caption{Detailed energy and force profiles for the steric clash anomaly in LiF along the $Pm$ transition path. Comparison of MACE-MP, MACE-OMAT, eSEN, and UMA (colored dashed lines) against the DFT reference (solid blue line).}
    \label{fig:lif_anomaly_pm}
\end{figure}

The DFT reference calculations reveal that at these very short interatomic distances, the PES does not show the expected steep Pauli repulsion. Instead, a ``softening'' occurs, resulting in significantly underestimated energy barriers compared to what would be expected from hard-sphere repulsion. 
We compared the performance of the MACE-MP model (trained primarily on near-equilibrium data) against the newer generation of models trained with abundant non-equilibrium data (MACE-OMAT, eSEN, and UMA). The quantitative results are summarized in Table \ref{tab:lif_anomaly}, and the energy/force profiles for the $Cm$, $P1$, and $Pm$ paths are visualized in Figures~\ref{fig:lif_anomaly_cm}, \ref{fig:lif_anomaly_p1}, and \ref{fig:lif_anomaly_pm}, respectively.

\begin{table}[H]
    \centering
    \caption{Comparison of energy barriers and maximum forces for LiF transformation paths across different MLIPs. The ``Anomaly'' refers to the DFT artifact where the barrier is unphysically low. MACE-MP fails to reproduce the artifact, resulting in large errors relative to DFT but a physically plausible repulsive barrier.}
    \label{tab:lif_anomaly}
    \resizebox{\textwidth}{!}{%
    \begin{tabular}{llrrrr}
        \toprule
        \textbf{Path} & \textbf{Model} & \textbf{Barrier} & \textbf{Barrier Error} & \textbf{Max Force} & \textbf{Force Error} \\
        & & (eV/atom) & (vs DFT) & (eV/\AA) & (vs DFT) \\
        \midrule
        \multirow{5}{*}{LiF\_Cm} 
        & DFT (Ref) & 8.98 & 0.00 & 125.00 & 0.00 \\
        & MACE-OMAT & 9.03 & 0.06 & 127.91 & 2.90 \\
        & eSEN & 9.37 & 0.39 & 104.25 & -20.75 \\
        & UMA & 9.87 & 0.89 & 81.72 & -43.29 \\
        & MACE-MP & 97.69 & 88.71 & 2546.49 & 2421.49 \\
        \midrule
        \multirow{5}{*}{LiF\_P1} 
        & DFT (Ref) & 4.31 & 0.00 & 275.63 & 0.00 \\
        & MACE-OMAT & 4.31 & 0.01 & 266.44 & -9.19 \\
        & eSEN & 4.40 & 0.10 & 275.64 & 0.01 \\
        & UMA & 4.42 & 0.11 & 215.23 & -60.40 \\
        & MACE-MP & 35.62 & 31.32 & 2341.07 & 2065.44 \\
        \midrule
        \multirow{5}{*}{LiF\_Pm} 
        & DFT (Ref) & 4.45 & 0.00 & 247.65 & 0.00 \\
        & MACE-OMAT & 4.53 & 0.08 & 213.99 & -33.66 \\
        & eSEN & 4.54 & 0.08 & 252.73 & 5.07 \\
        & UMA & 4.55 & 0.10 & 183.65 & -64.01 \\
        & MACE-MP & 18.52 & 14.06 & 407.51 & 159.85 \\
        \bottomrule
    \end{tabular}%
    }
\end{table}

The profiles in Figures~\ref{fig:lif_anomaly_cm}, \ref{fig:lif_anomaly_p1}, and \ref{fig:lif_anomaly_pm} reveal a stark contrast in model behavior. MACE-MP exhibits massive energy and force spikes (note the broken axes in the plots), consistent with a model predicting a steep repulsive wall in the absence of specific training data to the contrary. In contrast, MACE-OMAT, eSEN, and UMA reproduce the DFT barrier with high accuracy (errors $<$ 1 eV). These models have likely encountered similar high-energy or distorted configurations in their training sets (e.g., OMat24), allowing them to learn the specific topology of the DFT functional, including its artifacts.

This comparison serves as validation of the ``surrogate'' nature of modern foundation models. They act as faithful interpolators of the underlying electronic structure method. When the reference method contains artifacts, a sufficiently expressive and well-sampled MLIP will reproduce those artifacts, necessitating careful post-screening of results.

\subsection{Nudged elastic band methods in VibroML} 
\label{sec:neb_SI}

For solid-state transformations, which often involve changes in the unit cell volume and shape, it is crucial that the lattice vectors are allowed to relax along with the atomic positions. VibroML's implementation of the NEB method inherently supports this, performing a variable-cell calculation that optimizes a combined space of atomic and lattice degrees of freedom, which is essential for correctly describing reconstructive phase transitions. In detail, the algorithm works through the following automated steps:
\begin{itemize}
    \item \textbf{Structure Compatibility and Atom Mapping:} Before generating the transition path, VibroML automatically ensures physical compatibility between the endpoints. If the user provides geometrically incompatible unit cells (e.g., a 2x3x2 supercell versus a 3x2x3 supercell), the algorithm dynamically calculates the least common multiple for each element and generates identically sized, stoichiometric supercells. Furthermore, users do not need to manually map atomic indices. VibroML applies a periodic-boundary-aware global shift and employs the Hungarian algorithm~\cite{kuhnHungarianMethodAssignment1955} to pair atoms of the same species between the two structures, strictly minimizing the overall spatial displacement.
    \item \textbf{Path Discretization:} A discrete path consisting of a series of intermediate structures, or ``images'', is created between the initial and final states. To prevent unphysical atomic collisions during large cell shape deformations, this linear interpolation is performed on the fractional atomic coordinates alongside the lattice vectors, adhering to the minimum image convention.
    \item \textbf{Force Projections:} An optimization algorithm (steepest descent) is used to relax the chain of images. The total force on each image is composed of two components: the true force from the MLIP, projected perpendicular to the path, and a fictitious spring force between adjacent images, projected parallel to the path. This ``nudging'' procedure ensures that the images slide down the PES towards the MEP without cutting corners, while maintaining an even spacing along the path.
    \item \textbf{Climbing-Image NEB (\texttt{--method ci-neb} only):} To precisely locate the highest point on the MEP (the transition state or saddle point), a climbing-image modification can be employed. In this scheme, the image with the highest energy is no longer subjected to the spring force. Instead, the component of the true potential force parallel to the path is inverted, causing this image to be driven ``uphill'' along the path to converge exactly on the saddle point. This provides a more accurate value for the activation energy barrier.
    \item \textbf{Convergence Check:} After each optimization step, the algorithm checks if the convergence criterion has been met based on the maximum force magnitude across all intermediate images. If this maximum force falls below a user-defined threshold, the optimization terminates successfully.
\end{itemize}
\newpage

\subsection{Comparison of different potentials in energy, dynamical and thermal stability via VibroML}

\begin{figure}[H]
    \centering
    \includegraphics[width=\textwidth]{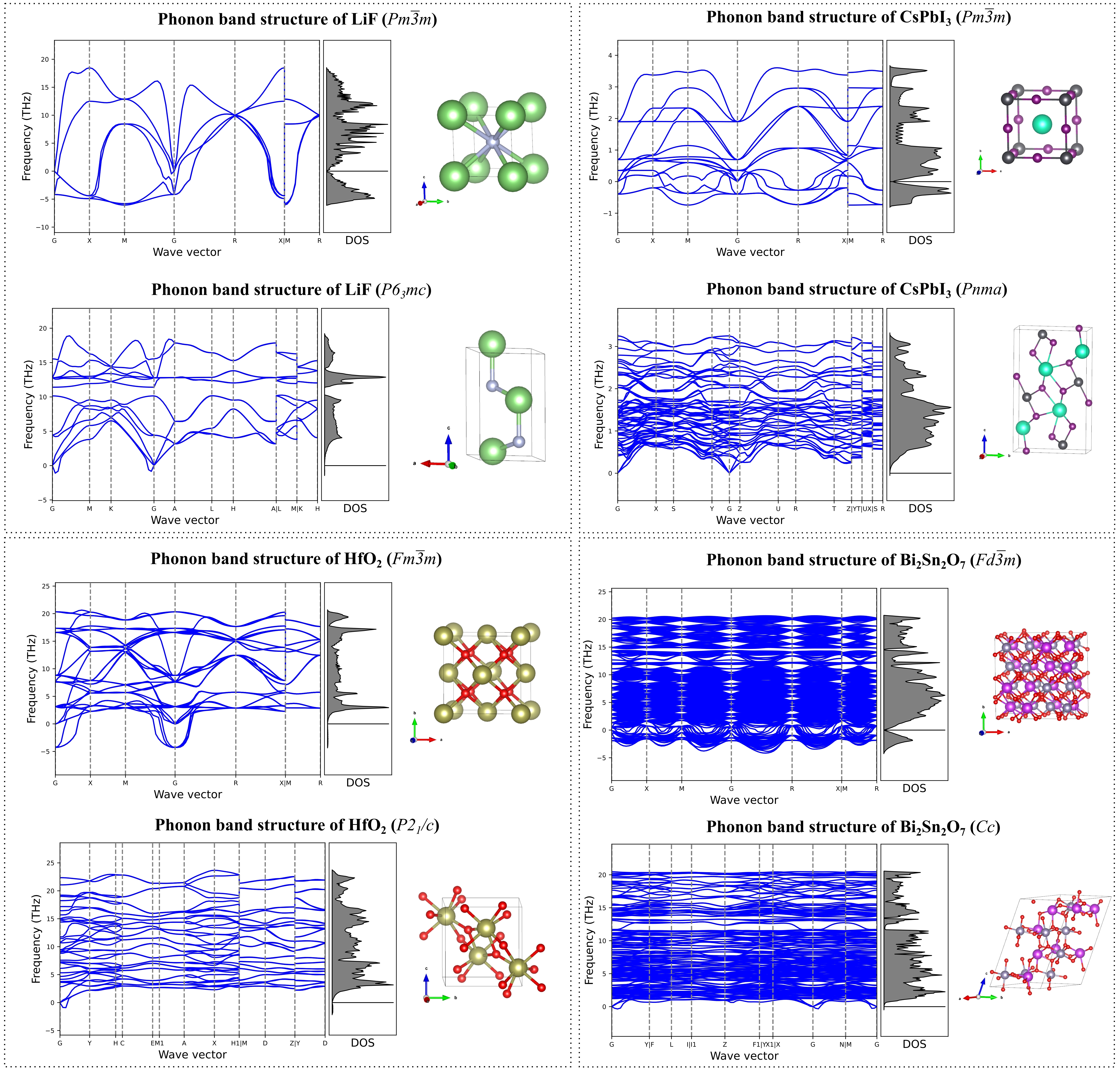}
    \caption{Phonon band structures, vibrational density of states (DOS), and corresponding crystal structures for the four benchmarked compounds evaluated by the MACE-OMAT-0 MLIP. For each material, the top panel displays the dynamically unstable, high-symmetry reference phase, characterized by prominent imaginary (plotted as negative) phonon frequencies. The bottom panel displays the dynamically stable, lowest-energy polymorph identified by VibroML's remediation workflows, demonstrating the successful elimination of the soft modes and the stabilization of the lattice.}
    \label{fig:phonon_dispersions_main}
\end{figure}

In this section we present a Table comparing different SOTA MLIPs with all relevant structures evaluated in VibroML for the 4 materials studied. We analyze the difference in relative energies between the polymorphs, their predicted dynamical stability based on the presence of soft phonon modes, and the assessment of thermal stability based on MD simulation. Furthermore, visual confirmation of the successful dynamical stabilization achieved by the VibroML workflows is provided in Figure~\ref{fig:phonon_dispersions_main}, which directly contrasts the phonon band structures of the initial high-symmetry phases against their completely stabilized polymorphs with lowest predicted energy.

\begin{landscape} 
\small 
\setlength{\tabcolsep}{5pt} 
\renewcommand{\arraystretch}{1.2} 
\setlength{\LTcapwidth}{\linewidth}
\begin{longtable}{ll l c c c >{\centering\arraybackslash}m{1cm} c >{\centering\arraybackslash}m{1.6cm} l}
\caption{\textbf{Comprehensive benchmarking of dynamical stability and structural relaxation across foundational MLIPs.} The data is grouped by compound, starting with the reference structure (Relative Energy = 0.0), followed by the remaining polymorphs sorted by MACE-OMAT relative energy from highest to lowest. The overall \textbf{Status} classifies a structure as \textbf{Stable} if ($\omega_{\text{min}} > -1.5$ THz AND Soft Frac $< 0.02$); \textbf{Quasi-Stable} if only one of the previous criteria is met; and \textbf{Unstable} if neither criterion is met. The evaluation metrics are defined as follows: 
(1) \textbf{Engine}: The Machine Learning Interatomic Potential used for geometric relaxation and force constant evaluation; 
(2) \textbf{Soft Frac (Softness Fraction)}: The proportion of imaginary (negative) phonon frequencies integrated across the sampled Brillouin zone; 
(3) \textbf{Soft Freq ($\omega_{\text{min}}$)}: The lowest (most negative) phonon frequency measured in THz, representing the magnitude of dynamical instability; 
(4) \textbf{HS Pt}: The high-symmetry point in the Brillouin zone where $\omega_{\text{min}}$ is localized; 
(5) \textbf{Rel. E}: The relaxed energy relative to the reference phase, measured in meV/atom; 
(6) \textbf{Phase Retained}: ``Yes'' indicates the initial crystallographic space group was preserved after structural relaxation, whereas ``No'' indicates the lattice underwent a spontaneous transition to a different local minimum (with the resulting space group annotated in parentheses); 
(7) \textbf{Cell Params}: The fully relaxed lattice vectors ($a, b, c$ in \AA) and angles ($\alpha, \beta, \gamma$ in degrees).
\label{tab:fullcomp_MLIPs_phonon_evals}} \\
\toprule
\textbf{Compound} & \textbf{Space Group} & \textbf{Engine} & \textbf{Status} & \textbf{Soft Frac} & \textbf{Soft Freq} & \textbf{HS Pt} & \textbf{Rel. E} & \textbf{Phase Retained} & \textbf{Cell Params} \\
& & & & & (THz) & & (meV/at) & & (a, b, c, $\alpha$, $\beta$, $\gamma$) \\
\midrule
\endfirsthead
\multicolumn{10}{c}{{\bfseries \tablename\ \thetable{} -- continued from previous page}} \\
\toprule
\textbf{Comp} & \textbf{SG} & \textbf{Engine} & \textbf{Status} & \textbf{Soft Frac} & \textbf{Soft Freq} & \textbf{HS Pt} & \textbf{Rel. E} & \textbf{Retained} & \textbf{Cell Params} \\
\midrule
\endhead
\bottomrule
\multicolumn{10}{r}{{Continued on next page}} \\
\endfoot
\bottomrule
\endlastfoot


\multirow{4}{*}{\large \textbf{LiF}}
& \multirow{4}{*}{\large $Pm\bar{3}m$} 
  & MACE-OMAT & \ust & 0.2883 & -6.07 & M & 0.0 & Yes & 2.55, 2.55, 2.55, 90, 90, 90 \\*
& & MACE-MP & \ust & 0.3133 & -6.67 & M & 0.0 & Yes & 2.56, 2.56, 2.56, 90, 90, 90 \\*
& & eSEN & \ust & 0.2833 & -6.15 & M & 0.0 & Yes & 2.56, 2.56, 2.56, 90, 90, 90 \\*
& & UMA & \ust & 0.2983 & -6.20 & M & 0.0 & Yes & 2.56, 2.56, 2.56, 90, 90, 90 \\
\cmidrule{2-10}

& \multirow{4}{*}{\large $Pnma$} 
  & MACE-OMAT & \stb & 0.0017 & -1.12 & - & -276.8 & Yes & 3.20, 5.23, 5.68, 90, 90, 90 \\*
& & MACE-MP & \stb & 0.0000 & -0.03 & - & -251.5 & Yes & 3.24, 5.22, 5.74, 90, 90, 90 \\*
& & eSEN & \stb & 0.0000 & -0.00 & - & -277.7 & Yes & 3.20, 5.24, 5.64, 90, 90, 90 \\*
& & UMA & \stb & 0.0017 & -1.47 & - & -275.2 & Yes & 3.20, 5.26, 5.61, 90, 90, 90 \\
\cmidrule{2-10}

& \multirow{4}{*}{\large $F\bar{4}3m$} 
  & MACE-OMAT & \stb & 0.0150 & -0.91 & - & -292.5 & Yes & 3.10, 3.10, 3.10, 60, 60, 60 \\*
& & MACE-MP & \stb & 0.0000 & -0.00 & - & -269.1 & Yes & 3.12, 3.12, 3.12, 60, 60, 60 \\*
& & eSEN & \ust & 0.0233 & -1.56 & - & -292.4 & Yes & 3.09, 3.09, 3.09, 60, 60, 60 \\*
& & UMA & \stb & 0.0150 & -1.02 & - & -288.8 & Yes & 3.09, 3.09, 3.09, 60, 60, 60 \\
\cmidrule{2-10}

& \multirow{4}{*}{\large $P4_2/mnm$} 
  & MACE-OMAT & \stb & 0.0029 & -1.23 & - & -296.4 & Yes & 5.29, 5.29, 3.12, 90, 90, 90 \\*
& & MACE-MP & \stb & 0.0004 & -0.17 & - & -269.4 & Yes & 5.33, 5.33, 3.14, 90, 90, 90 \\*
& & eSEN & \stb & 0.0000 & -0.00 & - & -296.8 & Yes & 5.28, 5.28, 3.12, 90, 90, 90 \\*
& & UMA & \qst & 0.0033 & -1.82 & - & -294.2 & Yes & 5.28, 5.28, 3.12, 90, 90, 90 \\
\cmidrule{2-10}

& \multirow{4}{*}{\large $P6_3mc$} 
  & MACE-OMAT & \qst & 0.0071 & -2.10 & - & -297.2 & Yes & 3.11, 3.11, 9.99, 90, 90, 120 \\*
& & MACE-MP & \stb & 0.0021 & -0.90 & - & -272.2 & Yes & 3.14, 3.14, 10.07, 90, 90, 120 \\*
& & eSEN & \stb & 0.0000 & -0.00 & - & -297.3 & Yes & 3.10, 3.10, 9.98, 90, 90, 120 \\*
& & UMA & \qst & 0.0075 & -2.03 & - & -294.0 & Yes & 3.11, 3.11, 9.97, 90, 90, 120 \\
\cmidrule{2-10}

& \multirow{4}{*}{\large $Fm\bar{3}m$} 
  & MACE-OMAT & \stb & 0.0008 & -0.73 & - & -297.2 & Yes & 4.98, 4.98, 2.87, 90, 90, 109.5 \\*
& & MACE-MP & \stb & 0.0000 & -0.04 & - & -279.4 & Yes & 5.01, 5.01, 2.89, 90, 90, 109.5 \\*
& & eSEN & \stb & 0.0025 & -1.00 & - & -297.9 & Yes & 4.98, 4.98, 2.88, 90, 90, 109.5 \\*
& & UMA & \qst & 0.0058 & -3.28 & G & -295.5 & Yes & 4.98, 4.98, 2.88, 90, 90, 109.5 \\
\cmidrule{2-10}

& \multirow{4}{*}{\large $P6_3/mmc$} 
  & MACE-OMAT & \stb & 0.0033 & -1.12 & - & -297.9 & Yes & 3.33, 3.33, 4.11, 90, 90, 120 \\*
& & MACE-MP & \stb & 0.0008 & -0.21 & - & -276.8 & Yes & 3.38, 3.38, 4.08, 90, 90, 120 \\*
& & eSEN & \qst & 0.0083 & -2.01 & G & -298.8 & Yes & 3.34, 3.34, 4.09, 90, 90, 120 \\*
& & UMA & \stb & 0.0058 & -1.42 & G & -296.5 & Yes & 3.34, 3.34, 4.09, 90, 90, 120 \\
\cmidrule{2-10}

& \multirow{4}{*}{\large $P6_3mc$} 
  & MACE-OMAT & \stb & 0.0033 & -1.10 & - & -303.1 & Yes & 3.14, 3.14, 4.89, 90, 90, 120 \\*
& & MACE-MP & \stb & 0.0000 & 0.01 & - & -276.5 & Yes & 3.16, 3.16, 4.92, 90, 90, 120 \\*
& & eSEN & \qst & 0.0050 & -1.70 & - & -303.1 & Yes & 3.13, 3.13, 4.91, 90, 90, 120 \\*
& & UMA & \stb & 0.0033 & -1.27 & - & -300.1 & Yes & 3.13, 3.13, 4.90, 90, 90, 120 \\
\cmidrule{2-10}

& \multirow{4}{*}{\large $P1$} 
  & MACE-OMAT & \ust & 0.0496 & -8.98 & R & -4088.4 & Yes & 2.94, 4.57, 5.49, 89.9, 90.7, 104.5 \\*
& & MACE-MP$^*$ & \qst & 0.0037 & -2.11 & G & -279.4 & No ($Fm\bar{3}m$) & 2.89, 5.01, 5.01, 80.4, 90, 106.8 \\*
& & eSEN$^*$ & \qst & 0.0029 & -1.96 & N & -297.9 & No ($Fm\bar{3}m$) & 2.88, 4.98, 4.98, 80.4, 90, 106.8 \\*
& & UMA$^*$ & \qst & 0.0071 & -2.62 & G & -295.5 & No ($Fm\bar{3}m$) & 2.88, 4.98, 4.98, 80.4, 90, 106.8 \\
\cmidrule{2-10}

& \multirow{4}{*}{\large $Pm$} 
  & MACE-OMAT & \ust & 0.0300 & -12.18 & Y & -7347.9 & Yes & 2.78, 5.60, 2.96, 90, 117.3, 90 \\*
& & MACE-MP$^*$ & \qst & 0.0033 & -1.54 & G & -279.4 & No ($Fm\bar{3}m$) & 2.89, 4.09, 4.09, 90, 135, 90 \\*
& & eSEN$^*$ & \qst & 0.0075 & -2.08 & G & -297.9 & No ($Fm\bar{3}m$) & 4.07, 4.07, 2.88, 90, 135, 90 \\*
& & UMA$^*$ & \qst & 0.0042 & -1.62 & G & -295.5 & No ($Fm\bar{3}m$) & 4.07, 4.07, 2.88, 90, 135, 90 \\
\cmidrule{2-10}

& \multirow{4}{*}{\large $Cm$} 
  & MACE-OMAT & \ust & 0.0450 & -17.00 & X & -14042.8 & Yes & 2.95, 2.95, 2.95, 75.8, 104.2, 112.5 \\*
& & MACE-MP$^*$ & \stb & 0.0000 & -0.00 & - & -279.4 & No ($Fm\bar{3}m$) & 2.89, 2.89, 4.09, 45, 135, 120 \\*
& & eSEN$^*$ & \qst & 0.0133 & -1.54 & - & -292.4 & No ($F\bar{4}3m$) & 3.09, 3.09, 3.09, 60, 120, 120 \\*
& & UMA$^*$ & \stb & 0.0150 & -1.03 & - & -288.8 & No ($F\bar{4}3m$) & 3.09, 3.09, 3.09, 60, 120, 120 \\
\midrule


\multirow{4}{*}{\large \textbf{CsPbI$_3$}} 
& \multirow{4}{*}{\large $Pm\bar{3}m$} 
  & MACE-OMAT & \ust & 0.2387 & -0.74 & M & 0.0 & Yes & 6.39, 6.39, 6.39, 90, 90, 90 \\*
& & MACE-MP & \ust & 0.0807 & -0.67 & R & 0.0 & Yes & 6.39, 6.39, 6.39, 90, 90, 90 \\*
& & eSEN & \ust & 0.2387 & -0.74 & R & 0.0 & Yes & 6.40, 6.40, 6.40, 90, 90, 90 \\*
& & UMA & \ust & 0.2467 & -0.70 & R & 0.0 & Yes & 6.40, 6.40, 6.40, 90, 90, 90 \\
\cmidrule{2-10}

& \multirow{4}{*}{\large $C2/m$} 
  & MACE-OMAT & \ust & 0.0283 & -0.82 & F & -18.0 & Yes & 8.90, 8.90, 8.93, 60.4, 119.6, 90.3 \\*
& & MACE-MP & \stb & 0.0077 & -0.35 & N & -18.6 & Yes & 8.84, 8.84, 9.00, 60.5, 119.5, 90.2 \\*
& & eSEN & \ust & 0.0300 & -0.99 & F & -19.2 & Yes & 8.89, 8.89, 8.98, 60.5, 119.5, 90.2 \\*
& & UMA & \ust & 0.0287 & -0.99 & F1 & -16.5 & Yes & 8.89, 8.89, 8.99, 60.5, 119.5, 90.2 \\
\cmidrule{2-10}

& \multirow{4}{*}{\large $Cm$} 
  & MACE-OMAT & \stb & 0.0028 & -0.16 & - & -23.0 & Yes & 10.96, 10.96, 12.54, 55.1, 124.9, 108.7 \\*
& & MACE-MP & \stb & 0.0005 & -0.11 & N & -18.9 & No ($I4mm$) & 10.95, 10.95, 12.55, 55.1, 124.9, 109.2 \\*
& & eSEN & \stb & 0.0062 & -0.36 & - & -21.4 & Yes & 10.97, 10.97, 12.54, 55.2, 124.8, 108.4 \\*
& & UMA & \stb & 0.0067 & -0.37 & - & -18.5 & Yes & 10.97, 10.97, 12.56, 55.2, 124.8, 108.5 \\
\cmidrule{2-10}

& \multirow{4}{*}{\large $P\bar{1}$} 
  & MACE-OMAT & \stb & 0.0048 & -0.21 & N & -28.7 & Yes & 8.03, 8.83, 14.97, 90.3, 95.9, 100.4 \\*
& & MACE-MP & \stb & 0.0000 & -0.00 & - & -24.4 & Yes & 7.83, 8.68, 14.99, 90.5, 95.0, 97.8 \\*
& & eSEN & \stb & 0.0012 & -0.11 & N & -27.5 & Yes & 8.07, 8.87, 15.03, 90.1, 96.1, 100.4 \\*
& & UMA & \stb & 0.0010 & -0.14 & N & -23.2 & Yes & 8.05, 8.82, 15.04, 90.2, 96.2, 100.1 \\
\cmidrule{2-10}

& \multirow{4}{*}{\large $Pnma$} 
  & MACE-OMAT & \stb & 0.0000 & -0.00 & - & -29.3 & Yes & 8.66, 9.15, 12.64, 90, 90, 90 \\*
& & MACE-MP & \stb & 0.0000 & -0.00 & - & -24.7 & Yes & 8.56, 9.22, 12.68, 90, 90, 90 \\*
& & eSEN & \stb & 0.0018 & -0.30 & S & -26.4 & No ($P2_12_12_1$) & 8.72, 9.12, 12.66, 90, 90, 90 \\*
& & UMA & \stb & 0.0020 & -0.29 & S & -22.9 & Yes & 8.74, 9.12, 12.66, 90, 90, 90 \\
\cmidrule{2-10}

& \multirow{4}{*}{\large $P\bar{1}$} 
  & MACE-OMAT & \stb & 0.0000 & -0.04 & - & -29.9 & Yes & 8.62, 10.80, 12.35, 83.2, 74.7, 70 \\*
& & MACE-MP & \stb & 0.0000 & -0.02 & - & -24.4 & Yes & 8.63, 10.88, 11.66, 83.9, 74, 73.1 \\*
& & eSEN & \stb & 0.0040 & -0.23 & N & -27.8 & No ($P1$) & 8.70, 10.78, 12.47, 82, 73.7, 69.5 \\*
& & UMA & \stb & 0.0040 & -0.22 & M & -23.3 & Yes & 8.69, 10.77, 12.47, 82, 73.5, 69.5 \\
\cmidrule{2-10}

& \multirow{4}{*}{\large $Pm$} 
  & MACE-OMAT & \stb & 0.0085 & -0.57 & - & -30.3 & Yes & 17.20, 4.90, 24.06, 90, 91.2, 90 \\*
& & MACE-MP & \stb & 0.0052 & -0.32 & - & -21.6 & Yes & 17.19, 4.87, 23.82, 90, 91.4, 90 \\*
& & eSEN & \stb & 0.0123 & -0.71 & - & -24.0 & No ($P1$) & 17.35, 4.87, 23.99, 90, 91.4, 90 \\*
& & UMA & \stb & 0.0117 & -0.71 & - & -19.8 & No ($P1$) & 17.35, 4.87, 23.96, 90, 91.4, 90 \\
\cmidrule{2-10}

& \multirow{4}{*}{\large $P2_1/m$} 
  & MACE-OMAT & \stb & 0.0000 & -0.00 & - & -33.2 & Yes & 10.35, 4.97, 20.93, 90, 102, 90 \\*
& & MACE-MP & \stb & 0.0000 & -0.00 & - & -21.2 & Yes & 10.31, 4.93, 20.87, 90, 101.4, 90 \\*
& & eSEN & \stb & 0.0000 & -0.00 & - & -25.6 & Yes & 10.45, 4.93, 20.92, 90, 101.8, 90 \\*
& & UMA & \stb & 0.0000 & -0.00 & - & -21.4 & Yes & 10.42, 4.93, 20.94, 90, 101.7, 90 \\
\cmidrule{2-10}

& \multirow{4}{*}{\large $Cmcm$} 
  & MACE-OMAT & \stb & 0.0107 & -1.06 & - & -34.9 & Yes & 8.45, 8.45, 12.37, 90, 90, 146.5 \\*
& & MACE-MP & \stb & 0.0030 & -0.51 & - & -25.6 & Yes & 8.40, 8.40, 12.47, 90, 90, 146.5 \\*
& & eSEN & \stb & 0.0097 & -1.11 & - & -25.0 & Yes & 8.47, 8.47, 12.42, 90, 90, 146.8 \\*
& & UMA & \stb & 0.0097 & -1.10 & - & -21.3 & Yes & 8.46, 8.46, 12.42, 90, 90, 146.8 \\
\cmidrule{2-10}

& \multirow{4}{*}{\large $Pnma$} 
  & MACE-OMAT & \stb & 0.0000 & -0.00 & - & -44.1 & Yes & 4.93, 10.86, 18.17, 90, 90, 90 \\*
& & MACE-MP & \stb & 0.0000 & -0.00 & - & -35.0 & Yes & 4.88, 10.91, 18.15, 90, 90, 90 \\*
& & eSEN & \stb & 0.0000 & -0.00 & - & -36.7 & Yes & 4.89, 10.92, 18.21, 90, 90, 90 \\*
& & UMA & \stb & 0.0000 & -0.00 & - & -32.6 & Yes & 4.89, 10.89, 18.22, 90, 90, 90 \\
\midrule


\multirow{4}{*}{\large \textbf{HfO$_2$}} 
& \multirow{4}{*}{\large $Fm\bar{3}m$} 
  & MACE-OMAT & \qst & 0.0153 & -4.27 & G & 0.0 & Yes & 5.05, 5.05, 5.05, 90, 90, 90 \\*
& & MACE-MP & \qst & 0.0153 & -3.81 & G & 0.0 & Yes & 5.06, 5.06, 5.06, 90, 90, 90 \\*
& & eSEN & \ust & 0.0575 & -6.56 & G & 0.0 & Yes & 5.07, 5.07, 5.07, 90, 90, 90 \\*
& & UMA & \ust & 0.0286 & -6.77 & G & 0.0 & Yes & 5.07, 5.07, 5.07, 90, 90, 90 \\
\cmidrule{2-10}

& \multirow{4}{*}{\large $Pmn2_1$} 
  & MACE-OMAT & \stb & 0.0000 & -0.00 & - & -11.4 & Yes & 3.47, 3.72, 5.20, 90, 90, 90 \\*
& & MACE-MP & \stb & 0.0000 & -0.00 & - & -25.6 & Yes & 3.40, 3.84, 5.17, 90, 90, 90 \\*
& & eSEN & \stb & 0.0000 & -0.00 & - & -39.7 & Yes & 3.44, 3.79, 5.18, 90, 90, 90 \\*
& & UMA & \stb & 0.0000 & -0.00 & - & -39.7 & Yes & 3.44, 3.79, 5.18, 90, 90, 90 \\
\cmidrule{2-10}

& \multirow{4}{*}{\large $P4_2/nmc$} 
  & MACE-OMAT & \stb & 0.0000 & -0.00 & - & -12.3 & Yes & 3.57, 3.57, 5.19, 90, 90, 90 \\*
& & MACE-MP & \ust & 0.0294 & -0.90 & A & -14.7 & Yes & 3.59, 3.59, 5.22, 90, 90, 90 \\*
& & eSEN & \ust & 0.0228 & -1.97 & M & -32.2 & Yes & 3.59, 3.59, 5.23, 90, 90, 90 \\*
& & UMA & \stb & 0.0044 & -0.51 & Z & -33.5 & Yes & 3.59, 3.59, 5.23, 90, 90, 90 \\
\cmidrule{2-10}

& \multirow{4}{*}{\large $Fdd2$} 
  & MACE-OMAT & \qst & 0.0047 & -2.25 & - & -15.5 & Yes & 6.54, 6.11, 5.25, 69.8, 61.3, 48.9 \\*
& & MACE-MP & \stb & 0.0028 & -1.07 & G & -33.1 & Yes & 6.56, 6.14, 5.25, 69.8, 61.5, 48.7 \\*
& & eSEN & \qst & 0.0161 & -2.82 & C & -45.8 & Yes & 6.56, 6.12, 5.26, 69.9, 61.2, 48.9 \\*
& & UMA & \qst & 0.0103 & -2.52 & - & -46.5 & Yes & 6.56, 6.12, 5.26, 69.9, 61.2, 48.9 \\
\cmidrule{2-10}

& \multirow{4}{*}{\large $P2_1/m$} 
  & MACE-OMAT & \stb & 0.0000 & -0.00 & - & -18.5 & Yes & 3.83, 3.47, 5.43, 90, 103.8, 90 \\*
& & MACE-MP & \stb & 0.0000 & -0.00 & - & -45.6 & Yes & 3.85, 3.45, 5.44, 90, 103.8, 90 \\*
& & eSEN & \stb & 0.0000 & -0.00 & - & -60.5 & Yes & 3.85, 3.47, 5.44, 90, 104.5, 90 \\*
& & UMA & \stb & 0.0000 & -0.00 & - & -61.9 & Yes & 3.85, 3.48, 5.45, 90, 104.5, 90 \\
\cmidrule{2-10}

& \multirow{4}{*}{\large $Pnma$} 
  & MACE-OMAT & \qst & 0.0019 & -2.42 & G & -21.8 & Yes & 3.42, 5.39, 7.70, 90, 90, 90 \\*
& & MACE-MP & \stb & 0.0011 & -1.11 & G & -40.3 & Yes & 3.41, 5.38, 7.73, 90, 90, 90 \\*
& & eSEN & \qst & 0.0039 & -2.19 & G & -59.1 & Yes & 3.42, 5.40, 7.68, 90, 90, 90 \\*
& & UMA & \qst & 0.0019 & -1.84 & - & -59.2 & Yes & 3.42, 5.41, 7.67, 90, 90, 90 \\
\cmidrule{2-10}

& \multirow{4}{*}{\large $P2_1/c$} 
  & MACE-OMAT & \stb & 0.0000 & -0.00 & - & -26.4 & Yes & 5.14, 10.28, 5.37, 90, 99.1, 90 \\*
& & MACE-MP & \stb & 0.0000 & -0.00 & - & -46.5 & Yes & 5.14, 10.29, 5.39, 90, 99.3, 90 \\*
& & eSEN & \qst & 0.0067 & -3.01 & G & -65.2 & Yes & 5.14, 10.33, 5.38, 90, 99.4, 90 \\*
& & UMA & \stb & 0.0000 & -0.00 & - & -65.5 & Yes & 5.14, 10.33, 5.39, 90, 99.4, 90 \\
\cmidrule{2-10}

& \multirow{4}{*}{\large $Pca2_1$} 
  & MACE-OMAT & \stb & 0.0011 & -1.41 & - & -28.4 & Yes & 5.04, 5.08, 5.25, 90, 90, 90 \\*
& & MACE-MP & \stb & 0.0006 & -0.38 & - & -41.3 & Yes & 5.05, 5.10, 5.26, 90, 90, 90 \\*
& & eSEN & \qst & 0.0058 & -2.51 & G & -59.5 & Yes & 5.05, 5.08, 5.26, 90, 90, 90 \\*
& & UMA & \qst & 0.0014 & -2.08 & - & -59.8 & Yes & 5.05, 5.08, 5.27, 90, 90, 90 \\
\cmidrule{2-10}

& \multirow{4}{*}{\large $P4_2/mnm$} 
  & MACE-OMAT & \stb & 0.0011 & -0.28 & - & -31.2 & Yes & 4.84, 4.84, 3.23, 90, 90, 90 \\*
& & MACE-MP & \stb & 0.0000 & -0.00 & - & -46.2 & Yes & 4.85, 4.85, 3.24, 90, 90, 90 \\*
& & eSEN & \qst & 0.0067 & -1.84 & R & -62.3 & Yes & 4.86, 4.86, 3.22, 90, 90, 90 \\*
& & UMA & \stb & 0.0022 & -1.02 & - & -64.3 & Yes & 4.86, 4.86, 3.22, 90, 90, 90 \\
\cmidrule{2-10}

& \multirow{4}{*}{\large $Pbcn$} 
  & MACE-OMAT & \stb & 0.0011 & -1.16 & - & -38.5 & Yes & 4.89, 5.24, 5.77, 90, 90, 90 \\*
& & MACE-MP & \stb & 0.0006 & -0.42 & - & -58.0 & Yes & 4.90, 5.23, 5.76, 90, 90, 90 \\*
& & eSEN & \qst & 0.0136 & -2.39 & G & -70.4 & Yes & 4.90, 5.25, 5.78, 90, 90, 90 \\*
& & UMA & \qst & 0.0014 & -2.18 & - & -72.4 & Yes & 4.90, 5.25, 5.79, 90, 90, 90 \\
\cmidrule{2-10}

& \multirow{4}{*}{\large $I4_1/amd$} 
  & MACE-OMAT & \stb & 0.0011 & -0.42 & - & -42.7 & Yes & 6.00, 6.00, 6.00, 141, 141, 56.3 \\*
& & MACE-MP & \stb & 0.0000 & -0.05 & - & -68.1 & Yes & 5.99, 5.99, 5.99, 140.8, 140.8, 56.6 \\*
& & eSEN & \qst & 0.0178 & -2.04 & Z & -93.0 & Yes & 6.00, 6.00, 6.00, 141, 141, 56.4 \\*
& & UMA & \stb & 0.0000 & -0.01 & - & -95.0 & Yes & 6.00, 6.00, 6.00, 141, 141, 56.3 \\
\cmidrule{2-10}

& \multirow{4}{*}{\large $P2_1/c$} 
  & MACE-OMAT & \stb & 0.0006 & -0.90 & - & -48.5 & Yes & 5.15, 5.16, 5.33, 90, 99.8, 90 \\*
& & MACE-MP & \stb & 0.0003 & -0.13 & - & -65.4 & Yes & 5.16, 5.17, 5.35, 90, 99.7, 90 \\*
& & eSEN & \stb & 0.0011 & -1.06 & G & -87.2 & Yes & 5.14, 5.18, 5.33, 90, 99.8, 90 \\*
& & UMA & \qst & 0.0008 & -1.53 & - & -86.1 & Yes & 5.14, 5.19, 5.33, 90, 99.7, 90 \\
\midrule


\multirow{4}{*}{\large \textbf{Bi$_2$Sn$_2$O$_7$}} 
& \multirow{4}{*}{\large $Fd\bar{3}m$} 
  & MACE-OMAT & \ust & 0.0669 & -4.15 & G & 0.0 & Yes & 10.73, 10.73, 10.73, 90, 90, 90 \\*
& & MACE-MP & \ust & 0.0772 & -3.04 & G & 0.0 & Yes & 10.73, 10.73, 10.73, 90, 90, 90 \\*
& & eSEN & \ust & 0.0682 & -2.05 & G & 0.0 & Yes & 10.74, 10.74, 10.74, 90, 90, 90 \\*
& & UMA & \ust & 0.0650 & -4.04 & G & 0.0 & Yes & 10.74, 10.74, 10.74, 90, 90, 90 \\
\cmidrule{2-10}

& \multirow{4}{*}{\large $P2_12_12_1$} 
  & MACE-OMAT & \stb & 0.0000 & -0.00 & - & 20.6 & Yes & 7.61, 7.79, 10.85, 90, 90, 90 \\*
& & MACE-MP & \stb & 0.0000 & -0.00 & - & 13.0 & Yes & 7.59, 7.79, 10.80, 90, 90, 90 \\*
& & eSEN & \qst & 0.0161 & -2.67 & G & 16.6 & Yes & 7.63, 7.79, 10.86, 90, 90, 90 \\*
& & UMA & \qst & 0.0164 & -2.67 & G & 18.1 & Yes & 7.63, 7.79, 10.85, 90, 90, 90 \\
\cmidrule{2-10}

& \multirow{4}{*}{\large $P2_1$} 
  & MACE-OMAT & \qst & 0.0195 & -2.76 & H & 12.7 & Yes & 7.67, 7.73, 11.33, 90, 109, 90 \\*
& & MACE-MP & \qst & 0.0132 & -1.66 & H1 & 12.1 & Yes & 7.67, 7.74, 11.29, 90, 108.8, 90 \\*
& & eSEN & \qst & 0.0199 & -2.78 & H & 10.9 & Yes & 7.68, 7.75, 11.34, 90, 108.9, 90 \\*
& & UMA & \qst & 0.0198 & -2.77 & H & 13.0 & Yes & 7.68, 7.75, 11.34, 90, 108.9, 90 \\
\cmidrule{2-10}

& \multirow{4}{*}{\large $P1$} 
  & MACE-OMAT & \stb & 0.0000 & -0.00 & - & 0.2 & Yes & 10.76, 10.83, 21.67, 90, 90, 90 \\*
& & MACE-MP & \stb & 0.0000 & -0.00 & - & -2.8 & Yes & 10.76, 10.84, 21.68, 90, 90, 90 \\*
& & eSEN & \stb & 0.0000 & -0.00 & - & -0.5 & Yes & 10.77, 10.85, 21.69, 90, 90, 90 \\*
& & UMA & \stb & 0.0000 & -0.00 & - & 1.2 & Yes & 10.77, 10.84, 21.69, 90, 90, 90 \\
\cmidrule{2-10}

& \multirow{4}{*}{\large $P1$} 
  & MACE-OMAT & \stb & 0.0000 & -0.00 & - & -12.2 & Yes & 10.80, 10.83, 43.35, 90, 90, 90 \\*
& & MACE-MP & \stb & 0.0000 & -0.00 & - & -14.0 & Yes & 10.80, 10.83, 43.33, 90, 90, 90 \\*
& & eSEN & \stb & 0.0000 & -0.00 & - & -14.0 & Yes & 10.81, 10.83, 43.37, 90, 89.9, 90 \\*
& & UMA & \stb & 0.0000 & -0.00 & - & -12.7 & Yes & 10.81, 10.83, 43.36, 90, 90, 90 \\
\cmidrule{2-10}

& \multirow{4}{*}{\large $Cc$} 
  & MACE-OMAT & \stb & 0.0038 & -0.16 & G & -12.6 & Yes & 15.31, 21.65, 7.64, 90, 90.4, 90 \\*
& & MACE-MP & \stb & 0.0000 & -0.00 & - & -14.4 & Yes & 15.31, 21.65, 7.64, 90, 90.4, 90 \\*
& & eSEN & \stb & 0.0000 & -0.00 & - & -16.5 & Yes & 15.31, 21.67, 7.67, 90, 90, 90 \\*
& & UMA & \stb & 0.0000 & -0.00 & - & -15.2 & Yes & 15.31, 21.67, 7.67, 90, 90, 90 \\
\cmidrule{2-10}

& \multirow{4}{*}{\large $Pna2_1$} 
  & MACE-OMAT & \stb & 0.0000 & 0.01 & - & -12.7 & Yes & 7.65, 7.66, 10.82, 90, 90, 90 \\*
& & MACE-MP & \stb & 0.0000 & -0.01 & - & -14.5 & Yes & 7.64, 7.66, 10.82, 90, 90, 90 \\*
& & eSEN$^{**}$ & \qst & 0.0173 & -3.08 & G & -14.7 & Yes & 7.65, 7.66, 10.83, 90, 90, 90 \\*
& & UMA$^{**}$ & \qst & 0.0172 & -2.98 & G & -13.3 & Yes & 7.65, 7.66, 10.83, 90, 90, 90 \\
\cmidrule{2-10}

& \multirow{4}{*}{\large $P2_1$} 
  & MACE-OMAT & \stb & 0.0000 & -0.00 & - & -13.1 & Yes & 7.67, 7.67, 10.79, 90, 90.1, 90 \\*
& & MACE-MP & \stb & 0.0000 & -0.00 & - & -14.8 & Yes & 7.67, 7.67, 10.78, 90, 90.1, 90 \\*
& & eSEN$^{**}$ & \qst & 0.0180 & -2.93 & H & -15.1 & Yes & 7.67, 7.67, 10.80, 90, 90.1, 90 \\*
& & UMA$^{**}$ & \qst & 0.0184 & -2.85 & H1 & -13.7 & Yes & 7.67, 7.67, 10.80, 90, 90, 90 \\
\cmidrule{2-10}

& \multirow{4}{*}{\large $P2_12_12_1$} 
  & MACE-OMAT & \stb & 0.0000 & -0.00 & - & -13.8 & Yes & 7.67, 7.67, 10.80, 90, 90, 90 \\*
& & MACE-MP & \stb & 0.0000 & -0.00 & - & -15.2 & Yes & 7.67, 7.67, 10.79, 90, 90, 90 \\*
& & eSEN$^{**}$ & \qst & 0.0181 & -3.40 & G & -15.6 & Yes & 7.68, 7.67, 10.80, 90, 90, 90 \\*
& & UMA$^{**}$ & \ust & 0.0235 & -3.69 & G & -14.2 & Yes & 7.67, 7.67, 10.80, 90, 90, 90 \\
\cmidrule{2-10}

& \multirow{4}{*}{\large $Cc$} 
  & MACE-OMAT & \stb & 0.0005 & -0.35 & - & -14.1 & Yes & 7.66, 7.66, 13.27, 73.1, 106.9, 120.1 \\*
& & MACE-MP & \stb & 0.0001 & -0.09 & - & -15.5 & Yes & 7.66, 7.66, 13.27, 73.1, 106.9, 120.1 \\*
& & eSEN$^{**}$ & \ust & 0.0295 & -3.77 & I1 & -16.5 & Yes & 7.66, 7.66, 13.29, 73.1, 106.9, 120.1 \\*
& & UMA$^{**}$ & \ust & 0.0298 & -3.63 & I1 & -15.1 & Yes & 7.66, 7.66, 13.28, 73.1, 106.9, 120.1 \\
\end{longtable}

\vspace{1ex}
\noindent\footnotesize{\textbf{$^{*}$Note on LiF anomalies:} The MACE-OMAT model identifies deep numerical artifacts for \ce{LiF} with highly negative energies (down to $-14042.8$~meV/atom for the $Cm$ phase), which correspond to unphysical steric clashes inherited from the DFT training data. During this specific relaxation protocol, MACE-MP, eSEN, and UMA avoid becoming trapped in these artificial basins, instead reverting to physically expected, higher-symmetry phases (such as $Fm\bar{3}m$ or $F\bar{4}3m$) with relative energies in the expected $-270$ to $-300$~meV/atom range.}
\vspace{1ex}
\noindent\footnotesize{\textbf{$^{**}$Note on \ce{Bi2Sn2O7} classification:} For these specific low-energy polymorphs, the ``Quasi-Stable'' and ``Unstable'' classifications generated by eSEN and UMA may be artifacts. Because these highly parameter-dense models require significant GPU memory, evaluating the finite-displacement phonon band structures for these large cells required reducing the supercell expansion size, which can artificially induce shallow imaginary frequencies.}
\end{landscape} 

\begin{table}[htbp]
    \centering
    \caption{\textbf{Relative energies (in meV/atom) of identified polymorphs across different MLIP engines and fully relaxed DFT.} The high-symmetry reference phase is set to 0.0. Anomalous steric clash phases for LiF have been excluded. Where multiple polymorphs share the same space group, they are distinguished by numerical indices corresponding to their order of appearance.}
    \label{tab:relative_energies_mlip_dft}
    \small
    \begin{tabular}{ll c c c c c}
        \toprule
        \textbf{Compound} & \textbf{Space Group} & \textbf{MACE-OMAT} & \textbf{MACE-MP} & \textbf{eSEN} & \textbf{UMA} & \textbf{DFT Relaxed} \\
        \midrule
        
        \multirow{6}{*}{\textbf{LiF}} 
        & $Pm\bar{3}m$ & 0.0 & 0.0 & 0.0 & 0.0 & 0.0 \\
        & $Pnma$ & -276.8 & -251.5 & -277.7 & -275.2 & -275.4 \\
        & $F\bar{4}3m$ & -292.5 & -269.1 & -292.4 & -288.8 & -289.9 \\
        & $P4_2/mnm$ & -296.4 & -269.4 & -296.8 & -294.2 & -294.5 \\
        & $Fm\bar{3}m$ & -297.2 & -279.4 & -297.9 & -295.5 & -296.1 \\
        & $P6_3mc$ & -303.1 & -276.5 & -303.1 & -300.1 & -300.6 \\
        \midrule
        
        \multirow{8}{*}{\textbf{CsPbI$_3$}} 
        & $Pm\bar{3}m$ & 0.0 & 0.0 & 0.0 & 0.0 & -5.4 \\
        & $C2/m$ & -18.1 & -18.6 & -19.2 & -16.5 & -21.5 \\
        & $Cm$ & -23.0 & -18.9 & -21.4 & -18.5 & -23.5 \\
        & $Pnma$ (1) & -29.3 & -24.7 & -26.4 & -22.9 & -27.9 \\
        & $P\bar{1}$ & -29.9 & -24.4 & -27.8 & -23.3 & -28.5 \\
        & $P2_1/m$ & -33.3 & -21.2 & -25.6 & -21.4 & -46.7 \\
        & $Cmcm$ & -34.9 & -25.6 & -25.0 & -21.3 & -26.8 \\
        & $Pnma$ (2) & -44.2 & -35.0 & -36.7 & -32.6 & -38.0 \\
        \midrule
        
        \multirow{6}{*}{\textbf{HfO$_2$}} 
        & $Fm\bar{3}m$ & 0.0 & 0.0 & 0.0 & 0.0 & -5.4 \\
        & $P4_2/nmc$ & -12.2 & -14.7 & -32.2 & -33.5 & -39.7 \\
        & $P2_1/m$ & -18.5 & -45.6 & -60.5 & -61.9 & -68.8 \\
        & $Pnma$ & -21.7 & -40.3 & -59.1 & -59.2 & -66.4 \\
        & $Pbcn$ & -38.5 & -58.0 & -70.4 & -72.4 & -79.7 \\
        & $P2_1/c$ & -48.5 & -65.4 & -87.2 & -86.1 & -95.1 \\
        \midrule
        
        \multirow{8}{*}{\textbf{Bi$_2$Sn$_2$O$_7$}} 
        & $Fd\bar{3}m$ & 0.0 & 0.0 & 0.0 & 0.0 & -10.2 \\
        & $P2_12_12_1$ (1) & +20.7 & +13.0 & +16.6 & +18.1 & +5.4 \\
        & $P2_1$ (1) & +12.7 & +12.1 & +10.9 & +13.0 & +0.5 \\
        & $Cc$ (1) & -12.6 & -14.4 & -16.5 & -15.2 & -24.1 \\
        & $Pna2_1$ & -12.7 & -14.5 & -14.7 & -13.3 & -24.0 \\
        & $P2_1$ (2) & -13.1 & -14.8 & -15.1 & -13.7 & -24.4 \\
        & $P2_12_12_1$ (2) & -13.7 & -15.2 & -15.6 & -14.2 & -24.8 \\
        & $Cc$ (2) & -14.1 & -15.5 & -16.5 & -15.1 & -25.8 \\
        \bottomrule
    \end{tabular}
\end{table}

\begin{table}[htbp]
    \centering
    \caption{\textbf{Phase stability ranking by space group from highest relative energy (least stable) to lowest relative energy (most stable).} Phases that relax to identical energetic depths are denoted with an equals sign ($=$).}
    \label{tab:phase_rankings}
    \footnotesize
    \begin{tabular}{l l p{10cm}}
        \toprule
        \textbf{Compound} & \textbf{Evaluator} & \textbf{Thermodynamic Hierarchy Ranking} \\
        \midrule
        
        \multirow{5}{*}{\textbf{LiF}} 
        & MACE-OMAT & $Pm\bar{3}m > Pnma > F\bar{4}3m > P4_2/mnm > Fm\bar{3}m > P6_3mc$ \\
        & MACE-MP & $Pm\bar{3}m > Pnma > F\bar{4}3m > P4_2/mnm > P6_3mc > Fm\bar{3}m$ \\
        & eSEN & $Pm\bar{3}m > Pnma > F\bar{4}3m > P4_2/mnm > Fm\bar{3}m > P6_3mc$ \\
        & UMA & $Pm\bar{3}m > Pnma > F\bar{4}3m > P4_2/mnm > Fm\bar{3}m > P6_3mc$ \\
        & DFT Relaxed & $Pm\bar{3}m > Pnma > F\bar{4}3m > P4_2/mnm > Fm\bar{3}m > P6_3mc$ \\
        \midrule
        
        \multirow{5}{*}{\textbf{CsPbI$_3$}} 
        & MACE-OMAT & $Pm\bar{3}m > C2/m > Cm > Pnma \text{ (1)} > P\bar{1} > P2_1/m > Cmcm > Pnma \text{ (2)}$ \\
        & MACE-MP & $Pm\bar{3}m > C2/m > Cm > P2_1/m > P\bar{1} > Pnma \text{ (1)} > Cmcm > Pnma \text{ (2)}$ \\
        & eSEN & $Pm\bar{3}m > C2/m > Cm > Cmcm > P2_1/m > Pnma \text{ (1)} > P\bar{1} > Pnma \text{ (2)}$ \\
        & UMA & $Pm\bar{3}m > C2/m > Cm > Cmcm > P2_1/m > Pnma \text{ (1)} > P\bar{1} > Pnma \text{ (2)}$ \\
        & DFT Relaxed & $Pm\bar{3}m > C2/m > Cm > Cmcm > Pnma \text{ (1)} > P\bar{1} > Pnma \text{ (2)} > P2_1/m$ \\
        \midrule
        
        \multirow{5}{*}{\textbf{HfO$_2$}} 
        & MACE-OMAT & $Fm\bar{3}m > P4_2/nmc > P2_1/m > Pnma > Pbcn > P2_1/c$ \\
        & MACE-MP & $Fm\bar{3}m > P4_2/nmc > Pnma > P2_1/m > Pbcn > P2_1/c$ \\
        & eSEN & $Fm\bar{3}m > P4_2/nmc > Pnma > P2_1/m > Pbcn > P2_1/c$ \\
        & UMA & $Fm\bar{3}m > P4_2/nmc > Pnma > P2_1/m > Pbcn > P2_1/c$ \\
        & DFT Relaxed & $Fm\bar{3}m > P4_2/nmc > Pnma > P2_1/m > Pbcn > P2_1/c$ \\
        \midrule
        
        \multirow{5}{*}{\textbf{Bi$_2$Sn$_2$O$_7$}} 
        & MACE-OMAT & $P2_12_12_1 \text{ (1)} > P2_1 \text{ (1)} > Fd\bar{3}m > Cc \text{ (1)} > Pna2_1 > P2_1 \text{ (2)} > P2_12_12_1 \text{ (2)} > Cc \text{ (2)}$ \\
        & MACE-MP & $P2_12_12_1 \text{ (1)} > P2_1 \text{ (1)} > Fd\bar{3}m > Cc \text{ (1)} > Pna2_1 > P2_1 \text{ (2)} > P2_12_12_1 \text{ (2)} > Cc \text{ (2)}$ \\
        & eSEN & $P2_12_12_1 \text{ (1)} > P2_1 \text{ (1)} > Fd\bar{3}m > Pna2_1 > P2_1 \text{ (2)} > P2_12_12_1 \text{ (2)} > Cc \text{ (1)} = Cc \text{ (2)}$ \\
        & UMA & $P2_12_12_1 \text{ (1)} > P2_1 \text{ (1)} > Fd\bar{3}m > Pna2_1 > P2_1 \text{ (2)} > P2_12_12_1 \text{ (2)} > Cc \text{ (2)} > Cc \text{ (1)}$ \\
        & DFT Relaxed & $P2_12_12_1 \text{ (1)} > P2_1 \text{ (1)} > Fd\bar{3}m > Pna2_1 > Cc \text{ (1)} > P2_1 \text{ (2)} > P2_12_12_1 \text{ (2)} > Cc \text{ (2)}$ \\
        
        \bottomrule
    \end{tabular}
\end{table}

\begin{landscape} 
\small 
\setlength{\tabcolsep}{4pt} 
\renewcommand{\arraystretch}{1.2} 
\setlength{\LTcapwidth}{\linewidth}

\begin{longtable}{ll l c c c c c c c c l}
\caption{\textbf{Comprehensive benchmarking of thermal stability via Molecular Dynamics (MD) across foundational MLIPs.} This table details the stability of generated polymorphs under finite-temperature NPT MD simulations (300 K). The overall \textbf{Verdict} classifies a structure as \stb\ (3 or 4 criteria passed) or \ust\ (2 or fewer criteria passed) based on four key trajectory metrics evaluated during the production phase: 
(1) \textbf{RMSD (\AA)}: Root-mean-square deviation of atomic positions from their initial lattice sites (Pass threshold: $< 1.0$~\AA); 
(2) \textbf{Vol (\%)}: Maximum cell volume fluctuation relative to the early production phase average (Pass threshold: $< 6\%$); 
(3) \textbf{RDF}: Pearson correlation coefficient between the initial Radial Distribution Function and the ensemble-averaged RDF of the final 20 frames, detecting melting or amorphization (Pass threshold: $> 0.5$); and 
(4) \textbf{Phase Retained}: A time-averaged structure is computed from the final 50 unwrapped frames to filter out thermal noise. This metric evaluates whether the lattice preserved its initial space group within a tolerance sweep (0.01 to 1.0~\AA). ``Yes'' indicates the initial symmetry was retained. ``No'' indicates the symmetry check failed, and the lattice either underwent a spontaneous solid-solid phase transition (with the new space group noted) or lost crystallinity entirely.
\label{tab:fullcomp_MLIPs_MD_evals}} \\

\toprule
\textbf{Comp} & \textbf{Init. SG} & \textbf{Engine} & \textbf{Verdict} & \textbf{E (eV/at)} & \textbf{RMSD (\AA)} & \textbf{R. Ver} & \textbf{Vol (\%)} & \textbf{V. Ver} & \textbf{RDF} & \textbf{RDF Ver} & \textbf{Phase Retained} \\
\midrule
\endfirsthead

\multicolumn{12}{c}{{\bfseries \tablename\ \thetable{} -- continued from previous page}} \\
\toprule
\textbf{Comp} & \textbf{Init. SG} & \textbf{Engine} & \textbf{Verdict} & \textbf{E (eV/at)} & \textbf{RMSD (\AA)} & \textbf{R. Ver} & \textbf{Vol (\%)} & \textbf{V. Ver} & \textbf{RDF} & \textbf{RDF Ver} & \textbf{Phase Retained} \\
\midrule
\endhead

\bottomrule
\multicolumn{12}{r}{{Continued on next page}} \\
\endfoot

\bottomrule
\endlastfoot


\multirow{4}{*}{\large \textbf{LiF}}
& \multirow{4}{*}{$Pm\bar{3}m$} 
  & MACE OMAT & \ust & -4.523 & 1.579 & Fail & 28.3 & Fail & -0.011 & Fail & No ($F\bar{4}3m$) \\*
& & MACE MP   & \ust & -4.523 & 1.318 & Fail & 13.7 & Fail & -0.013 & Fail & No ($P1$) \\*
& & eSEN      & \ust & -4.523 & 1.062 & Fail & 27.8 & Fail & -0.011 & Fail & No ($F\bar{4}3m$) \\*
& & UMA       & \ust & -4.523 & 1.906 & Fail & 9.2  & Fail & -0.011 & Fail & No ($F\bar{4}3m$) \\
\cmidrule{2-12}

& \multirow{4}{*}{$Pnma$} 
  & MACE OMAT & \ust & -4.800 & 2.258 & Fail & 48.0 & Fail & 0.001  & Fail & No ($P6_3/mmc$) \\*
& & MACE MP   & \ust & -4.800 & 2.258 & Fail & 38.4 & Fail & -0.024 & Fail & No ($Fm\bar{3}m$) \\*
& & eSEN      & \ust & -4.800 & 0.555 & Pass & 5.7  & Pass & 0.154  & Fail & No ($P6_3mc$) \\*
& & UMA       & \ust & -4.800 & 1.674 & Fail & 7.2  & Fail & 0.130  & Fail & No ($P6_3mc$) \\
\cmidrule{2-12}

& \multirow{4}{*}{$F\bar{4}3m$} 
  & MACE OMAT & \ust & -4.815 & 3.631 & Fail & 9.0  & Fail & -0.012 & Fail & No ($Fm\bar{3}m$) \\*
& & MACE MP   & \ust & -4.815 & \multicolumn{6}{c}{\textit{Explosion (FAILED\_UNSTABLE)}} & N/A \\*
& & eSEN      & \ust & -4.815 & 3.334 & Fail & 28.5 & Fail & -0.010 & Fail & No ($Fm\bar{3}m$) \\*
& & UMA       & \stb & -4.815 & 0.155 & Pass & 5.3  & Pass & 0.339  & Fail & Yes \\
\cmidrule{2-12}

& \multirow{4}{*}{$P4_2/mnm$} 
  & MACE OMAT & \ust & -4.819 & 0.464 & Pass & 31.4 & Fail & -0.027 & Fail & No ($Fm\bar{3}m$) \\*
& & MACE MP   & \ust & -4.819 & 0.916 & Pass & 35.3 & Fail & 0.093  & Fail & No ($P6_3/mmc$) \\*
& & eSEN      & \ust & -4.819 & 0.408 & Pass & 6.4  & Fail & 0.509  & Pass & No ($P6_3/mmc$) \\*
& & UMA       & \stb & -4.819 & 0.208 & Pass & 4.3  & Pass & 0.716  & Pass & Yes \\
\cmidrule{2-12}

& \multirow{4}{*}{$P6_3mc$} 
  & MACE OMAT & \ust & -4.820 & \multicolumn{6}{c}{\textit{Explosion (FAILED\_UNSTABLE)}} & N/A \\*
& & MACE MP   & \stb & -4.820 & 0.225 & Pass & 30.4 & Fail & 0.882  & Pass & Yes \\*
& & eSEN      & \stb & -4.820 & 0.223 & Pass & 3.9  & Pass & 0.529  & Pass & Yes \\*
& & UMA       & \stb & -4.820 & 0.238 & Pass & 4.6  & Pass & 0.675  & Pass & Yes \\
\cmidrule{2-12}

& \multirow{4}{*}{$Fm\bar{3}m$} 
  & MACE OMAT & \stb & -4.982 & 0.174 & Pass & 2.9  & Pass & 0.281  & Fail & Yes \\*
& & MACE MP   & \stb & -4.982 & 0.181 & Pass & 3.7  & Pass & 0.054  & Fail & Yes \\*
& & eSEN      & \stb & -4.982 & 0.177 & Pass & 4.4  & Pass & 0.290  & Fail & Yes \\*
& & UMA       & \stb & -4.982 & 0.177 & Pass & 4.0  & Pass & 0.219  & Fail & Yes \\
\cmidrule{2-12}

& \multirow{4}{*}{$P6_3/mmc$} 
  & MACE OMAT & \ust & -4.821 & 1.917 & Fail & 5.6  & Pass & 0.370  & Fail & No ($Fm\bar{3}m$) \\*
& & MACE MP   & \ust & -4.821 & 0.845 & Pass & 6.1  & Fail & 0.044  & Fail & No ($Fm\bar{3}m$) \\*
& & eSEN      & \ust & -4.821 & 0.276 & Pass & 9.4  & Fail & 0.438  & Fail & No ($P6_3mc$) \\*
& & UMA       & \stb & -4.821 & 0.200 & Pass & 7.1  & Fail & 0.579  & Pass & Yes \\
\cmidrule{2-12}

& \multirow{4}{*}{$P6_3mc$} 
  & MACE OMAT & \ust & -4.826 & 0.333 & Pass & 17.7 & Fail & 0.034  & Fail & No ($P6_3/mmc$) \\*
& & MACE MP   & \ust & -4.826 & 0.459 & Pass & 11.4 & Fail & 0.442  & Fail & No ($P6_3/mmc$) \\*
& & eSEN      & \stb & -4.826 & 0.184 & Pass & 5.6  & Pass & 0.465  & Fail & Yes \\*
& & UMA       & \stb & -4.826 & 0.185 & Pass & 4.7  & Pass & 0.393  & Fail & Yes \\
\cmidrule{2-12}

& \multirow{4}{*}{$P1$} 
  & MACE OMAT & \ust & -8.611 & \multicolumn{6}{c}{\textit{Explosion (FAILED\_UNSTABLE)}} & N/A \\*
& & MACE MP   & \stb & -8.611 & 0.717 & Pass & 3.8  & Pass & 0.075  & Fail & No ($Fm\bar{3}m$) \\*
& & eSEN      & \stb & -8.611 & 0.713 & Pass & 4.3  & Pass & 0.237  & Fail & No ($Fm\bar{3}m$) \\*
& & UMA       & \stb & -8.611 & 0.586 & Pass & 4.9  & Pass & 0.232  & Fail & No ($Fm\bar{3}m$) \\
\cmidrule{2-12}

& \multirow{4}{*}{$Pm$} 
  & MACE OMAT & \ust & -11.871 & \multicolumn{6}{c}{\textit{Explosion (FAILED\_UNSTABLE)}} & N/A \\*
& & MACE MP   & \stb & -11.871 & 0.166 & Pass & 4.7  & Pass & 0.115  & Fail & No ($Fm\bar{3}m$) \\*
& & eSEN      & \ust & -11.871 & 0.166 & Pass & 6.2  & Fail & 0.227  & Fail & No ($Fm\bar{3}m$) \\*
& & UMA       & \stb & -11.871 & 0.165 & Pass & 7.3  & Fail & 0.591  & Pass & No ($Fm\bar{3}m$) \\
\cmidrule{2-12}

& \multirow{4}{*}{$Cm$} 
  & MACE OMAT & \ust & -18.566 & \multicolumn{6}{c}{\textit{Explosion (FAILED\_UNSTABLE)}} & N/A \\*
& & MACE MP   & \ust & -18.566 & 0.163 & Pass & 9.2  & Fail & 0.344  & Fail & No ($Fm\bar{3}m$) \\*
& & eSEN      & \ust & -18.566 & 2.592 & Fail & 26.2 & Fail & -0.011 & Fail & No ($Fm\bar{3}m$) \\*
& & UMA       & \ust & -18.566 & 1.955 & Fail & 15.2 & Fail & -0.012 & Fail & No ($Fm\bar{3}m$) \\
\midrule


\multirow{4}{*}{\large \textbf{CsPbI$_3$}} 
& \multirow{4}{*}{$Pm\bar{3}m$} 
  & MACE OMAT & \ust & -2.796 & 0.793 & Pass & 4.9 & Pass & 0.332 & Fail & No ($P1$) \\*
& & MACE MP   & \ust & -2.796 & 0.768 & Pass & 5.5 & Pass & 0.136 & Fail & No ($Pnma$) \\*
& & eSEN      & \ust & -2.796 & 0.694 & Pass & 4.0 & Pass & 0.385 & Fail & No ($Pnma$) \\*
& & UMA       & \ust & -2.796 & 0.648 & Pass & 3.9 & Pass & 0.412 & Fail & No ($P2_1/m$) \\
\cmidrule{2-12}

& \multirow{4}{*}{$P\bar{1}$} 
  & MACE OMAT & \ust & -2.824 & 2.989 & Fail & 8.1  & Fail & 0.670 & Pass & Yes \\*
& & MACE MP   & \ust & -2.824 & 2.097 & Fail & 7.2  & Fail & 0.687 & Pass & Yes \\*
& & eSEN      & \ust & -2.824 & 2.011 & Fail & 4.0  & Pass & 0.489 & Fail & No ($P1$) \\*
& & UMA       & \ust & -2.824 & 2.760 & Fail & 15.3 & Fail & 0.355 & Fail & No ($P1$) \\
\cmidrule{2-12}

& \multirow{4}{*}{$Pnma$} 
  & MACE OMAT & \ust & -2.825 & 1.176 & Fail & 12.4 & Fail & 0.335 & Fail & Yes \\*
& & MACE MP   & \ust & -2.825 & 1.061 & Fail & 14.5 & Fail & 0.347 & Fail & Yes \\*
& & eSEN      & \stb & -2.825 & 0.929 & Pass & 5.3  & Pass & 0.522 & Pass & No ($P\bar{1}$) \\*
& & UMA       & \ust & -2.825 & 1.088 & Fail & 7.8  & Fail & 0.545 & Pass & No ($Pm$) \\
\cmidrule{2-12}

& \multirow{4}{*}{$Pm$} 
  & MACE OMAT & \stb & -2.826 & 0.834 & Pass & 3.8  & Pass & 0.692 & Pass & Yes \\*
& & MACE MP   & \stb & -2.826 & 0.839 & Pass & 3.9  & Pass & 0.799 & Pass & Yes \\*
& & eSEN      & \stb & -2.826 & 0.871 & Pass & 7.2  & Fail & 0.793 & Pass & No ($P1$) \\*
& & UMA       & \stb & -2.826 & 0.891 & Pass & 9.4  & Fail & 0.760 & Pass & No ($P1$) \\
\cmidrule{2-12}

& \multirow{4}{*}{$P2_1/m$} 
  & MACE OMAT & \ust & -2.829 & 5.136 & Fail & 12.9 & Fail & 0.539 & Pass & Yes \\*
& & MACE MP   & \ust & -2.829 & 4.960 & Fail & 6.3  & Fail & 0.580 & Pass & Yes \\*
& & eSEN      & \ust & -2.829 & 5.236 & Fail & 6.4  & Fail & 0.565 & Pass & No ($P1$) \\*
& & UMA       & \ust & -2.829 & 5.188 & Fail & 4.8  & Pass & 0.498 & Fail & No ($P1$) \\
\cmidrule{2-12}

& \multirow{4}{*}{$Cmcm$} 
  & MACE OMAT & \ust & -2.831 & 1.123 & Fail & 22.2 & Fail & 0.356 & Fail & No ($P2_1/m$) \\*
& & MACE MP   & \stb & -2.831 & 0.404 & Pass & 16.2 & Fail & 0.513 & Pass & Yes \\*
& & eSEN      & \ust & -2.831 & 0.375 & Pass & 9.4  & Fail & 0.380 & Fail & Yes \\*
& & UMA       & \ust & -2.831 & 0.415 & Pass & 9.2  & Fail & 0.342 & Fail & Yes \\
\cmidrule{2-12}

& \multirow{4}{*}{$Pnma$} 
  & MACE OMAT & \ust & -2.840 & 0.387 & Pass & 30.0 & Fail & 0.368 & Fail & Yes \\*
& & MACE MP   & \ust & -2.840 & 0.507 & Pass & 16.1 & Fail & 0.460 & Fail & Yes \\*
& & eSEN      & \stb & -2.840 & 0.382 & Pass & 8.3  & Fail & 0.501 & Pass & Yes \\*
& & UMA       & \ust & -2.840 & 0.492 & Pass & 14.8 & Fail & 0.476 & Fail & No ($Pc$) \\
\midrule


\multirow{4}{*}{\large \textbf{HfO$_2$}} 
& \multirow{4}{*}{$Fm\bar{3}m$} 
  & MACE OMAT & \ust & -10.117 & 0.453 & Pass & 3.1  & Pass & 0.189 & Fail & No ($Pca2_1$) \\*
& & MACE MP   & \ust & -10.117 & 0.919 & Pass & 13.9 & Fail & 0.155 & Fail & No ($P2_1/c$) \\*
& & eSEN      & \ust & -10.117 & 0.506 & Pass & 2.0  & Pass & 0.154 & Fail & No ($Pca2_1$) \\*
& & UMA       & \ust & -10.117 & 0.576 & Pass & 4.2  & Pass & 0.317 & Fail & No ($P2_1/c$) \\
\cmidrule{2-12}

& \multirow{4}{*}{$Pmn2_1$} 
  & MACE OMAT & \ust & -10.128 & 0.519 & Pass & 9.2  & Fail & 0.387 & Fail & No ($P2_1/c$) \\*
& & MACE MP   & \ust & -10.128 & 0.845 & Pass & 22.0 & Fail & 0.415 & Fail & No ($P4_2/mnm$) \\*
& & eSEN      & \ust & -10.128 & 0.393 & Pass & 4.1  & Pass & 0.139 & Fail & No ($Pca2_1$) \\*
& & UMA       & \stb & -10.128 & 0.390 & Pass & 2.9  & Pass & 0.564 & Pass & No ($P2_1/m$) \\
\cmidrule{2-12}

& \multirow{4}{*}{$P4_2/nmc$} 
  & MACE OMAT & \ust & -10.129 & 0.530 & Pass & 8.4  & Fail & 0.564 & Pass & No ($P2_1/c$) \\*
& & MACE MP   & \ust & -10.129 & 0.464 & Pass & 5.3  & Pass & 0.321 & Fail & No ($P2_1/c$) \\*
& & eSEN      & \ust & -10.129 & 0.429 & Pass & 2.5  & Pass & -0.046& Fail & No ($Pca2_1$) \\*
& & UMA       & \ust & -10.129 & 0.431 & Pass & 2.1  & Pass & -0.046& Fail & No ($Pca2_1$) \\
\cmidrule{2-12}

& \multirow{4}{*}{$Fdd2$} 
  & MACE OMAT & \stb & -10.132 & 0.098 & Pass & 1.0  & Pass & 0.777 & Pass & Yes \\*
& & MACE MP   & \ust & -10.132 & 6.325 & Fail & 17.6 & Fail & 0.496 & Fail & No ($P1$) \\*
& & eSEN      & \ust & -10.132 & 12.981& Fail & 16.4 & Fail & 0.651 & Pass & No ($P1$) \\*
& & UMA       & \stb & -10.132 & 0.097 & Pass & 1.1  & Pass & 0.804 & Pass & Yes \\
\cmidrule{2-12}

& \multirow{4}{*}{$P2_1/m$} 
  & MACE OMAT & \ust & -10.135 & 1.756 & Fail & 9.0  & Fail & 0.518 & Pass & Yes \\*
& & MACE MP   & \ust & -10.135 & 3.065 & Fail & 22.2 & Fail & 0.396 & Fail & No ($Pnma$) \\*
& & eSEN      & \stb & -10.135 & 1.750 & Fail & 5.1  & Pass & 0.601 & Pass & Yes \\*
& & UMA       & \ust & -10.135 & 1.744 & Fail & 6.3  & Fail & 0.544 & Pass & Yes \\
\cmidrule{2-12}

& \multirow{4}{*}{$Pnma$} 
  & MACE OMAT & \stb & -10.139 & 0.101 & Pass & 9.9  & Fail & 0.772 & Pass & Yes \\*
& & MACE MP   & \stb & -10.139 & 0.133 & Pass & 11.4 & Fail & 0.787 & Pass & Yes \\*
& & eSEN      & \stb & -10.139 & 0.138 & Pass & 3.1  & Pass & 0.791 & Pass & Yes \\*
& & UMA       & \stb & -10.139 & 0.147 & Pass & 5.1  & Pass & 0.808 & Pass & Yes \\
\cmidrule{2-12}

& \multirow{4}{*}{$P2_1/c$} 
  & MACE OMAT & \stb & -10.143 & 1.216 & Fail & 4.0  & Pass & 0.511 & Pass & Yes \\*
& & MACE MP   & \ust & -10.143 & 1.491 & Fail & 3.2  & Pass & 0.644 & Pass & No ($Pbca$) \\*
& & eSEN      & \ust & -10.143 & 1.275 & Fail & 6.3  & Fail & 0.517 & Pass & Yes \\*
& & UMA       & \stb & -10.143 & 1.262 & Fail & 2.9  & Pass & 0.528 & Pass & Yes \\
\cmidrule{2-12}

& \multirow{4}{*}{$Pca2_1$} 
  & MACE OMAT & \ust & -10.145 & 0.427 & Pass & 11.7 & Fail & -0.039& Fail & No ($Pbcn$) \\*
& & MACE MP   & \ust & -10.145 & 0.350 & Pass & 14.2 & Fail & 0.333 & Fail & No ($Pbcn$) \\*
& & eSEN      & \stb & -10.145 & 0.107 & Pass & 2.6  & Pass & 0.741 & Pass & Yes \\*
& & UMA       & \ust & -10.145 & 0.455 & Pass & 4.5  & Pass & 0.024 & Fail & No ($Pbcn$) \\
\cmidrule{2-12}

& \multirow{4}{*}{$P4_2/mnm$} 
  & MACE OMAT & \stb & -10.148 & 0.151 & Pass & 4.9  & Pass & 0.768 & Pass & Yes \\*
& & MACE MP   & \stb & -10.148 & 0.194 & Pass & 9.4  & Fail & 0.771 & Pass & Yes \\*
& & eSEN      & \stb & -10.148 & 0.125 & Pass & 1.4  & Pass & 0.679 & Pass & Yes \\*
& & UMA       & \stb & -10.148 & 0.125 & Pass & 1.9  & Pass & 0.684 & Pass & Yes \\
\cmidrule{2-12}

& \multirow{4}{*}{$Pbcn$} 
  & MACE OMAT & \ust & -10.155 & 1.051 & Fail & 17.6 & Fail & 0.606 & Pass & No ($P2_1/c$) \\*
& & MACE MP   & \ust & -10.155 & 0.488 & Pass & 13.5 & Fail & 0.373 & Fail & No ($P2_1/c$) \\*
& & eSEN      & \ust & -10.155 & 0.403 & Pass & 4.0  & Pass & 0.429 & Fail & No ($P2_1/c$) \\*
& & UMA       & \stb & -10.155 & 0.099 & Pass & 6.0  & Pass & 0.765 & Pass & Yes \\
\cmidrule{2-12}

& \multirow{4}{*}{$I4_1/amd$} 
  & MACE OMAT & \ust & -10.160 & 5.877 & Fail & 9.1  & Fail & 0.414 & Fail & No ($P\bar{1}$) \\*
& & MACE MP   & \ust & -10.160 & 7.621 & Fail & 22.6 & Fail & 0.837 & Pass & No ($C2/m$) \\*
& & eSEN      & \ust & -10.160 & 6.331 & Fail & 45.8 & Fail & 0.305 & Fail & No ($P1$) \\*
& & UMA       & \ust & -10.160 & 7.544 & Fail & 78.8 & Fail & 0.719 & Pass & No ($P1$) \\
\cmidrule{2-12}

& \multirow{4}{*}{$P2_1/c$} 
  & MACE OMAT & \stb & -10.165 & 1.163 & Fail & 2.4  & Pass & 0.642 & Pass & Yes \\*
& & MACE MP   & \ust & -10.165 & 1.208 & Fail & 3.5  & Pass & 0.435 & Fail & No ($Pbcn$) \\*
& & eSEN      & \stb & -10.165 & 1.156 & Fail & 2.1  & Pass & 0.721 & Pass & Yes \\*
& & UMA       & \stb & -10.165 & 1.157 & Fail & 1.6  & Pass & 0.722 & Pass & Yes \\
\midrule


\multirow{4}{*}{\large \textbf{Bi$_2$Sn$_2$O$_7$}} 
& \multirow{4}{*}{$Fd\bar{3}m$} 
  & MACE OMAT & \stb & -6.017 & 0.423 & Pass & 4.5 & Pass & 0.463 & Fail & Yes \\*
& & MACE MP   & \stb & -6.017 & 0.351 & Pass & 2.4 & Pass & 0.444 & Fail & Yes \\*
& & eSEN      & \stb & -6.017 & 0.229 & Pass & 1.4 & Pass & 0.502 & Pass & Yes \\*
& & UMA       & \stb & -6.017 & 0.232 & Pass & 1.8 & Pass & 0.441 & Fail & Yes \\
\cmidrule{2-12}

& \multirow{4}{*}{$P2_12_12_1$} 
  & MACE OMAT & \stb & -5.997 & 0.359 & Pass & 5.1 & Pass & 0.675 & Pass & No ($Imma$) \\*
& & MACE MP   & \stb & -5.997 & 0.324 & Pass & 4.9 & Pass & 0.710 & Pass & No ($Imma$) \\*
& & eSEN      & \stb & -5.997 & 0.461 & Pass & 1.9 & Pass & 0.790 & Pass & No ($Pnma$) \\*
& & UMA       & \stb & -5.997 & 0.460 & Pass & 2.2 & Pass & 0.793 & Pass & No ($Pnma$) \\
\cmidrule{2-12}

& \multirow{4}{*}{$P2_1$} 
  & MACE OMAT & \ust & -6.004 & 6.865 & Fail & 9.8 & Fail & 0.705 & Pass & No ($P1$) \\*
& & MACE MP   & \stb & -6.004 & 0.228 & Pass & 1.0 & Pass & 0.721 & Pass & No ($Cmc2_1$) \\*
& & eSEN      & \ust & -6.004 & 4.716 & Fail & 3.9 & Pass & 0.700 & Pass & No ($P1$) \\*
& & UMA       & \ust & -6.004 & 4.561 & Fail & 1.8 & Pass & 0.669 & Pass & No ($P1$) \\
\cmidrule{2-12}

& \multirow{4}{*}{$P1$} 
  & MACE OMAT & \stb & -6.017 & 0.375 & Pass & 1.0 & Pass & 0.856 & Pass & No ($P2$) \\*
& & MACE MP   & \stb & -6.017 & 0.325 & Pass & 0.9 & Pass & 0.856 & Pass & Yes \\*
& & eSEN      & \stb & -6.017 & 0.259 & Pass & 1.0 & Pass & 0.903 & Pass & No ($Fd\bar{3}m$) \\*
& & UMA       & \stb & -6.017 & 0.405 & Pass & 0.7 & Pass & 0.837 & Pass & No ($P2$) \\
\cmidrule{2-12}

& \multirow{4}{*}{$P1$} 
  & MACE OMAT & \stb & -6.029 & 0.358 & Pass & 0.6 & Pass & 0.865 & Pass & No ($Fd\bar{3}m$) \\*
& & MACE MP   & \stb & -6.029 & 0.321 & Pass & 0.8 & Pass & 0.867 & Pass & No ($Fd\bar{3}m$) \\*
& & eSEN      & \stb & -6.029 & 0.266 & Pass & 0.9 & Pass & 0.956 & Pass & No ($Fd\bar{3}m$) \\*
& & UMA       & \stb & -6.029 & 0.248 & Pass & 0.6 & Pass & 0.923 & Pass & No ($Fd\bar{3}m$) \\
\cmidrule{2-12}

& \multirow{4}{*}{$Cc$} 
  & MACE OMAT & \stb & -6.030 & 0.242 & Pass & 0.7 & Pass & 0.874 & Pass & No ($Fd\bar{3}m$) \\*
& & MACE MP   & \stb & -6.030 & 0.270 & Pass & 1.2 & Pass & 0.813 & Pass & No ($Fd\bar{3}m$) \\*
& & eSEN      & \stb & -6.030 & 0.194 & Pass & 1.0 & Pass & 0.943 & Pass & No ($Fd\bar{3}m$) \\*
& & UMA       & \stb & -6.030 & 0.272 & Pass & 0.7 & Pass & 0.932 & Pass & No ($Fd\bar{3}m$) \\
\cmidrule{2-12}

& \multirow{4}{*}{$Pna2_1$} 
  & MACE OMAT & \stb & -6.030 & 0.350 & Pass & 6.2 & Fail & 0.713 & Pass & No ($Fd\bar{3}m$) \\*
& & MACE MP   & \stb & -6.030 & 0.203 & Pass & 5.6 & Pass & 0.682 & Pass & No ($Fd\bar{3}m$) \\*
& & eSEN      & \stb & -6.030 & 0.288 & Pass & 1.7 & Pass & 0.738 & Pass & No ($Fd\bar{3}m$) \\*
& & UMA       & \stb & -6.030 & 0.268 & Pass & 2.9 & Pass & 0.715 & Pass & No ($Fd\bar{3}m$) \\
\cmidrule{2-12}

& \multirow{4}{*}{$P2_1$} 
  & MACE OMAT & \stb & -6.030 & 0.320 & Pass & 2.7 & Pass & 0.763 & Pass & No ($Fd\bar{3}m$) \\*
& & MACE MP   & \stb & -6.030 & 0.352 & Pass & 3.1 & Pass & 0.765 & Pass & No ($Fd\bar{3}m$) \\*
& & eSEN      & \stb & -6.030 & 0.303 & Pass & 1.9 & Pass & 0.739 & Pass & No ($Fd\bar{3}m$) \\*
& & UMA       & \stb & -6.030 & 0.338 & Pass & 2.2 & Pass & 0.747 & Pass & No ($Fd\bar{3}m$) \\
\cmidrule{2-12}

& \multirow{4}{*}{$P2_12_12_1$} 
  & MACE OMAT & \stb & -6.031 & 0.250 & Pass & 0.8 & Pass & 0.729 & Pass & No ($Fd\bar{3}m$) \\*
& & MACE MP   & \stb & -6.031 & 0.211 & Pass & 5.1 & Pass & 0.684 & Pass & No ($Fd\bar{3}m$) \\*
& & eSEN      & \stb & -6.031 & 0.304 & Pass & 1.8 & Pass & 0.713 & Pass & No ($Fd\bar{3}m$) \\*
& & UMA       & \stb & -6.031 & 0.250 & Pass & 2.4 & Pass & 0.721 & Pass & No ($Fd\bar{3}m$) \\
\cmidrule{2-12}

& \multirow{4}{*}{$Cc$} 
  & MACE OMAT & \ust & -6.031 & 6.107 & Fail & 2.9 & Pass & 0.658 & Pass & No ($P1$) \\*
& & MACE MP   & \ust & -6.031 & 5.990 & Fail & 5.5 & Pass & 0.701 & Pass & No ($P1$) \\*
& & eSEN      & \ust & -6.031 & 6.010 & Fail & 4.8 & Pass & 0.610 & Pass & No ($P1$) \\*
& & UMA       & \ust & -6.031 & 6.069 & Fail & 7.1 & Fail & 0.619 & Pass & No ($P1$) \\

\end{longtable}
\end{landscape}

\newpage
\subsection{Benchmarking MLIPs for phase transition path finding with VibroML}
\label{sec:neb_results}
VibroML contains a nudge elastic band (NEB) module to accelerate the search for minimum energy paths (MEP) in solid-state phase transitions. We evaluate the performance of three state-of-the-art MLIP architectures, MACE, eSEN and UMA, acting as surrogate models to drive NEB calculations.
The resulting paths generated by the MLIPs were subsequently evaluated using DFT single-point calculations on the images. This ``MLIP-driven, DFT-verified'' approach allows us to quantify the accuracy of the potentials in describing complex energy landscapes and their ability to locate physically viable transition mechanisms.

For each compound (\ce{LiF}, \ce{CsPbI3}, \ce{HfO2}, and \ce{Bi2Sn2O7}), initial and final crystallographic phases were defined, based on DFT evaluation of the results in Table~\ref{tab:stability_remediation}. Specifically, we utilized the high-symmetry starting structure and the corresponding lowest-energy polymorph identified by VibroML and verified by DFT. The NEB optimization was performed entirely using the MLIP forces. The quality of the resulting path is assessed by comparing the MLIP predictions against the ground-truth DFT values for the exact same geometries.

We defined the following metrics to quantify performance:
\begin{itemize}
    \item Reaction energy error ($\epsilon_{reaction}$): The absolute difference in the relative energy of the final state ($|E_{final}^{DFT}-E_{final}^{MLIP}|$). This measures thermodynamic accuracy.
    \item Barrier energy error ($\epsilon_{barrier}$): The absolute difference in the maximum energy along the path ($|E_{max}^{DFT}-E_{max}^{MLIP}|$). This measures kinetic accuracy.
    \item Path quality ($\overline{E}_{path}$): The average DFT energy of the images along the path. A lower value indicates the MLIP successfully found a more stable, physically favorable transition mechanism.
    \item Global energy accuracy ($MAE_{energy}$): The mean absolute error of relative energies (excluding the reference image 0).
    \item Force accuracy ($MAE_{forces}$): The mean absolute error of atomic forces across all images.
    \item Endpoint stability ($\mu_{endpoints}$): The sum of force errors at the initial and final images. High values indicate ``hallucinated minima'', where the MLIP predicts a stable structure that DFT still finds residual forces.
\end{itemize}

\begin{figure}[H]
    \centering
    \includegraphics[width=\textwidth]{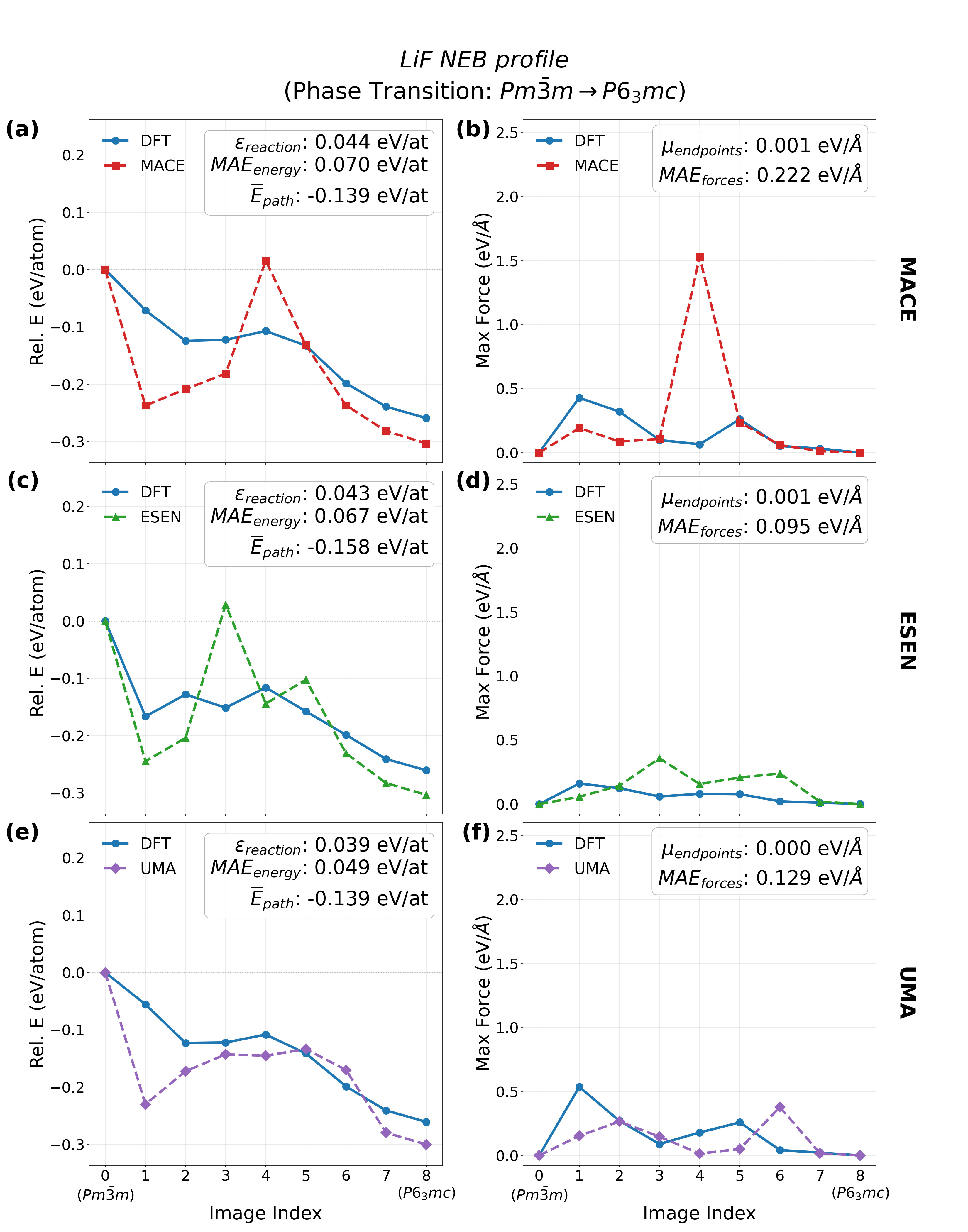}
    \caption{Comparative NEB profiles for lithium fluoride (\ce{LiF}). Energy (left) and maximum force (right) profiles for the transition from $Pm\overline{3}m$ to $P6_3mc$ with different MLIPs, (a, b) MACE, (c, d) eSEN, and (e, f) UMA compared to DFT. Solid blue lines represent DFT verification, while dashed colored lines represent MLIP predictions.}
    \label{fig:sx1}
\end{figure}

We now proceed to the system-by-system analysis. The transition of lithium fluoride (\ce{LiF}) from cubic ($Pm\overline{3}m$) to hexagonal ($P6_3mc$), presented in Figure~\ref{fig:sx1}, highlights potential pitfalls in NEB calculations regarding force outliers. MACE exhibited a massive force error spike mid-transition, resulting in a high of 0.222 eV/\AA. While it captured the general energy trend, this suggests an overestimation of force gradients in intermediate configurations. eSEN identified the lowest energy path with an average energy ($\overline{E}_{path}$) of -0.158 eV/atom, implying it found a highly stable transition mechanism. However, UMA offered the most balanced performance for this system, achieving the lowest energy error (0.049 eV/atom) and the best endpoint stability, indicating excellent agreement with DFT regarding the stability of the phases.

The halide perovskite cesium lead iodide (\ce{CsPbI3}) presents the most complex case in this benchmark, as depicted in Figure~\ref{fig:sx2}. Unlike the displacive transitions observed in the oxides, this system undergoes a profound structural reconstruction, transforming from the 3D corner-sharing connectivity of the cubic black phase ($Pm\overline{3}m$) to the 1D edge-sharing columnar structure found by VibroML ($P2_1/m$). This connectivity change necessitates significant bond breaking and reorganization, resulting in a distinct activation energy barrier that is absent in the other compounds.

The choice of MLIP had a decisive impact on the transition mechanism found for this reconstruction. UMA successfully identified the most kinetically favorable pathway, locating a mechanism with a DFT-verified barrier of only 0.092 eV/atom. In stark contrast, MACE identified a path traversing a much steeper energetic saddle point, with a barrier of 0.175 eV/atom—nearly double that found by UMA. eSEN occupied the middle ground, finding a barrier of 0.117 eV/atom. This comparison highlights a trade-off: while MACE demonstrated superior numerical stability on the path it generated ($MAE_{energy}=0.019$ eV/atom), it failed to locate the optimal low-energy transition channel that UMA navigated. UMA resolved the mechanism problem better, finding a lower barrier, even though its point-wise energy predictions were numerically less precise than MACE.

\begin{figure}[H]
    \centering
    \includegraphics[width=\textwidth]{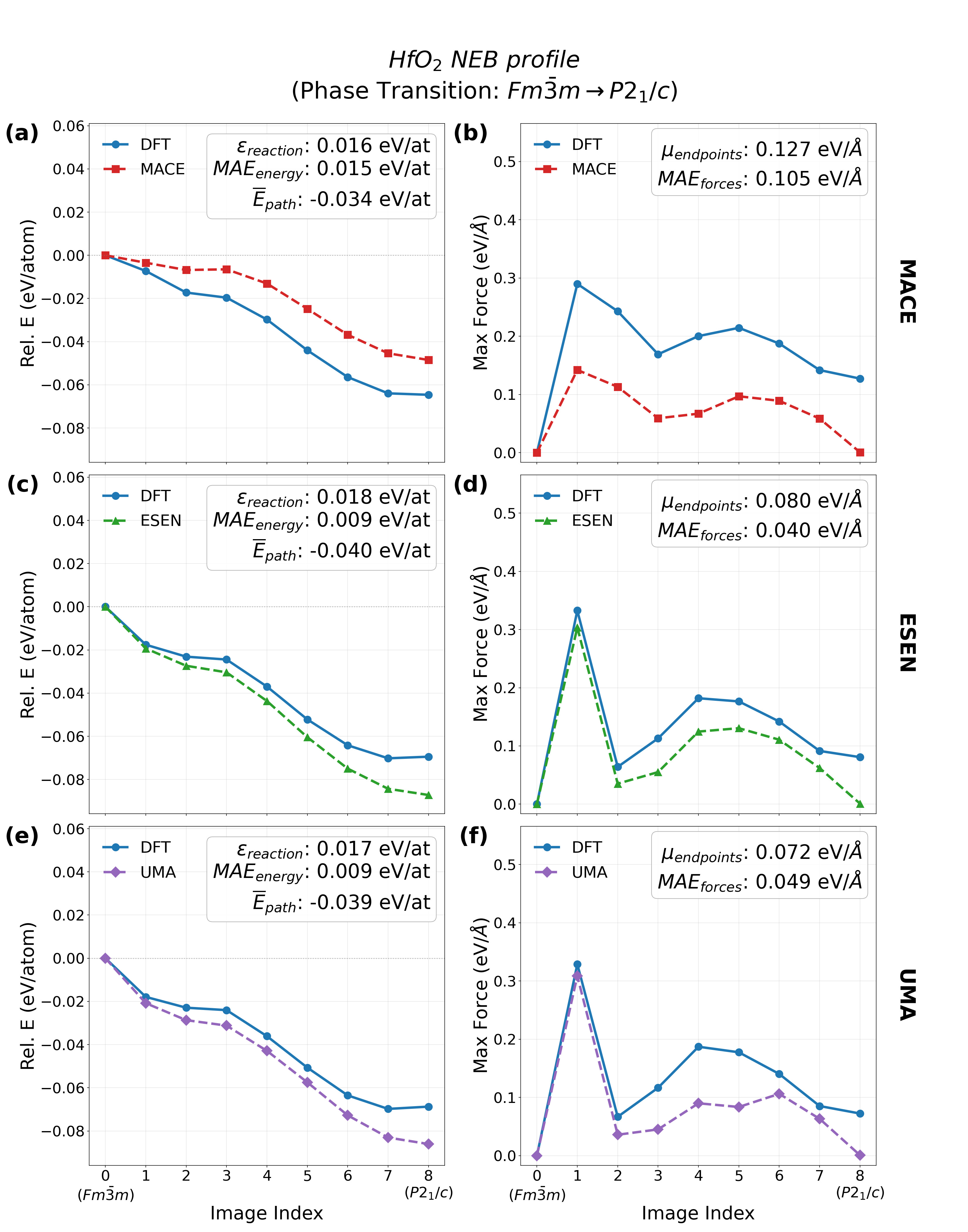}
    \caption{Comparative NEB profiles for hafnium oxide (\ce{HfO2}). Energy (left) and maximum force (right) profiles for the transition from $Fm\overline{3}m$ to $P2_1/c$ with different MLIPs, (a, b) MACE, (c, d) eSEN, and (e, f) UMA compared to DFT. Solid blue lines represent DFT verification, while dashed colored lines represent MLIP predictions.}
    \label{fig:sx3}
\end{figure}

For hafnium oxide (\ce{HfO2}), we analyzed the barrier-less downhill transition from cubic ($Fm\overline{3}m$) to monoclinic ($P2_1/c$), shown in Figure~\ref{fig:sx3}. The metrics indicate a split performance between UMA and eSEN. eSEN located the optimal transition path with the lowest average energy ( $\overline{E}_{path} = -0.0398$ eV/atom) and significantly outperformed the other architectures in force prediction, achieving an outstanding MAE$_{forces}$ of 0.04 eV/\AA\ compared to 0.105 eV/\AA\ for MACE. This indicates that eSEN is particularly robust in capturing the local gradients of the hafnia force fields. UMA performed similarly well in path finding and achieved the best overall energy agreement with an $MAE_{energy}=8.7$ meV/atom, confirming that both models handle this oxide system effectively.

\begin{figure}[H]
    \centering
    \includegraphics[width=\textwidth]{CsPbI3_COMBINED_panel.jpg}
    \caption{Comparative NEB profiles for cesium lead iodide (\ce{CsPbI3}). Energy (left) and maximum force (right) profiles for the transition from $Pm\overline{3}m$ to $P2_1/m$ with different MLIPs, (a, b) MACE, (c, d) eSEN, and (e, f) UMA compared to DFT. Solid blue lines represent DFT verification, while dashed colored lines represent MLIP predictions.}
    \label{fig:sx2}
\end{figure}

\begin{figure}[H]
    \centering
    \includegraphics[width=\textwidth]{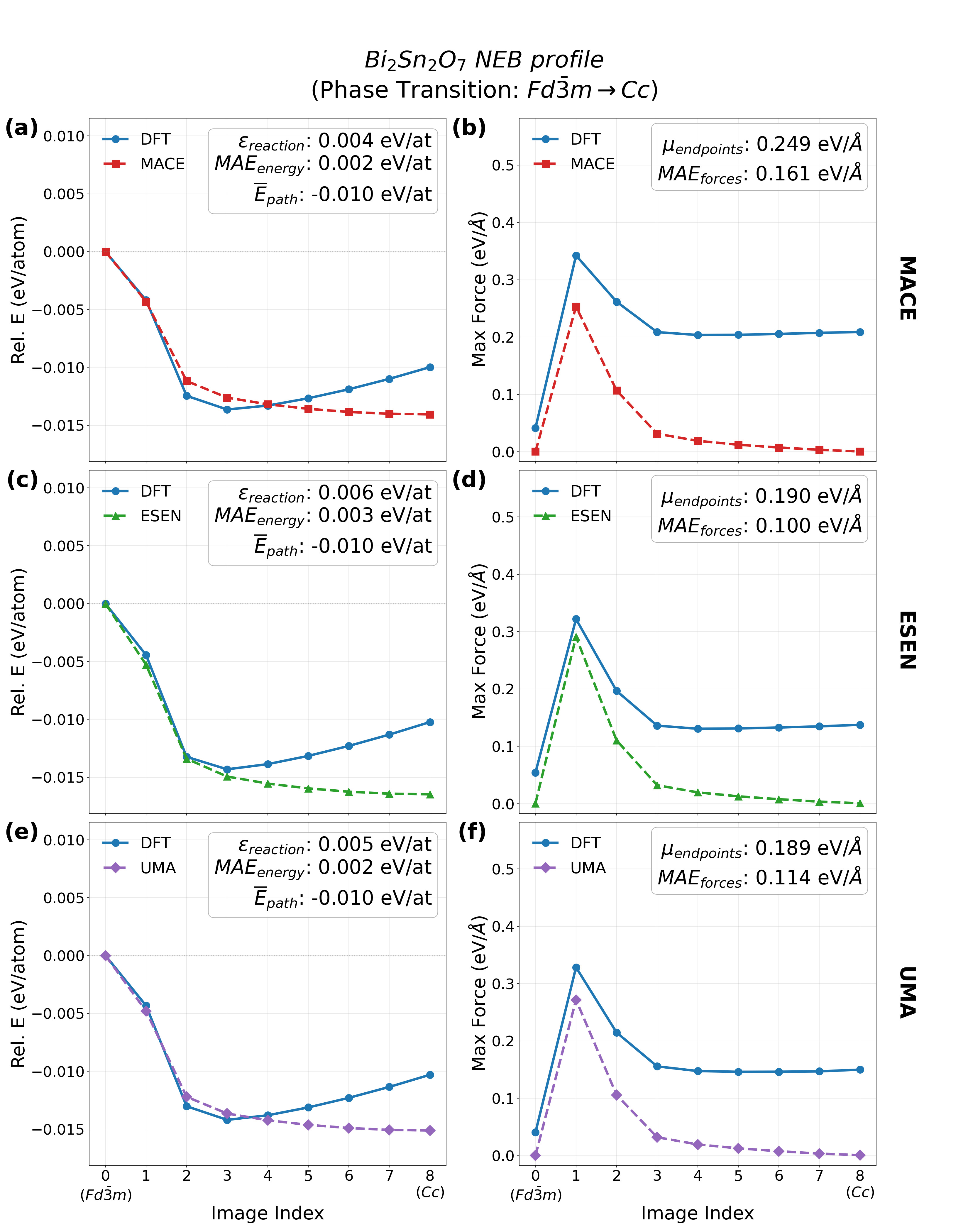}
    \caption{Comparative NEB profiles for bismuth stannate (\ce{Bi2Sn2O7}). Energy (left) and maximum force (right) profiles for the transition from $Fd\overline{3}m$ to $Cc$ with different MLIPs, (a, b) MACE, (c, d) eSEN, and (e, f) UMA compared to DFT. Solid blue lines represent DFT verification, while dashed colored lines represent MLIP predictions.}
    \label{fig:sx4}
\end{figure}

Finally, the profiling of bismuth stannate (\ce{Bi2Sn2O7}) for the transition pyrochlore ($Fd\overline{3}m$) $\rightarrow$ monoclinic ($Cc$) is shown in Figure~\ref{fig:sx4}. This system exhibits a ``downhill'' transition character where the energy decreases monotonically. All three models performed exceptionally well, showing that the PES for this oxide is well-captured by current MLIPs. eSEN found the most favorable path with a DFT average energy of -0.0103 eV/atom, marginally beating UMA and MACE. MACE demonstrated the highest energy fidelity with an $MAE_{energy}$ of only 1.6 meV/atom, capturing the DFT profile almost perfectly. eSEN however provided the most accurate forces for this system ($MAE_{forces}\approx0.10$ eV/\AA).

A critical observation across the benchmarked systems, specifically \ce{CsPbI3}, \ce{HfO2}, and \ce{Bi2Sn2O7}, is a recurring discrepancy regarding the stability of the final endpoints. This is evidenced by high force errors at the final image, ranging from 0.07 to 0.21 eV/\AA. While the MLIPs frequently predict the final structures found by VibroML as relaxed minima with near-zero forces, DFT verification reveals significant residual forces. 

This discrepancy suggests that while the initial high-symmetry phases are well-represented in the training data, the specific low-symmetry polymorphs identified by VibroML likely represent out-of-distribution samples. Crucially, this generalization failure manifests as a systematic underprediction of the PES curvature---a phenomenon known as ``softening'', which has been well-documented in early foundation models
\cite{fuLearningSmoothExpressive2025,anamComprehensiveAssessmentBenchmark2025}. When the model underestimates the stiffness of the bond stretching or bending modes in these novel phases, it effectively flattens the potential well. As a result, the geometric configuration that satisfies $\nabla E_{\text{MLIP}} \approx 0$ is spatially displaced from the true DFT minimum. Consequently, when validated against DFT, this MLIP-optimized geometry corresponds to a point on the steeper, true potential surface where the gradient is non-zero, appearing as a relatively high residual force.

In conclusion, the benchmarking of VibroML across these four diverse systems reveals that no single MLIP architecture dominates every metric. MACE consistently provides excellent energy matching but occasionally struggles with force spikes or locating the global lowest-energy path, as seen in the barrier-limited transition. eSEN excels at force prediction and frequently identifies the thermodynamically most favorable transition mechanisms. UMA serves as a robust all-rounder, often balancing high-quality path finding, with quite low errors in the energy and forces.
\newpage

\subsection{Computational Cost Comparison of MLIPs}
\label{sec:mlip_cost_comparison}

To quantify the computational overhead of the evaluated MLIPs, we benchmarked their execution times for phonon dispersion and molecular dynamics (MD) stability evaluations using a single NVIDIA A100 GPU. The total execution times (including structural relaxation) and relative costs are summarized in Table~\ref{tab:mlip_cost}.

\begin{table}[H]
    \centering
    \caption{Execution times and relative computational costs for the four MLIP architectures evaluating \ce{HfO2} ($Fm\overline{3}m$). Phonon dispersion calculations utilized a $4 \times 4 \times 4$ supercell, while the finite-temperature molecular dynamics (MD) evaluations utilized a $3 \times 3 \times 3$ supercell. Relative costs are normalized against the MACE-OMAT-0 baseline.}
    \label{tab:mlip_cost}
    \renewcommand{\arraystretch}{1.2}
    \begin{tabular}{@{}lcccc@{}}
        \toprule
        \multirow{2}{*}{\textbf{MLIP Engine}} & \multicolumn{2}{c}{\textbf{Phonon Dispersion ($4 \times 4 \times 4$)}} & \multicolumn{2}{c}{\textbf{Molecular Dynamics ($3 \times 3 \times 3$)}} \\
        \cmidrule(lr){2-3} \cmidrule(l){4-5}
        & \textbf{Time (sec)} & \textbf{Relative Cost} & \textbf{Time (sec)} & \textbf{Relative Cost} \\
        \midrule
        \textbf{MACE-OMAT-0} & 63 & 1.0$\times$ & 358 & 1.0$\times$ \\
        \textbf{MACE-MP} & 66 & 1.05$\times$ & 336 & 0.93$\times$ \\
        \textbf{eSEN-30M} & 176 & 2.8$\times$ & 2777 & 7.7$\times$ \\
        \textbf{UMA-m-1p1} & 278 & 4.4$\times$ & 4055 & 11.3$\times$ \\
        \bottomrule
    \end{tabular}
\end{table}

For static phonon evaluations on the $4 \times 4 \times 4$ supercell, MACE and MACE-MP exhibit the highest efficiency ($\sim$1 minute). UMA requires 4.4$\times$ more time, while eSEN requires 2.8$\times$ more time. 

During finite-temperature MD simulations on the $3 \times 3 \times 3$ supercell, scaling differences become heavily pronounced. MACE and MACE-MP complete the trajectory in roughly 6 minutes, whereas eSEN and UMA require $\sim$46 minutes (7.7$\times$) and $\sim$68 minutes (11.3$\times$), respectively. This confirms MACE as the most cost-effective engine for the extensive sampling and dynamic evaluations required in VibroML.

\subsection{Algorithm Settings, Computational Efficiency and Relaxation Heuristics}
\label{sec:comp_efficiency}

The specific parameters utilized for each remediation workflow are detailed in Table~\ref{tab:sets_remedalgorithms}. To ensure a fair baseline for performance comparison on the massive \ce{Bi2Sn2O7} supercells, the maximum number of atoms per supercell was left unconstrained. Instead, memory exhaustion during the phonon evaluations was prevented by fixing the supercell size to $2 \times 2 \times 2$ for phonon dispersion calculation and lowering the maximum allowed volume expansion in the relaxation of distorted structures to 1.5.

\begin{table}[H]
    \centering
    \caption{Key algorithmic parameters used for the dynamical instability remediation workflows. To prevent memory exhaustion during the MLIP phonon evaluations of the large \ce{Bi2Sn2O7} system (88 atoms/cell), its settings for phonon computation, maximum volume expansion, and iteration thresholds were specifically constrained. All other global search parameters remained consistent across all compounds.}
    \label{tab:sets_remedalgorithms}
    \resizebox{\textwidth}{!}{%
    \begin{tabular}{@{}lcccc@{}}
        \toprule
        \textbf{Parameter} & \textbf{GA} & \textbf{TRAD} & \textbf{ALL} & \textbf{RAND} \\
        \midrule
        \multicolumn{5}{@{}l}{\textbf{Global Search Parameters}} \\[0.5ex]
        Displacement Scales & [0.25, 1.0, 4.0] & [0.25, 1.0, 4.0] & [0.25, 1.0, 4.0] & N/A \\
        Cell Scale Factors & [-0.07, 0.0, 0.07] & [-0.07, 0.0, 0.07] & [-0.07, 0.0, 0.07] & N/A \\
        Mode 2 Ratio Scales & [-0.25, 0.0, 0.25] & N/A & [-0.25, 0.0, 0.25] & N/A \\
        Population Size & 100 & N/A & N/A & N/A \\
        Generations & 3 & N/A & N/A & N/A \\
        New Points per Iteration & 50 & N/A & N/A & N/A \\
        Max Atoms per Supercell & None & None & None & None \\
        \midrule
        \multicolumn{5}{@{}l}{\textbf{Standard Constraints (\ce{LiF}, \ce{CsPbI3}, \ce{HfO2})}} \\[0.5ex]
        Supercell, $\delta$, $F_{max}$ & Auto\textsuperscript{a} & Auto\textsuperscript{a} & Auto\textsuperscript{a} & Auto\textsuperscript{a} \\
        Max Soft Mode Iterations & 5 & 5 & 5 & 5 \\
        Max Volume Expansion & 2.5 & 2.5 & 2.5 & 2.5 \\
        \midrule
        \multicolumn{5}{@{}l}{\textbf{\ce{Bi2Sn2O7} Specific Constraints}} \\[0.5ex]
        Supercell Selection & 2$\times$2$\times$2 & 2$\times$2$\times$2 & 2$\times$2$\times$2 & 2$\times$2$\times$2 \\
        $\delta$ (step size, \AA) & 0.03 & 0.03 & 0.03 & 0.03 \\
        $F_{max}$ tolerance (eV/\AA) & 0.001 & 0.001 & 0.001 & 0.001 \\
        Max Soft Mode Iterations & 4 & 4 & 4 & 4 \\
        Max Volume Expansion & 1.5 & 1.5 & 1.5 & 1.5 \\
        \bottomrule
        \multicolumn{5}{@{}p{\textwidth}@{}}{\footnotesize \textsuperscript{a} The \texttt{--auto} setting performs a sequential parameter sweep to find the optimal relaxation and phonon evaluation settings. It tests sets of parameters simultaneously, defaulting to [2$\times$2$\times$2, $\delta=0.05$, $F_{max}=0.001$], then [3$\times$3$\times$3, $\delta=0.03$, $F_{max}=0.0005$], and [4$\times$4$\times$4, $\delta=0.01$, $F_{max}=0.0001$], terminating early if the improvement in the negative frequency drops below half of the specified threshold.}
    \end{tabular}%
    }
\end{table}

To validate the effectiveness and computational cost of the implemented remediation algorithms, we tracked the actual execution times across the four benchmarked compounds. The results, executed sequentially on a single NVIDIA A100 (40GB) GPU, are summarized in Table~\ref{tab:execution_times}. It is important to note that because VibroML's algorithms evaluate batches of independent candidate structures at each generation, the workflows are inherently parallelizable. While the current benchmarks reflect single-GPU sequential execution, deploying VibroML across multi-GPU architectures would yield near-linear speedups, reducing the polymorph mapping time substantially.

\begin{table}[H]
    \centering
    \caption{Actual execution times (in hours) for the dynamical instability remediation workflows across the benchmarked compounds. All calculations were performed sequentially on a single NVIDIA A100 (40GB) GPU.}
    \label{tab:execution_times}
    \renewcommand{\arraystretch}{1.2}
    \begin{tabular}{@{}lcccc@{}}
        \toprule
        \textbf{Compound} & \textbf{GA} & \textbf{TRAD} & \textbf{ALL} & \textbf{RAND} \\
        \textbf{(\# atoms)} & (hours) & (hours) & (hours) & (hours) \\
        \midrule
        \textbf{\ce{LiF} (2)} & 1.03 & 0.03 & 0.08 & 0.06 \\
        \textbf{\ce{CsPbI3} (5)} & 0.25 & 0.65 & 1.17 & 2.46 \\
        \textbf{\ce{HfO2} (12)} & 1.39 & 0.18 & 0.43 & 3.97 \\
        \textbf{\ce{Bi2Sn2O7} (88)} & 19.73 & 63.31 & 4.16 & 1.93 \\
        \bottomrule
    \end{tabular}
\end{table}

The execution time of 63.31 hours of the \texttt{TRAD} algorithm on \ce{Bi2Sn2O7} is driven by two compounding factors: the inherent computational scaling of massive supercells and structural relaxation stagnation. In highly distorted, low-symmetry configurations, MLIP potential energy surfaces often exhibit flat gradients or unphysical local features. Standard optimizers such as BFGS struggle to navigate these regions, typically requiring hundreds of micro-steps to achieve force convergence ($F_{max}$). Multiplying this large number of iterative steps by the computational cost of evaluating a massive 1408-atom cell which was found during the optimization leads to the extended execution times observed.

To mitigate these struggling relaxations and prevent the computational pipeline from hanging endlessly on non-viable configurations, VibroML features a custom, MLIP-aware relaxation module. Rather than relying solely on force convergence, VibroML implements a dynamic \texttt{EnergyVolumeStopper}. This module continuously monitors the thermodynamic state during optimization and forcefully terminates the relaxation if it detects unphysical behavior, such as a drastic energy drop (e.g., $< -5.0$~eV/atom relative to the initial state) or a volume expansion exceeding a user-defined safety threshold (e.g., $>1.5\times$ the initial volume). To balance this strict filtering with the initially steep gradients of displaced supercells, a ``relaxation patience'' parameter enforces a minimum number of iterations before these termination criteria are applied. 

While these integrated heuristics successfully prevent infinite optimization loops, the performance of the \texttt{TRAD} method on large systems demonstrates that further adaptations to the optimization module may be necessary to fully tame the relaxation of massive, low-symmetry unit cells. Ultimately, this underscores the superiority of the Genetic Algorithm (\texttt{GA}), which naturally avoids these pathological regions of the PES by selecting for smaller, thermodynamically competitive cells, drastically reducing the overall compute time (19.73 hours for \ce{Bi2Sn2O7}).

\newpage
\subsection{Dynamical and Thermal Stability Validation of Alloyed Candidates}
\label{sec:final_validation}

To confirm that the optimal alloyed compositions identified by the ProtoCSP-VibroML pipeline successfully relieve the geometric frustration of their unalloyed parent structures, we subjected the final predicted ground states---\ce{Cs2K_{0.9}Na_{0.1}InI3Br3} and \ce{K_{0.6}Ba_{0.4}Ta_{0.6}Zr_{0.4}Se_{2.1}O_{0.9}}---to dynamical and thermal stability assessments.

\begin{figure}[H]
    \centering
    \includegraphics[width=\textwidth]{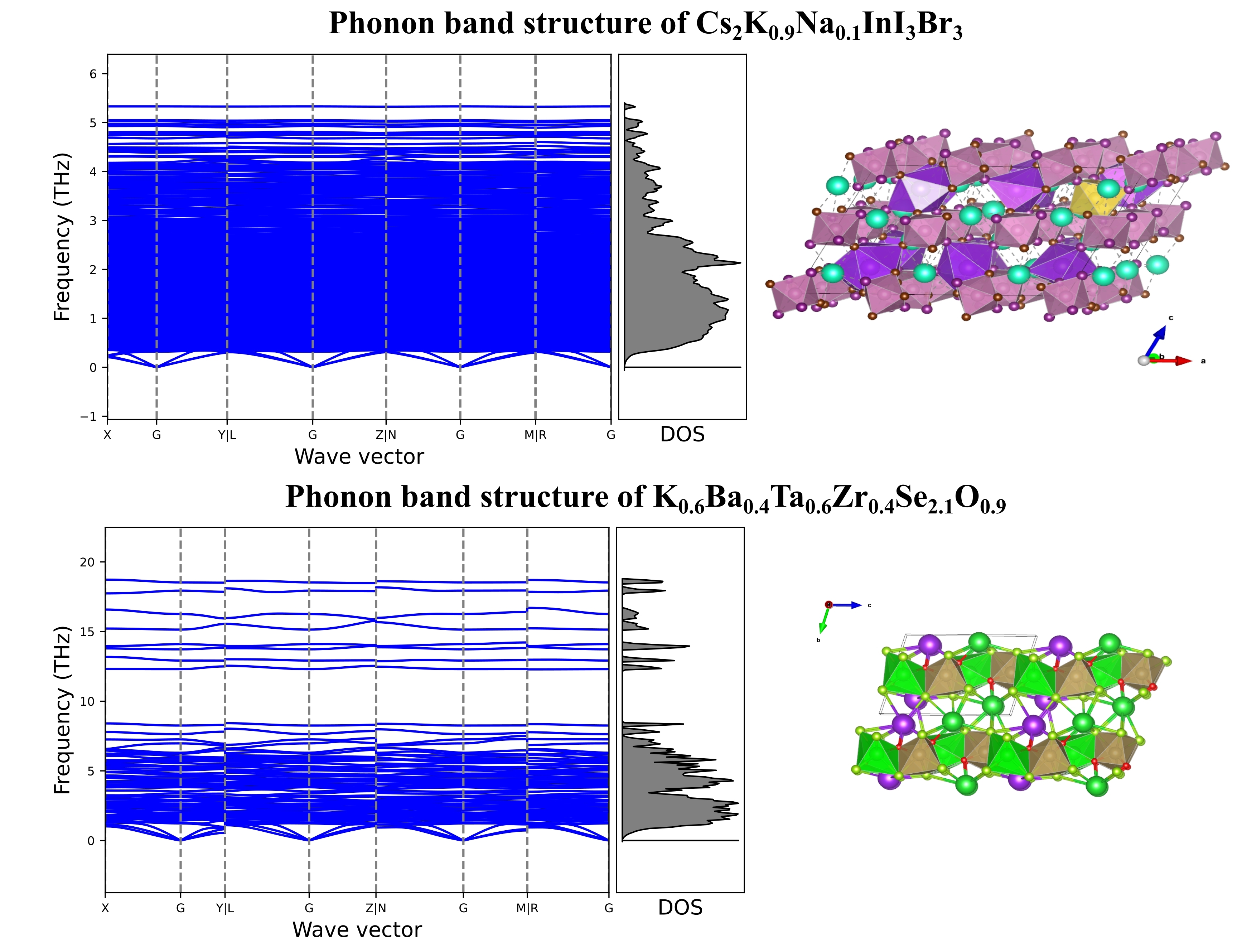}
    \caption{Phonon band structures and vibrational density of states (DOS) for the optimized alloyed candidates: the 3D double perovskite \ce{Cs2K_{0.9}Na_{0.1}InI3Br3} (top) and the 2D layered network \ce{K_{0.6}Ba_{0.4}Ta_{0.6}Zr_{0.4}Se_{2.1}O_{0.9}} (bottom). The absence of imaginary frequencies below the zero-axis confirms the successful remediation of the dynamical instabilities present in the baseline unalloyed compounds.}
    \label{fig:sup_finalperovsk_phononspectra}
\end{figure}

First, the harmonic phonon dispersions were evaluated using the MACE-OMAT-0 MLIP (Figure~\ref{fig:sup_finalperovsk_phononspectra}). In stark contrast to the pure \ce{Cs2KInI6} and \ce{KTaSe3} baseline structures (which exhibited deeply imaginary optical modes exceeding $-0.7$~THz and $-2.3$~THz, respectively), both optimized alloy networks display completely real phonon frequencies across the entire Brillouin zone, confirming their dynamical stability at 0~K with the foundational model.

\begin{figure}[H]
    \centering
    \includegraphics[width=\textwidth]{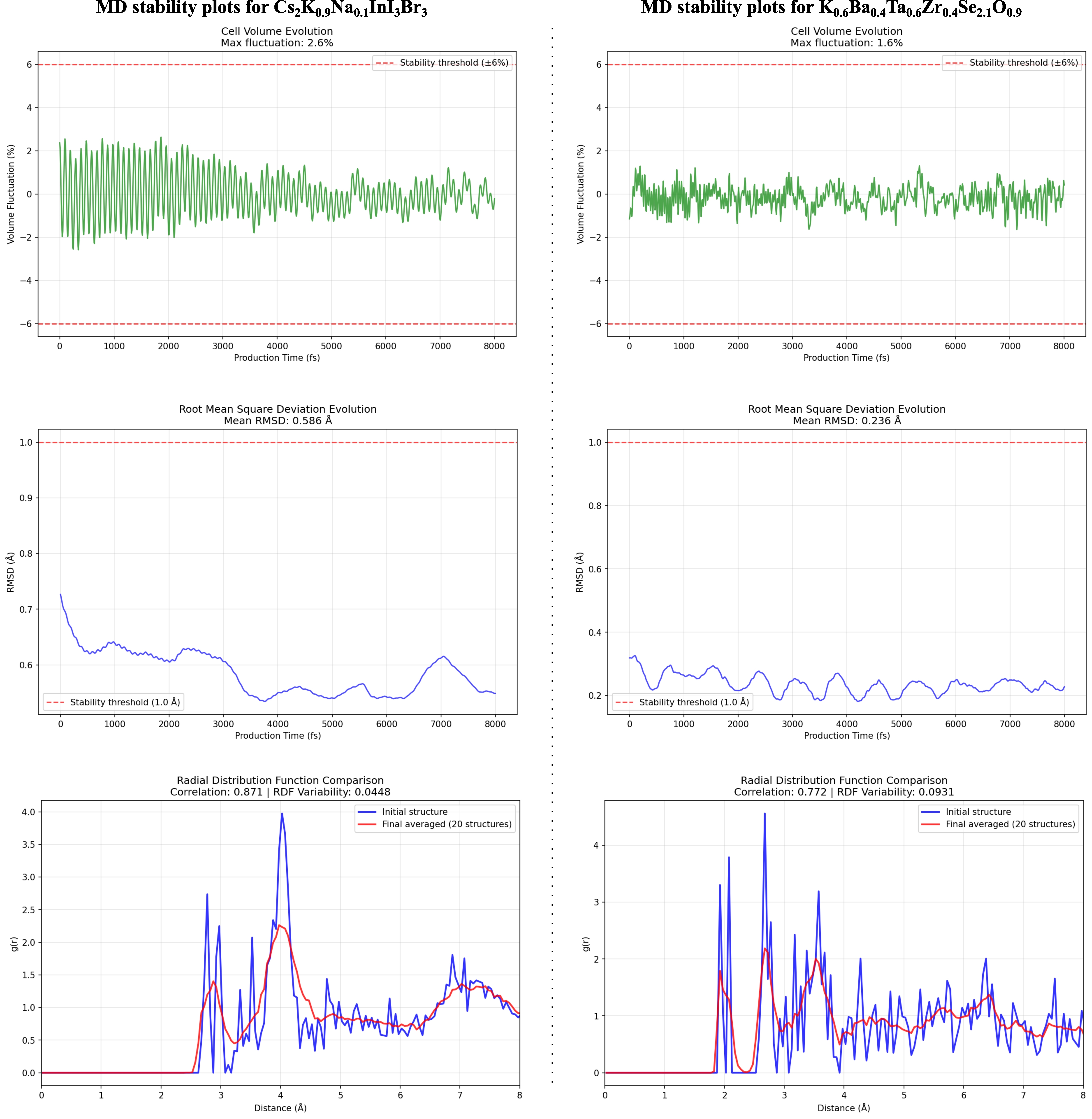}
    \caption{Molecular dynamics (MD) stability analysis at 300~K for the alloyed candidates \ce{Cs2K_{0.9}Na_{0.1}InI3Br3} (left column) and \ce{K_{0.6}Ba_{0.4}Ta_{0.6}Zr_{0.4}Se_{2.1}O_{0.9}} (right column). The panels display the time evolution of the unit cell volume fluctuation, the root-mean-square deviation (RMSD) of atomic positions, and the initial versus final averaged radial distribution functions (RDF). Both materials remain well within the designated stability thresholds (dashed red lines), demonstrating robust thermal resilience and structural retention.}
    \label{fig:sup_finalperovsk_md}
\end{figure}

Second, we evaluated their resilience to thermal fluctuations by performing 10~ps molecular dynamics (NPT ensemble, 300~K) simulations with the same model. The stability metrics (Figure~\ref{fig:sup_finalperovsk_md}) suggest that both structures are robust under ambient conditions. The 3D elpasolite network of \ce{Cs2K_{0.9}Na_{0.1}InI3Br3} maintained an average RMSD of just 0.586~\AA\ and a maximum volume fluctuation of 2.6\%. Similarly, the 2D layered motif of \ce{K_{0.6}Ba_{0.4}Ta_{0.6}Zr_{0.4}Se_{2.1}O_{0.9}} proved exceptionally rigid, yielding a mean RMSD of 0.236~\AA\ and a maximum volume fluctuation of 1.6\%. Both configurations successfully retained their initial space group symmetries ($P1$ primitive cell representations) across the production trajectory and exhibited high RDF correlations with their 0~K reference structures, ruling out spontaneous amorphization or dimensional collapse.

\subsection{Comprehensive White Space Remediation Data}
\label{sec:whitespace_si}

Table~\ref{tab:full_whitespace_data} presents the complete dataset for the 50 sparsely populated and absolute ``white space'' multi-component systems sampled from the Alexandria database and the LeMat-Bulk ProtoCSP generation workflow.

\begin{longtable}{@{}l l c c c c c@{}}
\caption{Complete structural and dynamical remediation data for all sampled multi-component systems.}
\label{tab:full_whitespace_data} \\
\toprule
\textbf{Category} & \textbf{Composition} & \textbf{Unique} & \textbf{$\Delta E$} & \textbf{$\omega_{\text{init}}$} & \textbf{$\zeta_{\text{init}}$} & \textbf{$\omega_{\text{final}}$} \\
& & \textbf{Phases} & \textbf{(meV/at)} & \textbf{(THz)} & \textbf{(\%)} & \textbf{(THz)} \\
\midrule
\endfirsthead

\multicolumn{7}{c}%
{{\bfseries \tablename\ \thetable{} -- continued from previous page}} \\
\toprule
\textbf{Category} & \textbf{Composition} & \textbf{Unique} & \textbf{$\Delta E$} & \textbf{$\omega_{\text{init}}$} & \textbf{$\zeta_{\text{init}}$} & \textbf{$\omega_{\text{final}}$} \\
& & \textbf{Phases} & \textbf{(meV/at)} & \textbf{(THz)} & \textbf{(\%)} & \textbf{(THz)} \\
\midrule
\endhead

\midrule
\multicolumn{7}{r}{{Continued on next page}} \\
\endfoot

\bottomrule
\endlastfoot

Alexandria (4-element) & \ce{Al12Sn4AsH3} & 74 & -88.6 & -16.19 & 12.58 & 0.00 \\
Alexandria (4-element) & \ce{Cs2AlRuI6} & 103 & -159.2 & -0.30 & 0.73 & -0.08 \\
Alexandria (4-element) & \ce{Cs2CaZrBr6} & - & - & 0.00 & 0.20 & - \\
Alexandria (4-element) & \ce{Cs2CaZrI6} & 20 & -57.8 & -0.18 & 1.47 & -0.01 \\
Alexandria (4-element) & \ce{Cs2FeSiI6} & 72 & -121.1 & -2.63 & 9.20 & -0.19 \\
Alexandria (4-element) & \ce{Cs2MgGeF6} & 84 & -78.9 & -3.97 & 8.30 & -0.80 \\
Alexandria (4-element) & \ce{Cs2MoHI6} & 106 & -221.2 & -10.33 & 10.20 & 0.00 \\
Alexandria (4-element) & \ce{Cs2MoWF6} & 8 & -20.9 & -2.00 & 7.80 & -0.10 \\
Alexandria (4-element) & \ce{Cs2NbCoI6} & 75 & -184.9 & -1.27 & 12.03 & 0.00 \\
Alexandria (4-element) & \ce{Cs2NbGeBr6} & - & - & 0.00 & 0.13 & - \\
Alexandria (4-element) & \ce{Cs2RbSiI6} & 81 & -111.0 & -0.74 & 18.70 & -0.05 \\
Alexandria (4-element) & \ce{Cs2TaAlI6} & 70 & -136.7 & -0.35 & 4.10 & 0.00 \\
Alexandria (4-element) & \ce{Cs2TiMoCl6} & - & - & 0.00 & 0.13 & - \\
Alexandria (4-element) & \ce{Cs2TiMoI6} & 77 & -208.0 & -0.51 & 2.93 & -0.02 \\
Alexandria (4-element) & \ce{Cs2ZnMoBr6} & - & - & 0.00 & 0.20 & - \\
Alexandria (4-element) & \ce{Cs2ZrGaF6} & - & - & -2.02 & 20.87 & - \\
Alexandria (4-element) & \ce{K2RbYH6} & 48 & -114.7 & -11.22 & 33.77 & 0.00 \\
Alexandria (4-element) & \ce{KRb2MoH6} & - & - & 0.00 & 0.20 & - \\
Alexandria (4-element) & \ce{Rb2BaNbI6} & 83 & -129.7 & -0.79 & 21.97 & 0.00 \\
Alexandria (4-element) & \ce{Rb2BaZrI6} & 104 & -119.4 & -0.86 & 27.87 & 0.00 \\
Alexandria (4-element) & \ce{Rb2CaNiI6} & 50 & -56.5 & -0.76 & 6.53 & -0.15 \\
Alexandria (4-element) & \ce{Rb2CoMoI6} & 99 & -132.8 & -1.06 & 12.37 & 0.00 \\
Alexandria (4-element) & \ce{Rb2CoSnI6} & 67 & -78.3 & -0.78 & 6.83 & 0.00 \\
Alexandria (4-element) & \ce{Rb2NaSiI6} & 47 & -65.4 & -0.74 & 12.90 & -0.64 \\
Alexandria (4-element) & \ce{Rb2SiNiI6} & 82 & -120.8 & -2.01 & 11.87 & -0.02 \\
Alexandria (4-element) & \ce{Rb2SrTiI6} & 32 & -45.8 & -0.78 & 10.23 & -0.05 \\
Alexandria (4-element) & \ce{Cs3K3Rb2Na} & 36 & -24.3 & -0.59 & 10.74 & -0.04 \\
Alexandria (4-element) & \ce{Sr2Ca2Ti3MnO12} & 4 & -13.6 & -4.32 & 4.10 & 0.00 \\

\midrule
Alexandria (5-element) & \ce{BaSrVPO6} & 114 & -264.3 & -0.42 & 0.43 & -0.01 \\
Alexandria (5-element) & \ce{BaYScGeO6} & 81 & -138.9 & -7.36 & 17.67 & 0.00 \\
Alexandria (5-element) & \ce{BaYTaNbO6} & 133 & -260.5 & -6.05 & 19.07 & 0.00 \\
Alexandria (5-element) & \ce{Cs2FeCo(CN)6} & - & - & 0.00 & 0.12 & - \\
Alexandria (5-element) & \ce{Cs2LiMn(CN)6} & - & - & 0.00 & 0.12 & - \\
Alexandria (5-element) & \ce{Cs2MgCo(CN)6} & - & - & 0.00 & 0.12 & - \\
Alexandria (5-element) & \ce{K2MnNi(CN)6} & 18 & -67.7 & -0.80 & 5.77 & -0.26 \\
Alexandria (5-element) & \ce{NaCaZrScO6} & 124 & -207.9 & -5.72 & 21.40 & -1.79 \\
Alexandria (5-element) & \ce{NaFeH8(NF3)2} & 18 & -26.9 & -9.72 & 11.65 & 0.00 \\
Alexandria (5-element) & \ce{Rb2CdNi(CN)6} & 3 & -3.2 & -2.02 & 3.38 & 0.00 \\
Alexandria (5-element) & \ce{Rb2FeCo(CN)6} & - & - & 0.00 & 0.08 & - \\
Alexandria (5-element) & \ce{Rb2FeCu(CN)6} & 25 & -39.5 & -2.12 & 5.90 & 0.00 \\
Alexandria (5-element) & \ce{Rb2MnCr(CN)6} & 3 & -1.4 & -0.44 & 0.75 & 0.00 \\
Alexandria (5-element) & \ce{Rb2MnNi(CN)6} & 14 & -74.7 & -0.23 & 0.52 & 0.00 \\

\midrule
White Space (Cubic) & \ce{Cs2GaNbF6} & - & - & 0.00 & 0.02 & - \\
White Space (Cubic) & \ce{K2BiSbF6} & 83 & -332.1 & -3.68 & 40.02 & -1.09 \\
White Space (Cubic) & \ce{K2ZnFeI6} & 82 & -167.7 & -1.16 & 14.47 & 0.00 \\
White Space (Cubic) & \ce{Li2CdFeBr6} & 54 & -300.3 & -5.65 & 38.54 & -0.30 \\
White Space (Cubic) & \ce{Li2CoZnI6} & 76 & -380.1 & -5.11 & 38.81 & 0.00 \\
White Space (Cubic) & \ce{Li2CuCrCl6} & 86 & -367.6 & -5.68 & 32.98 & -0.40 \\
White Space (Cubic) & \ce{Li2PdZnCl6} & 78 & -335.9 & -5.65 & 35.97 & -0.62 \\
White Space (Cubic) & \ce{Na2BiNbI6} & 100 & -326.3 & -2.38 & 39.88 & -0.27 \\
White Space (Cubic) & \ce{Na2CoCrBr6} & 92 & -138.4 & -2.20 & 31.03 & 0.00 \\
White Space (Cubic) & \ce{Na2CrFeCl6} & 66 & -130.0 & -2.83 & 29.79 & 0.00 \\

\midrule
White Space (Transmuted) & \ce{Cs2GaNbF6} & - & - & 0.00 & 0.20 & - \\
White Space (Transmuted) & \ce{K2BiSbF6} & 107 & -166.0 & -0.33 & 1.33 & -0.85 \\
White Space (Transmuted) & \ce{K2ZnFeI6} & - & - & -0.06 & 0.38 & - \\
White Space (Transmuted) & \ce{Li2CdFeBr6} & - & - & -0.08 & 0.20 & - \\
White Space (Transmuted) & \ce{Li2CoZnI6} & - & - & -0.04 & 0.53 & - \\
White Space (Transmuted) & \ce{Li2CuCrCl6} & 54 & -53.9 & -2.30 & 3.37 & -0.45 \\
White Space (Transmuted) & \ce{Li2PdZnCl6} & 13 & -12.1 & -0.67 & 1.40 & -0.25 \\
White Space (Transmuted) & \ce{Na2BiNbI6} & - & - & 0.00 & 0.13 & - \\
White Space (Transmuted) & \ce{Na2CoCrBr6} & - & - & 0.00 & 0.05 & - \\
White Space (Transmuted) & \ce{Na2CrFeCl6} & 38 & -35.9 & -0.13 & 0.50 & 0.00 \\

\end{longtable}

\end{document}